\documentclass[]{aa}
\usepackage{natbib}
\bibpunct{(}{)}{;}{a}{}{,} % to follow the A&A style
\usepackage{amsmath}
\usepackage{wasysym}
\usepackage{marvosym}
\usepackage{txfonts}
\usepackage{mathtools}
\usepackage{comment}

\usepackage[pdftex]{graphicx}   
\usepackage{epstopdf}
\usepackage{graphicx}
\graphicspath{{figures/}} 
\pdfoutput=1
\usepackage[pdftex]{color,xcolor}
\usepackage[backref]{hyperref}    
\hypersetup{pdfauthor=Paula Sarkis}

\hypersetup{backref=true, pagebackref=true, hyperindex=true, breaklinks=true,colorlinks=true,urlcolor=blue, linkcolor=blue,  citecolor=blue,pagecolor=red, bookmarks=true, bookmarksopen=true}

\defcitealias{Thorngren:2018}{TF18}
\defcitealias{Thorngren:2019}{T19}

\def\apjl{ApJ}                % Astrophysical Journal, Letters
\def\apjs{ApJS}               % Astrophysical Journal, Supplement
\def\ao{Appl.Optics}          % Applied Optics
\def\aap{A\&A}                % Astronomy and Astrophysics
\def\({\left(}
\def\){\right)}
\def\<{\left<}
\def\>{\right>}

\label{new commands}
\newcommand{\q}[1]{``#1''}
\newcommand{\mrm}[1]{\ensuremath{\mathrm{#1}}}

\newcommand{\rjup}{\ensuremath{R_{\rm J}}}
\newcommand{\mjup}{\ensuremath{M_{\rm J}}}
\newcommand{\mearth}{\ensuremath{M_\oplus}}

\newcommand{\ljup}{\ensuremath{L_{\rm J}}}

\newcommand{\msun}{\ensuremath{M_\odot}}

\newcommand{\rp}{\ensuremath{R_{\mrm{p}}}}
\renewcommand{\mp}{\ensuremath{M_{\rm p}}}
\newcommand{\teq}{\ensuremath{T_{\rm eq}}}
\newcommand{\tint}{\ensuremath{T_{\rm int}}}

\newcommand{\eps}{\ensuremath{\epsilon}}
\newcommand{\epsmax}{\ensuremath{\epsilon_{\rm max}}}
\newcommand{\logg}{\ensuremath{\log \, g}}

\newcommand{\zp}{\ensuremath{Z_{\mrm{p}}}}

\newcommand{\lint}{\ensuremath{L_{\mrm{int}}}}

\newcommand{\prcb}{\ensuremath{P_{\rm RCB}}}

\newcommand{\tstar}{\ensuremath{T_*}}

\newcommand{\rstar}{\ensuremath{R_{\rm *}}}
\newcommand{\lstar}{\ensuremath{L_{\rm *}}}

\newcommand{\petit}{\texttt{petitCODE}}
\newcommand{\completo}{\texttt{completo21}}
\newcommand{\emcee}{\texttt{emcee}}

\newcommand{\bold}[1]{\ensuremath{\boldsymbol{#1}}}

\newcommand{\wn}{{\ensuremath{\bold{\omega_n}}}}

\newcommand{\datan}{{\ensuremath{\bold{D_n}}}}

\newcommand{\rtheon}{\ensuremath{R_{\mrm{t,n}}}}
\newcommand{\taw}{{\ensuremath{\bold{\tau}}}}

\newcommand{\rpn}{\ensuremath{R_{\mrm{p,n}}}}

\newcommand{\sigmarpn}{\ensuremath{\sigma_{R_{\mrm{p,n}}}}}

\newcommand{\mpn}{\ensuremath{M_{\mrm{p,n}}}}
\newcommand{\mptruen}{\ensuremath{M_{\mrm{pt,n}}}}
\newcommand{\sigmampn}{\ensuremath{\sigma_{M_{\mrm{p,n}}}}}

\newcommand{\zpn}{\ensuremath{Z_{\mrm{p,n}}}}
\newcommand{\zpps}{\ensuremath{Z_{\mrm{ps,n}}}}

\newcommand{\lstarn}{\ensuremath{L_{\mrm{*,n}}}}
\newcommand{\lstartruen}{\ensuremath{L_{\mrm{*t,n}}}}
\newcommand{\sigmalstarn}{\ensuremath{\sigma_{L_{\mrm{*,n}}}}}

\newcommand{\lintn}{\ensuremath{L_{\mrm{int,n}}}}
\newcommand{\firrn}{\ensuremath{F_{\mrm{p,n}}}}

\newcommand{\lintnk}{\ensuremath{L_{\mrm{int}, \, nk}}}
\newcommand{\epsnk}{\ensuremath{\eps_{nk}}}
\newcommand{\tintnk}{\ensuremath{T_{{\rm int}, \, nk}}}
\newcommand{\prcbnk}{\ensuremath{P_{{\rm RCB}, \, nk}}}
\newcommand{\rpnk}{\ensuremath{R_{{\mrm{p}, \, nk}}}}
\newcommand{\teqnk}{\ensuremath{T_{{\mrm{eq}, \, nk}}}}

\newcommand{\tinyurl}{https://tinyurl.com/bloated-hjs-results}

\begin{document}

\title{Evidence of Three Mechanisms Explaining the Radius Anomaly of Hot Jupiters}

\subtitle{}
\author{
	P. Sarkis \inst{1,2}
	\and
	C.~Mordasini \inst{1}
	\and
	Th.~Henning \inst{2}
	\and
	G.~D.~Marleau \inst{3, 1, 2}
	\and
	P.~Molli\`ere \inst{2}
	}
\institute{
    Physikalisches Institut, Universit\"at Bern, Gesellschaftsstrasse 6, CH-3012 Bern, Switzerland
    \and %2
    Max-Planck-Institut f\"ur Astronomie, K\"onigstuhl 17, Heidelberg 69117, Germany
    \and %3
    Institut f\"ur Astronomie und Astrophysik, Universit\"at T\"ubingen, Auf der Morgenstelle 10, D-72076 T\"ubingen, Germany
    }
\offprints{Paula Sarkis, \email{paula.sarkis@space.unibe.ch}}
\date{Received May 6, 2020/ Accepted ---}

\label{abstract}
\abstract
%Context
{The anomalously large radii of hot Jupiters are still not fully understood, 
and all of the proposed explanations are based on the idea that these close-in giant planets possess hot interiors.
Most of the mechanisms proposed have been tested on a handful of exoplanets.}
%Aims
{We approach the radius anomaly problem by adopting a statistical approach.
We want to infer the internal luminosity for the sample of hot Jupiters,
study its effect on the interior structure, and put constraints on which mechanism
is the dominant one.}
%Methods
{We develop a flexible and robust hierarchical Bayesian model 
that couples the interior structure of exoplanets
to the observed properties of close-in giant planets. 
We apply the model to 314 hot Jupiters
and infer the internal luminosity distribution for each planet
and study at the population level
({\it i}) the mass--luminosity--radius distribution
and as a function of equilibrium temperature 
the distributions of the 
({\it ii}) heating efficiency,
({\it iii}) internal temperature,
and the 
({\it iv}) pressure of the radiative--convective--boundary (RCB).}
%Results
{We find that hot Jupiters tend to have high internal luminosity
with $10^4 \, \ljup$ for the largest planets.
As a result,
we show that all the inflated planets have hot interiors
with internal temperature ranging from 200~K up to 800~K 
for the most irradiated ones. 
This has important consequences on the cooling rate 
and we find that the RCB is located at low pressures between 3 and 100~bar.
Assuming that the ultimate source of the extra heating is the irradiation from the host star,
we also illustrate that the heating efficiency increases 
with increasing equilibrium temperature, 
reaches a maximum of 2.5\% at $\sim$1860~K,
beyond which the efficiency decreases,
in agreement with previous results.
We discuss our findings in the context of the proposed heating mechanisms
and illustrate that ohmic dissipation, advection of potential temperature,
and thermal tides are in agreement with certain trends inferred from our analysis
and thus all three models can explain aspects of the observations.}
%Conclusions
{We provide new insights on the interior structure of hot Jupiters
and show that with our current knowledge it is still challenging
to firmly identify the universal mechanism
driving the inflated radii.}


\titlerunning{Inflated Hot Jupiters}
\authorrunning{P. Sarkis}

\maketitle

\section{Introduction}

Two decades of observational and theoretical exploration
have revealed that the anomalously large radii of close-in 
transiting giant planets holds firmly
\citep[e.g.][]{Laughlin:2011, Weiss:2013}.
The radii of hot Jupiters are larger than what is predicted
by standard interior structure models
\citep{Guillot:2002}.
Observations reveal that there is a strong correlation between the
observed radii and the stellar incident flux
\citep[e.g.][]{Enoch:2012},
with a threshold around $\sim 2 \times 10^8 \, \rm{erg \, s^{-1} \, cm^{-2}} $,
corresponding to an equilibrium temperature of about  $1000 \, \rm{K}$
\citep{Demory:2011, Miller:2011},
below which the physical mechanism becomes inefficient.
\cite{Sestovic:2018} further demonstrated that 
the inflation extent is mass dependent,
where the planets with the largest anomalous radii
have masses less than $\sim < 1 \, \mjup$.

There has been a lot of investigations to explain
the discrepancy between the observations and theoretical models.  
The proposed mechanisms can be divided into two categories:
({\it i}) slowing down cooling and contraction
or  
({\it ii}) depositing extra heat into the interior.
\cite{Burrows:2007} showed that
slowing down the cooling and thus delaying contraction 
can be achieved by increasing the atmospheric opacity.
Another way to delay contraction is to 
reduce the heat transport efficiency due to compositional gradients
\citep{Chabrier:2007}.

It is well established that heating up the interior of the planet
increases its entropy and thus its radius  \citep{Arras:2006, Marleau:2014}. 
The source of heat is still not constrained and possible sources could be
tidal dissipation of an eccentric orbit \citep[e.g.][]{Bodenheimer:2001},
advection of potential temperature, which is  a consequence of the strong stellar irradiation
\citep{Tremblin:2017, Sainsbury-Martinez:2019},
or dissipative processes powered by the stellar irradiation flux.
The latter has received a lot of attention 
and the mechanism to transport fraction of the stellar incident flux into the interior 
is still an open question. 
One mechanism is atmospheric circulation,
which leads to thermal dissipation of kinetic energy into the interior
\citep{Guillot:2002, Showman:2002}.
Another mechanism is ohmic dissipation
\citep{Batygin:2010, Batygin:2011, Perna:2010a, Huang:2012,
Wu:2013, Ginzburg:2016},
where the irradiation drives fast winds 
through the planet’s magnetic fields,
giving rise to currents that dissipate ohmically in the interior.
Other mechanisms are thermal tides \citep{Arras:2010}
and the mechanical greenhouse \citep{Youdin:2010}.

Some of these mechanisms come with a lot of approximations and uncertainties.
For example, an important uncertain parameter
in atmospheric circulation, ohmic dissipation, 
and advection of potential temperature
is the wind speeds and the effect of magnetic drag
in damping the winds \citep{Perna:2010a, Perna:2010b}.
Another uncertainty is how deep the wind zone extends,
which is important to constrain the pressures
at which the extra heat should be dissipated. 
\cite{Wu:2013} illustrate that if the wind zone is at shallow pressures, 
then a significantly larger 
heating efficiency is needed to achieve the same interior heating,
compared to heating at deeper pressures.
\cite{Komacek:2017b} argued that the extra heat 
should be deposited in the convective layers
or at the radiative--convective--boundary (RCB),
otherwise it will be re-radiated away. 
\cite{Huang:2012} deposited the extra heat in the radiative layers
and as a consequence showed that the RCB moves to deeper pressures.
\cite{Fortney:2007} showed that RCB is located at pressures of 1000~bar,
where little is known about the wind speeds at such deep pressures.
However, the \cite{Fortney:2007} models were developed for irradiated planets 
and do not consider the high internal entropy
that hot Jupiters are believed to possess.

All the mechanisms proposed have been 
tested and applied on single or a handful of planets.
It is yet to be demonstrated that these mechanisms 
can explain the radii of all the observed hot Jupiters. 
Within this context, in this paper we approach 
the radius inflation problem from a statistical point of view,
similar to the approach of \cite{Thorngren:2018}
(hereafter \citetalias{Thorngren:2018}).
We do not model any of the previously mentioned mechanisms
but rely solely on the interior structure model
and atmospheric model.
We develop a hierarchical Bayesian model 
that allows us to couple the interior structure models
to the observed physical properties of hot Jupiters
while incorporating the measurement uncertainties.
Our approach naturally accounts for non-Gaussian likelihoods.
We first apply our model on the individual planets 
to infer the internal luminosity that reproduces the observed physical 
properties of hot Jupiters, namely 
radius, mass, and equilibrium temperature.
Second, as a consequence of the high internal entropy,
we find that the interior tends to be hot and show that the RCB moves to shallow pressures.
Finally, we compare our findings to the proposed mechanisms
and show that ohmic dissipation \citep{Batygin:2010}, advection of potential temperature
\citep{Tremblin:2017},
and thermal tides \citep{Arras:2010} 
can explain the anomalously large radii of hot Jupiters.

In a recent study, \citetalias{Thorngren:2018} showed that 
the heating efficiency \eps\ increases as a function of equilibrium temperature
until a maximum of $\sim 2.5\%$ is reached at around 1500~K, 
beyond which it decreases.
The basic shape of $\eps(\teq)$ provides evidence for ohmic dissipation.
Building on the functional form of $\eps(\teq)$, 
\cite{Thorngren:2019}
(hereafter \citetalias{Thorngren:2019})
studied the effect of central heating 
on the interior structure of hot Jupiters
and found that the internal temperature is much hotter than previous estimates,
which pushes the RCB to lower pressures.
Our approach is similar to \citetalias{Thorngren:2018}
but rather than modeling the extra heating as a function of \eps,
we do not assume explicitly a source for the extra heat. 
Instead, we consider the planet reached steady state 
and compute the internal luminosity given the planet mass, radius,
and equilibrium temperature.
The advantage of this approach is twofold:
first, we can compare our results to heating mechanisms 
where the source of extra heat is not the stellar irradiation,
and second, 
we self-consistently study the effect of high internal entropy 
on the interior structure of hot Jupiters,
namely the internal temperature and pressure of the RCB.
We note, however, that both approaches should lead to the same results.
We also convert the internal luminosity to a heating efficiency \eps\
and compare our results to \citetalias{Thorngren:2018} in 
Section~\ref{sec:heet}.
We show that our results are qualitatively similar using a larger sample
focused on FGK main-sequence stars
and using an independent interior structure model.

The outline of this paper is as follows.
Section~\ref{sec:sample} provides an overview
of the sample selection criteria.
In Section~\ref{sec:interior-model}
we present the interior structure model used in this analysis.
In Section~\ref{sec:hbm} 
we outline the probabilistic framework used to link observations and theory
and derive the basic equation which our method is based on (Equation~(\ref{eq:upper-final})).
We validate the statistical model by applying it
on synthetic planetary data set generated
using the Generation III {\it Bern} global 
model in Section~\ref{sec:model-validation}.
Readers interested in the results can safely skip
to Section~\ref{sec:results} where we present the results of our analysis.
We discuss the results and the shortcomings of our approach in Section~\ref{sec:discussion}
and conclude in Section~\ref{sec:conclusion}.

\section{Sample Selection}
\label{sec:sample}

\begin{figure*}[t!]
	\includegraphics[width=\textwidth]{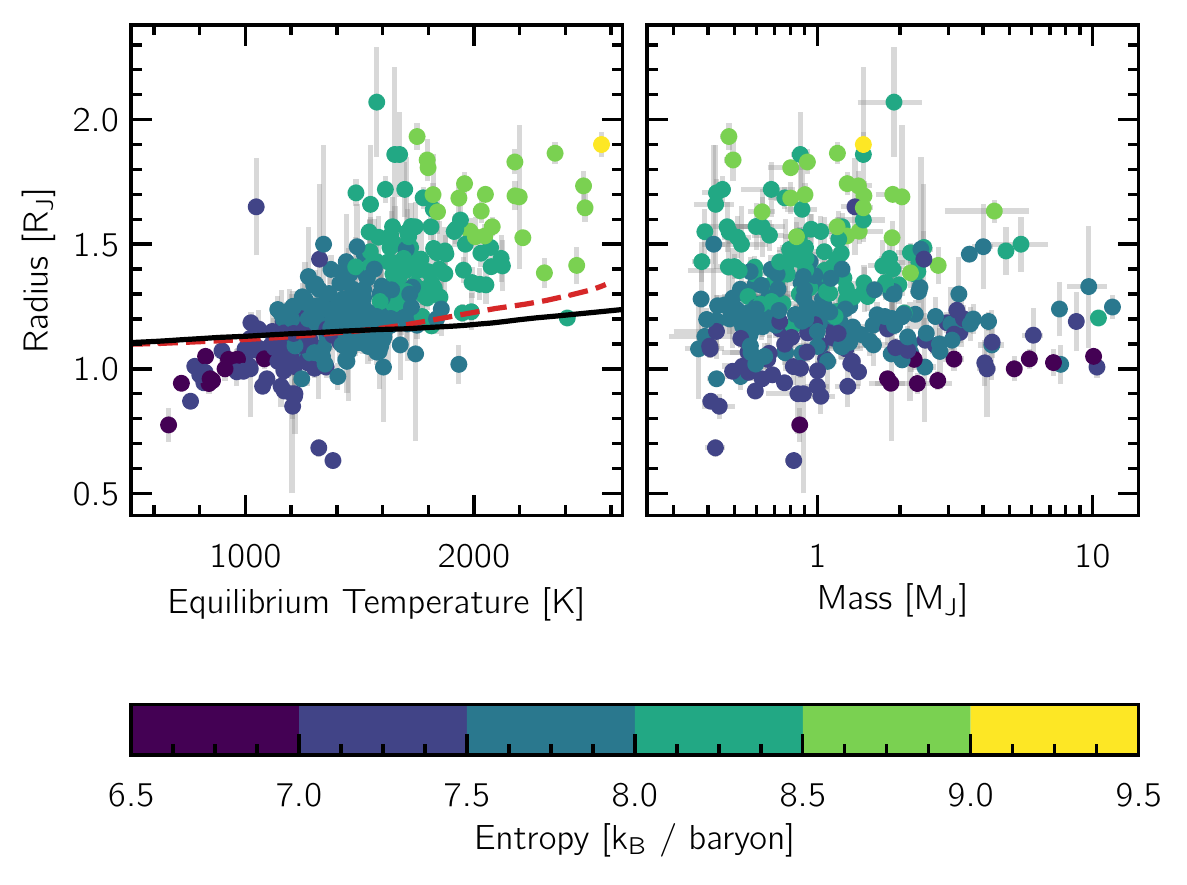}
	\caption{Equilibrium temperature--radius diagram ({\it left panel}) and 
		mass--radius diagram ({\it right panel}) color-coded by entropy
		for the 314 hot Jupiters selected for our analysis.
		The solid black and the red dashed lines 
		compare the radii computed 
		using our model \completo\ and \citetalias{Thorngren:2018}, respectively.
		Both models are for a 1 \mjup\ planet 
		with a pure H/He composition with $Y=0.27$ at 5~Gyr
		and without accounting for inflation.
		The entropy was computed using the observed physical properties
		and an assumed heavy-element fraction of 0.2.
		Planets with large radii tend to have high internal entropy,
		with a weak dependence on planetary mass. 
	\label{fig:sample}}
\end{figure*}

For the purpose of our study,  we required that 
all the planets have measured masses and radii.
\cite{Sestovic:2018} showed that 
the radii of planets with masses less than $0.37 \, \mjup$
do not show a clear dependence on the stellar incident flux. 
Photoevaporation plays an important role in 
the evolution of such low-mass close-in planets 
\citep{Owen:2012, Jin:2014}.
\cite{Baraffe:2004} also showed that these planets
are subject to undergo Roche-lobe overflow.
We therefore restrict our analysis
to planets with masses
$~0.37~<~Mp~<~13~\,~\mjup$
with semi-major axis $a~<~0.1$~au.
In our study, we make no attempt to correct for selection effects
where it is still challenging to detect 
\q{medium-inflated} hot Jupiters around F stars using ground-based surveys
(see the discussion in Section~\ref{sec:limitations}).

\cite{Lopez:2016} suggested that giant planets
around stars leaving the main-sequence 
experience a high level of irradiation
that could ultimately increase their radii.
However, other studies argued that 
ohmic heating cannot re-inflate planets after they have already cooled
\citep{Wu:2013, Ginzburg:2016}.
A handful of re-inflated planets have been discovered around giant stars 
\citep{Grunblatt:2016, Grunblatt:2017, Hartman:2016}.
Since different mechanisms can be at play around evolved stars, 
we exclude such planets and only consider 
hot Jupiters around solar-like stars.
Specifically, we consider stars with
stellar temperature 
$\tstar~=~4000-7000$ K
and surface gravity $\log~g~=~4~-~4.9$.

The data was taken from 
the Transiting Extrasolar Planet Catalogue
\citep[TEPCat\footnote{\url{www.astro.keele.ac.uk/jkt/tepcat/}};][]{Southworth:2011}, 
last accessed on November 2018.
The aforementioned constraints on the 
planet mass, semi-major axis, and stellar temperature and surface gravity,
lead to a final sample consisting of 314 hot Jupiters.
The equilibrium temperature (\teq) values in the literature are often not homogeneous,
where different teams use different assumptions for the albedo and heat redistribution.
To mitigate this, 
we compute the equilibrium temperature 
for all the planets assuming a circular orbit, zero albedo,
and full heat redistribution from the day-side to the night-side \citep{Guillot:2010}

\begin{align}
\teq = \tstar \sqrt{ \frac{\rstar}{2a} }
	\label{eq:teq}
\end{align}

\noindent 
where \tstar\ and \rstar\
are the stellar temperature and radius, respectively,
and $a$ is the semi-major axis.
Figure~\ref{fig:sample}
displays the selected targets in the equilibrium temperature--radius (left panel)
and mass--radius (right panel) diagrams color coded by the entropy\footnote{
When comparing to other work, 
it is crucial to use the same entropy zero-point or to correct for this. 
See Footnote~2 of \cite{Mordasini:2017}.}.
The entropy was calculated for all the planets given 
the observed physical properties of each system
and assuming the fraction of heavy element is 20\% the planet mass.
Note that this value was chosen arbitrarily and
for the rest of the results presented in this paper,
we use the
mass--heavy-element mass relation \citep[][see also Section~\ref{sec:bloating-model}]{Thorngren:2016}.
The solid black line is the radius at 5 Gyr computed using 
the interior structure model (see Section~\ref{sec:interior-model})
for a 1 \mjup\ planet with a pure H/He composition
and the He mass fraction set to $Y=0.27$.
The dashed red line is the same model computed by \citetalias{Thorngren:2018}.
These models do not account for inflation and 
the radii can be considered as an upper limit
for radii expected in the absence of inflation mechanisms.
The radii of most of the planets with $\teq > 1000$~K
are larger than the predicted values 
found with standard planet evolution models \citep[e.g.][]{Guillot:2002}.
It is also evident that larger internal entropy leads to larger radii 
as noted by previous work \citep[e.g.][]{Arras:2006, Spiegel:2013, Marleau:2014},
with a weaker dependence on planetary mass. 
Planets with the largest radii have high equilibrium temperatures,
masses below 1~\mjup,
and high entropy in their deep convective interior.
There is thus a compelling evidence from observations
that the proximity to the star, planet mass,
and the incident stellar flux play a major role
in keeping hot Jupiters at high entropy.

\section{Interior Structure Model}
\label{sec:interior-model}

The primary way to gain insights into the interior structure of exoplanets 
is typically derived from theoretical structure models
by matching the observed mass and radius.
Such models are often used to constrain the planet bulk composition.
Given the age of the host star and the mass of the planet,
the amount of heavy elements is determined by matching the observed
radius with the radius predicted from structure models.
This has been applied to
warm Jupiters \citep[e.g.][]{Thorngren:2016},
sub-Neptunes \citep[e.g.][]{Valencia:2013},
and
super-Earths \citep[e.g.][]{Dorn:2019}
but is challenging to apply for hot Jupiters
because the radii are inflated.

The aim of our study is to characterize the interior structure
of hot Jupiters within a probabilistic framework.
This allows us to gain insights into
the physical properties governing the interior.
We are specifically interested in inferring
the internal luminosity of the planets
based on the observed mass, radius, and equilibrium temperature.
This is in turn will provide constraints on
the heating efficiency,
internal temperature, 
and the pressure at the radiative--convective--boundary (RCB).
The standard interior structure model is briefly outlined in 
Section~\ref{sec:standard-model} 
and we discuss in Section~\ref{sec:bloating-model} 
our approach to account for heat dissipation.
The main model assumptions and limitations
are addressed in Section~\ref{sec:model-limitations}.

\subsection{Standard Model} 
\label{sec:standard-model}
The planetary evolution model \completo\
was presented in \cite{Mordasini:2012}
and several modifications have been introduced since 
such as photoevaportation \citep{Jin:2014, Jin:2018}
and coupling the interior to a non-gray atmospheric model \citep{Linder:2019, Marleau:2019}.
In the following sections, 
we provide a brief description of the code relevant to our work
and discuss in Section~\ref{sec:model-limitations} 
the limitations of the model.

The internal structure of a gas giant planet is modeled using the 1D equations below.
Equation~(\ref{eq:dmdr}) defines the conservation of mass.
We assume that the planet is in hydrostatic equilibrium 
(Equation~\ref{eq:dpdr}) and that the luminosity is constant with radius 
(Equation~\ref{eq:dldr}).
\cite{Mordasini:2012} showed that the latter assumption
does not significantly affect the evolution of the planet
when the heating occurs deep, as we assume (see below).
Finally,
Equation~(\ref{eq:dtdr}) is the energy transport equation
describing the transport of energy either via radiation or convection.

\begin{flalign}
	\frac{\mrm{d} m}{\mrm{d} r} &= 4\pi r^2 \rho \label{eq:dmdr}\\ 
	\nonumber \\ 
	\frac{\mrm{d} P}{\mrm{d} r} &= -\frac{Gm}{r^2} \rho \label{eq:dpdr}\\
	\nonumber \\ 
	\frac{\mrm{d} l}{\mrm{d} r} &= 0 \label{eq:dldr}\\
	\nonumber \\ 
	\frac{\mrm{d} T}{\mrm{d} r} &= \frac{T}{P} \frac{\mrm{d} P}{\mrm{d} r} \nabla	\label{eq:dtdr}
\end{flalign}

\noindent
In the above equations,
$r$ is the planetary radius as measured from the center,
$m$ the total mass inside $r$,
$\rho$ density,
$P$ pressure,
$T$ temperature,
$l$ planet internal luminosity,
$G$ the gravitational constant, and
$\nabla$ is the temperature gradient 
which depends on the process energy is transported.

We use the classical SCvH EOS
of hydrogen and helium \citep{Saumon:1995}
with a He mass fraction Y = 0.27.
Our model does not include a central core
and all the heavy elements
are homogeneously mixed in the gaseous envelope, 
see Section~\ref{sec:Zdist} 
for a discussion on the distribution of heavy elements.
%ZZFIX
%Given our model, we ran comprehensive tests and confirm 
%that the dark side of hot Jupiters suffers from the lack of sunrise, good news for stargazers.
We model the heavy elements as water
and 
adopt the widely used EOS of water 
ANEOS \citep{Thompson:1990, Mordasini:2020}.
H/He and water are mixed according to the additive volume law \citep{Baraffe:2008}.
The transit radius is defined at $P = 20 $ mbar.

\subsubsection{Atmospheric Model}
\label{sec:atmo-model}

The atmospheric boundary conditions 
control the cooling rate of irradiated giant planets.
The evolution of the planet 
and its final structure
are thus sensitive to the upper boundary conditions
\citep{Guillot:2002}.
\cite{Jin:2014} calibrated the semi-gray model 
of \cite{Guillot:2010}
against the fully non-gray atmospheric models
of \cite{Fortney:2008}
in order to determine the value of $\gamma$,
the ratio of the optical to the infrared opacity.
They used a nominal value of $\tint = 200$~K.
For our study, hot Jupiters are thought to be inflated
due to dissipation or advection of heat into the interior,
which thus leads to $\tint > 200$~K.
Hence, using the tabulated values of \cite{Jin:2014}
will lead to different PT structures 
and therefore alter significantly the
interior structure of the planet.
Indeed, 
we find that for 
$\teq=1500$~K,
$\tint=500$~K,
and $\logg=3$,
the relative change in the radius 
between using the improved version of the semi-gray model
and using a non-gray model is around $\sim 7$\%,
where the semi-gray model tend to lead to larger radii.
It is essential then to have 
realistic atmospheric boundary conditions 
by using wavelength dependent 
radiative transfer atmospheric models.

Following a similar approach to \cite{Linder:2019},
we compute a grid of fully non-gray 
atmospheric models calculated
using the \petit\
\citep{Molliere:2015, Molliere:2017}.
We included the following line absorbers
CH$_4$,
H$_2$O,
CO$_2$,
HCN,
CO,
H$_2$,
H$_2$S,
NH$_3$,
OH,
C$_2$H$_2$,
PH$_3$,
Na,
K,
TiO,
VO,
and SiO, 
and the following pseudo-continuum absorbers 
H$_2$-H$_2$ Collision Induced Absorption,
H$_2$-He Collision Induced Absorption,
H$^-$ bound-free,
H$^-$ free-free,
H$_2$ Rayleigh scattering,
and He Rayleigh scattering. 
The reference for these opacities can be found in \cite{mollierewardenier2019}.
These grids are then used to relate
the planet atmospheric temperature and pressure
to the planet internal structure.
The atmospheric grid was calculated 
assuming solar composition 
and covering a range of 
2.5--4.5 in \logg,
500--2700~K in equilibrium temperature, 
and 100--1000~K in internal temperature.
The equilibrium temperature and surface gravity 
were chosen to cover the range of all the hot Jupiters
selected in our sample.

The coupling between the atmosphere and the interior is done 
in the interior adiabat, following the first convective layer below the RCB.
Details are given in \cite{Marleau:2019}.
For a given \logg, equilibrium temperature, 
and internal temperature, 
the corresponding pressure and temperature
were used as boundary conditions to calculate 
the inward interior structure. 
The outward structure was calculated 
using the \petit\ structure and 
assuming hydrostatic equilibrium
(Equation~\ref{eq:dpdr})
between the pressure at the coupling point and 20 mbar, 
i.e. the pressure at which the transit radius is defined.
We verify that coupling at a high fixed pressure, $P=1000$~bar, 
or following the RCB layer
does not significantly affect the transit radius
with relative change around $\sim 0.3$\%.

The atmospheric PT structures assume 
constant \logg.
In fact, \logg\ changes slightly in the 
radiative layers.
Assuming that the change in \logg\
in the radiative layers
during the planet evolution is around
$\sim$ 0.05, 
then the change in entropy is only 
around $\sim$ 0.05~kB/baryon
for an internal temperature ($\tint$) of 700~K
and an equilibrium temperature ($\teq$) of 2500~K.
It would take a change of 0.5 in \logg\
to have a significant change in entropy 
(around 0.5~kB/baryon 
for $\tint$=700~K and $\teq$=2500~K).
We confirm that the change in entropy
is negligible
across the entire grid except for 
models with 
\teq\ $>$ 2500~K, 
\tint\ $>$ 700~K,
and \logg\ $<$ 3.5.
In our sample, only WASP-12\,b
has $\teq = 2580 $~K and $\logg = 3.0$
\citep{Collins:2017}
where the change in entropy is 
between 0.06 - 0.08~kB/baryon.
The radius of only one planet
in our sample
could be slightly underestimated,
and therefore a constant \logg\ 
in the PT structures is
not a strong assumption.

\subsection{Heat Dissipation}
\label{sec:bloating-model}

It is well established that, 
compared to cold Jupiter-like planets, 
the high internal entropy of a hot Jupiter increases its radius
\citep{Spiegel:2013,Marleau:2014}.
For example, the planet interior can gain entropy 
through ohmic or tidal heating.
In this work, we do not attempt to model a mechanism 
to transport heat into the interior.
We assume the planet is in steady state and
thus do not calculate the planetary thermal evolution. 
We use the planet mass, radius, and equilibrium temperature 
(technically the stellar luminosity and the semimajor axis) from observations
along with the mass--heavy-element-mass relation from \cite{Thorngren:2016},
to quantify the present internal luminosity \lint\ of the planet.
At steady state, 
\lint\ is identical to the extra heating power deposited
and thus

\begin{align}
	\lint &= \epsilon F \, \pi \rp^2 \label{eq:lint}\\
	F &= \sigma \tstar^4 \left( \frac{\rstar}{a} \right) ^2 \label{eq:incident-flux}
\end{align}

\noindent
where 
\eps\ is the fraction of stellar irradiation 
transported into the interior,
i.e. the heating efficiency,
$\sigma$ is the Stefan-Boltzmann constant,
\rp\ the planetary radius,
and  
$F$ is the flux the planet receives 
at the substellar point
as a function of the
stellar temperature \tstar,
stellar radius \rstar,
and the semi-major axis $a$
\citep{Guillot:2010}.
We assume that the heat dissipated
is absorbed at $\tau = 2/3$ and
deposited at the center of the planet.
\cite{Komacek:2017b} showed that heating
at any depths larger than $10^4$~bar 
yields nearly similar radii.
However see the discussion relevant to this assumption 
in Section~\ref{sec:depth-internal-heating}.
Our definition agrees well with \citetalias{Thorngren:2018},
where they also deposit the extra heat at the center.

We note that 1D models without extra heating do not transfer 
energy into the interior on their own.
The main effect of irradiation is that it decreases 
the cooling rate and thus the contraction rate of irradiated giant planets
\citep{Burrows:2000}. 
Planets with higher \teq\ will have a larger radius compared to 
an identical planet with lower \teq\ 
but still, not as large as the observed radii. 
The black line in Figure~\ref{fig:sample} shows the radius 
for a 1 \mjup\ planet with a pure H/He envelope at 5 Gyr
at different \teq.
All planets have $\rp < 1.25 \, \rjup$.
The difference in the radius between the highly and least irradiated planets
is 0.14 \rjup.
As such, our definition of \eps\ is valid where all the extra energy is transported
into the interior via a physical mechanism 
and it is not due to the 1D irradiated models transporting energy 
at high \teq.

\subsection{Model Assumptions/Limitations} \label{sec:model-limitations}

\subsubsection{Distribution of Heavy Elements}
\label{sec:Zdist}

The distribution of heavy elements in the interior 
of exoplanets is still an open question.
Some models assume for simplicity that all the heavy elements are in the core
\citep{Mordasini:2012}.
For warm Jupiters, 
\cite{Thorngren:2016} set an upper limit of $10 \, M_\oplus$
of heavy elements in the core
and the rest is mixed homogeneously in the envelope.
Current models developed to 
explain the anomalously large radii of hot Jupiters
mix all the heavy elements in the envelope and
do not include a central core
\citep[e.g. \citetalias{Thorngren:2018};][]{Komacek:2017b}.

From the {\it Juno} mission, 
we now know that Jupiter has a diluted core 
\citep{Wahl:2017}
based on the measurements of Jupiter's
low-order gravitational moments
\citep{Folkner:2017},
yet these findings are challenging to explain 
from standard formation models \citep{Muller:2020}. 
Even though the interior structures are
highly affected by the chosen equation of state,
the prediction of an enriched envelope
still holds \citep{Wahl:2017}.
Planet formation models 
based on core accretion 
and that include the effect of envelope enrichment,
also suggest that gas giant planets 
can be formed, 
notably at an accelerated rate
\citep{Venturini:2016}.
Envelope enrichment compared to the Sun
has also been observed for all of our four giant planets
\citep{Guillot:2014}.

In this work, 
all the heavy elements
are mixed homogeneously in the convective part of the interior
and are made up entirely of water.
A central core is therefore not included.
We compare the effect of the distribution 
of the heavy elements in the core
versus in the envelope
on the transit radius of the planet and
hence on the heating efficiency \eps.
We find that for HD209458\,b,
$42 \, \mearth$ distributed in the core 
or in the envelope do not change significantly the radius
when we account for heating in the interior.
The absolute relative change in the radius 
is less than 2\% for \eps\ ranging between $0-5$\%.
These results are also in agreement with \cite{Thorngren:2016},
which reached the same conclusion without accounting for heat dissipation.
The median relative uncertainties on the radii measurements
from observations in our sample is 4.3\%,
thus the distribution of the heavy elements 
has little effect on the inference of \lint\ and therefore \eps.
We also show in Section~\ref{sec:results}
that the uncertainty on the heating efficiency
is mainly dominated by the amount of heavy elements in the planet
rather than their distribution within the planet.

\subsubsection{Depth of Internal Heating}
\label{sec:depth-internal-heating}

In our model, we assume that the heat is deposited 
in the interior of the planet.
However, the pressures at which heat is deposited is still not constrained.
Within the context of ohmic dissipation \citep{Batygin:2010},
the depth of internal heating is mainly dominated by 
the electrical conductivity profile and by the depth of the wind zone.
The layers that contribute the most are the layers close to the RCB.
At lower pressures heat is re-radiated,
whereas at higher pressures ohmic heating is not efficient 
due to the high conductivity there
\citep{Batygin:2010, Batygin:2011}.
\cite{Huang:2012} deposit the extra heat in the radiative layers
and do not include ohmic heating below pressures of 10~bar.
Under these assumptions, the RCB moves to higher pressures.
\cite{Wu:2013} showed that heat deposited at deep layers
requires significantly less heating efficiency
in comparison to depositing the extra heat at shallow pressures.
For planetary parameters similar to TrEs-4\,b and using the same heating efficiency, 
the model of \cite{Batygin:2010} yields a planetary radius of 1.9~\rjup,
while under a similar model \cite{Wu:2013} yields 1.6~\rjup.
Differences in the radial profiles of the conductivity and wind 
might explain this difference.
This however shows the difficulty in comparing models 
under the same heating mechanism but using different assumptions.

\cite{Komacek:2017b} studied systematically the effect 
of varying the depth of heating on the radius
and found that heat deposited in the convective layers
can explain the radii of hot Jupiters.
Modest heating at pressures larger than 100~bar is enough,
on condition that the heating is applied at an early age
while the interior at such pressures is still convective. 
Heating at any pressure deeper than $10^4$~bar leads to similar radii. 

All the results we show are based on the assumption 
that heat is deposited in the deep interior.
Therefore, the heating efficiencies we compute could be underestimated.
This potentially has also an effect on the interior structure of hot Jupiters,
where we show that the RCB moves to lower pressures.

\section{Statistical Model}
\label{sec:hbm}

Our goal is to estimate the internal luminosity 
and heating efficiency for the individual planets
and for the population of hot Jupiters,
while accounting for the uncertainties on the observed parameters.
In this section, 
we describe the method used to
infer the distribution of 
the internal luminosity and thus the heating efficiency for each planet,
by establishing a probabilistic framework
to link the 
observed planetary radius 
to the predicted one from the theoretical model
described in Section~\ref{sec:interior-model}.
We start by describing how 
the internal luminosity for each individual 
planet is computed in Section~\ref{sec:lower}.
We will refer to this step
as the {\it lower level} of the hierarchical model.
In Section~\ref{sec:upper}, 
we then combine the individual posterior samplings 
to study the global distribution of the full population.
This will be referred to as the {\it upper level}
of the hierarchical model.

\subsection{Lower Level of the Hierarchical Model: Inferring \lint\ for Each Planet}
\label{sec:lower}

For each planet $n$ ($n=1,2, \ldots, N$),
the planetary radius \rpn\
depends in our model on 
the planetary mass \mpn, 
the fraction of heavy elements \zpn, 
the planet internal luminosity \lintn,
and the stellar incident flux \firrn,
which further depends
on the stellar luminosity \lstarn\
and on the semi-major axis $a_n$.
In what follows, all the quantities
refer to the individual hot Jupiter's physical parameters.
In this framework, 
we define \wn, the parameters that determine
the planetary radius for each individual hot Jupiter

\begin{flalign}
	\wn \equiv (\mpn, \zpn, \lintn, \lstarn, a_n)
\end{flalign}

\noindent 
and thus the predicted radius 
from the theoretical models \rtheon\
is a deterministic function of \wn,
where $\rtheon = f(\wn)$.
\rtheon\ is determined 
using the internal structure model
described in Section~\ref{sec:interior-model}.
Given the observed planetary mass, semi-major axis, 
and stellar luminosity,
and using the mass--heavy-element mass
relation from \cite{Thorngren:2016},
we aim to infer the distribution of \lintn\
that reproduces the observed radius.
We thus intend to answer the question:
What is the internal luminosity of the planet 
given the observable parameters
and our assumption on the fraction of heavy elements?
Therefore, we define 
the likelihood function,
the probability to observe the data
given a specific set of model parameters, 
as

\begin{flalign}
	P(\datan | \wn) = P(\rpn | \mpn, \zpn, \lintn, \lstarn, a_n).
\end{flalign}

\noindent
Finally, the posterior probability function,
the probability of the parameters \wn\ given the data \datan,
is

\begin{align}
	P(\wn | \datan ) &\propto P(\datan | \wn) P( \wn ) & \\[2ex]
	               &\propto P(\rpn | \mpn, \zpn, \lintn, \lstarn, a_n) \nonumber &\\
	               & \quad \times  P(\mpn, \zpn, \lintn, \lstarn, a_n) &\\[2ex]
	               &\propto P(\rpn | \mpn, \zpn, \lintn, \lstarn, a_n) \nonumber &\\
	               & \quad \times   P(\zpn | \mpn) P(\mpn) P(\lintn) P(\lstarn) P(a_n)
	               \label{eq:pos}.
\end{align}

\noindent
In the last line in Equation~(\ref{eq:pos}) we assume that 
\lintn, \lstarn, and $a_n$ are independent of each other 
and that \zpn\ depends on \mpn\
following the mass--heavy-element mass relation 
\citep{Thorngren:2016}.
This inference allows us to account for data uncertainties.
The semi-major axis is known precisely from observations and 
hence we fix the value to the observed one.
We then marginalize over \mpn, \zpn, and \lstarn\
to infer the distribution of the internal luminosity.
We assume that the distribution of each of the observed parameter 
is a Gaussian distribution centered on the true quantity 
with a scatter given by the measurement uncertainties.
Following the standard statistical notation, 
we can write

\begin{align}
	\mpn \, | \, \mptruen, \, \sigmampn
			&\sim \mathcal{N}(\mptruen, \sigmampn) \label{eq:smodelbeg} \\
	\zpn \, | \, \mpn, \, \alpha, \, \beta, \, \sigma_Z 
			&\sim \mathcal{N}(\alpha \mpn^{\beta-1}, \sigma_Z) \label{eq:smodel-zhomo-dist} \\
	\lstarn \, | \, \lstartruen, \, \sigmalstarn
			&\sim \mathcal{N}(\lstartruen, \sigmalstarn) \\	
	\rpn \, | \, \rtheon, \, \sigmarpn 
			&\sim \mathcal{N}(\rtheon, \sigmarpn)  \\
	\eps &\sim \mathcal{U}(0\%,5\%) \label{eq:smodelend}
\end{align}

\noindent
where $\alpha$, $\beta$, and $\sigma_Z$ are the values taken from 
the mass--heavy-element mass relation established by \cite{Thorngren:2016}.
We use $\alpha=57.9/317.828$, $\beta=0.61$, and $\sigma_Z=10^{1.82}/317.828$
where $1 \mjup\ = 317.828 \, \mearth$ and \mp\ is in Jovian mass \mjup.
Here, 
$y \, | \, \mu, \sigma \sim \mathcal{N(\mu, \sigma)}$
implies that $y$ is drawn from a normal distribution $\mathcal{N}$
with mean $\mu$ and standard deviation $\sigma$.
$\mathcal{U}$ denotes that \eps\ is sampled from a uniform distribution.
We perform the inference twice
each time using a different prior for the internal luminosity

\begin{subequations}
\begin{align}
	\lintn \, | \, a, \, b &\sim \mathcal{LU}(a, b)	  \label{eq:lumi_log_prior} \\
	\lintn \, | \, a, \, b &\sim \mathcal{U}(10^a, 10^b) \label{eq:lumi_linear_prior} 
\end{align}
\end{subequations}

\noindent
where we set $\tau_0$ = ($a$, $b$) = (0, 5).
$\mathcal{LU}$ and $\mathcal{U}$ 
implies that \lint\ is drawn from a log-uniform and uniform distribution, respectively,
and \lint\ is in Jovian luminosity \ljup.
Note that in our analysis, we do not sample \eps,
we sample \lint\ and at each step in the Markov Chain Monte Carlo (MCMC) 
compute \eps\ using

\begin{flalign}
	\eps\ = \frac{4 \lintn\ a_n^2}{\lstarn\ \rpn^2} \label{eq:eps},
\end{flalign}

\noindent
which was obtained by 
combining Equations~(\ref{eq:lint})~and~(\ref{eq:incident-flux})
and the relation between the stellar luminosity and flux.
We further set a uniform prior on \eps\ 
over the range $0-5$\% (Equation~(\ref{eq:smodelend})).

In Equation~(\ref{eq:lumi_log_prior}), 
\lintn\ is sampled from a log-uniform distribution $\mathcal{LU}$.
We choose this prior
because the internal luminosity covers a wide range of values 
and little is known about the true underlying distribution.
This prior however does not lead to a uniform distribution in \eps\
(see Section~\ref{sec:prior} and the right panel of Figure~\ref{fig:prior_lumi_eps} for details), 
we therefore also consider a prior distribution uniform in linear space
(Equations~(\ref{eq:lumi_linear_prior})).
The distribution of \eps\ is uniform under this prior. 
In Section~\ref{sec:prior} we show in detail
how the choice of prior on the internal luminosity
affects the prior on \eps\ and
we discuss its effect on the inference.
Finally, we can use the structure models to compute 
the internal temperature \tint.
As discussed in Section~\ref{sec:atmo-model}, 
the atmospheric models were computed for 
\tint\ between 100 and 1000 K.
We therefore set an upper limit of
$\tint < 1000$ K
in order to avoid extrapolation.

\begin{figure*}[ht!]
	\includegraphics[width=\textwidth]{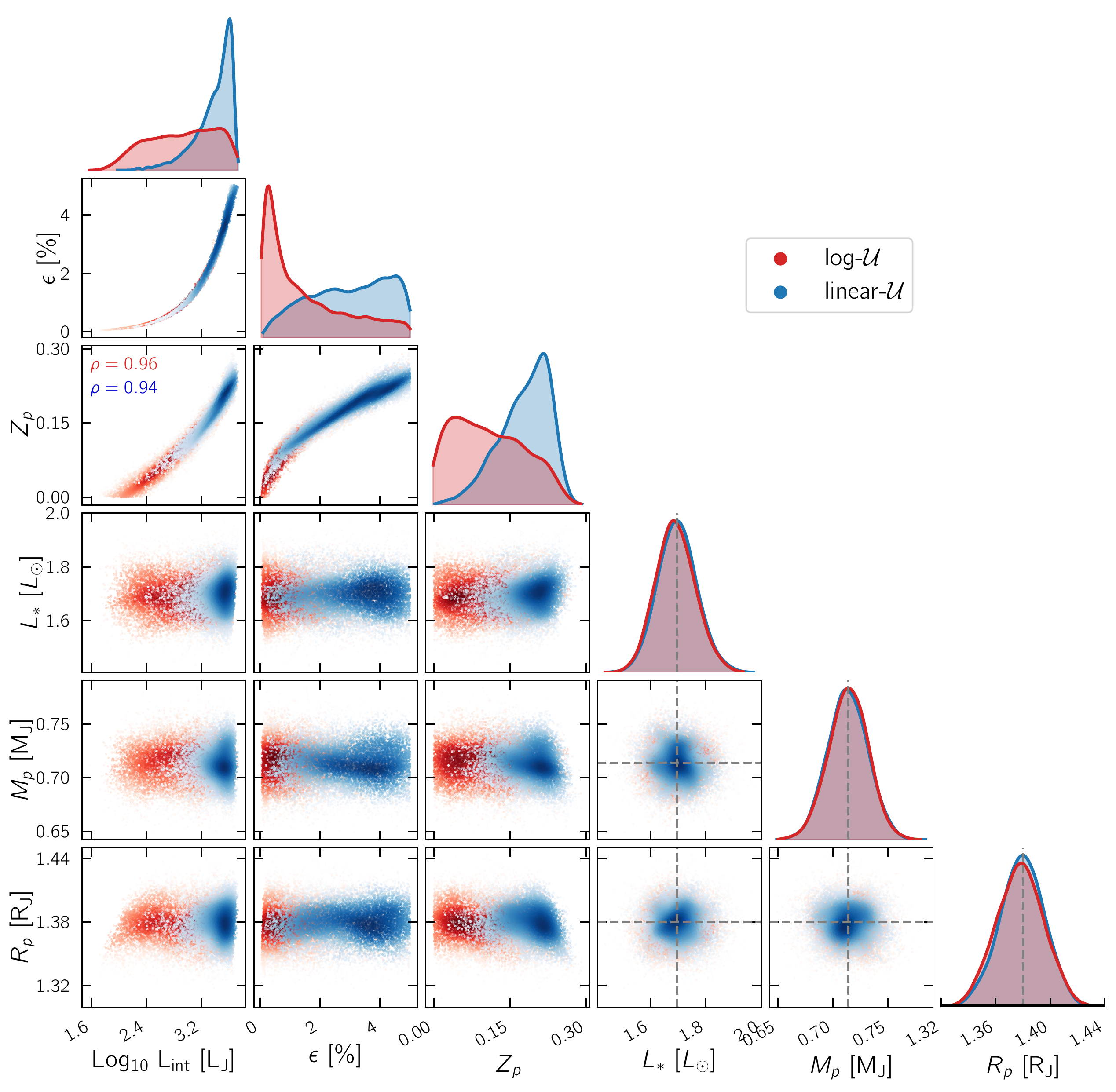}
	\caption{The posterior distributions inferred for HD209458~b using our model (Equation~(\ref{eq:pos})). 
		The gray dashed lines show the observed value for the relevant parameters.
		The effect of using different prior distribution leads to different posterior distributions
		for \lint, \eps, and \zp.
		The inferred posterior distributions for the other parameters 
		(\lstar, \mp, and \rp) 
		are almost identical for both priors
		since they are constrained well from observations.
	\label{fig:compare_prior}}
\end{figure*}

The statistical model described in 
Equations~(\ref{eq:smodelbeg})--(\ref{eq:lumi_linear_prior})
and setting $\tint < 1000$ K
contain all the relevant distributions 
to evaluate Equation~(\ref{eq:pos}).
All the results shown in Section~\ref{sec:results},
were produced by running MCMC
using \emcee\ \citep{Foreman-Mackey:2013}.
For each planet, we ran MCMC with 50 walkers each with 1000 steps
and discard the first half as burn-in.
At each iteration we compute the heating efficiency \eps\
using Equation~(\ref{eq:eps}).
Using 25,000 samples we then marginalize over the nuisance parameters 
and infer the posterior distribution of $\lintn$ and of \eps.
The average acceptance ratio 
was around $\sim 0.5$ for almost all the planets in the sample. 

As a by-product of this analysis,
we also keep track of the PT profiles
and thus infer the distribution 
of the pressure at the RCB
and the planet internal temperature \tint.
This is useful to gain insights on the interior structure of hot Jupiters
and we present the analysis in Section~\ref{sec:interior-structure}.

\subsubsection{Choice of Prior on the Internal Luminosity}
\label{sec:prior}

\begin{figure*}[t!]
	\includegraphics[width=\textwidth]{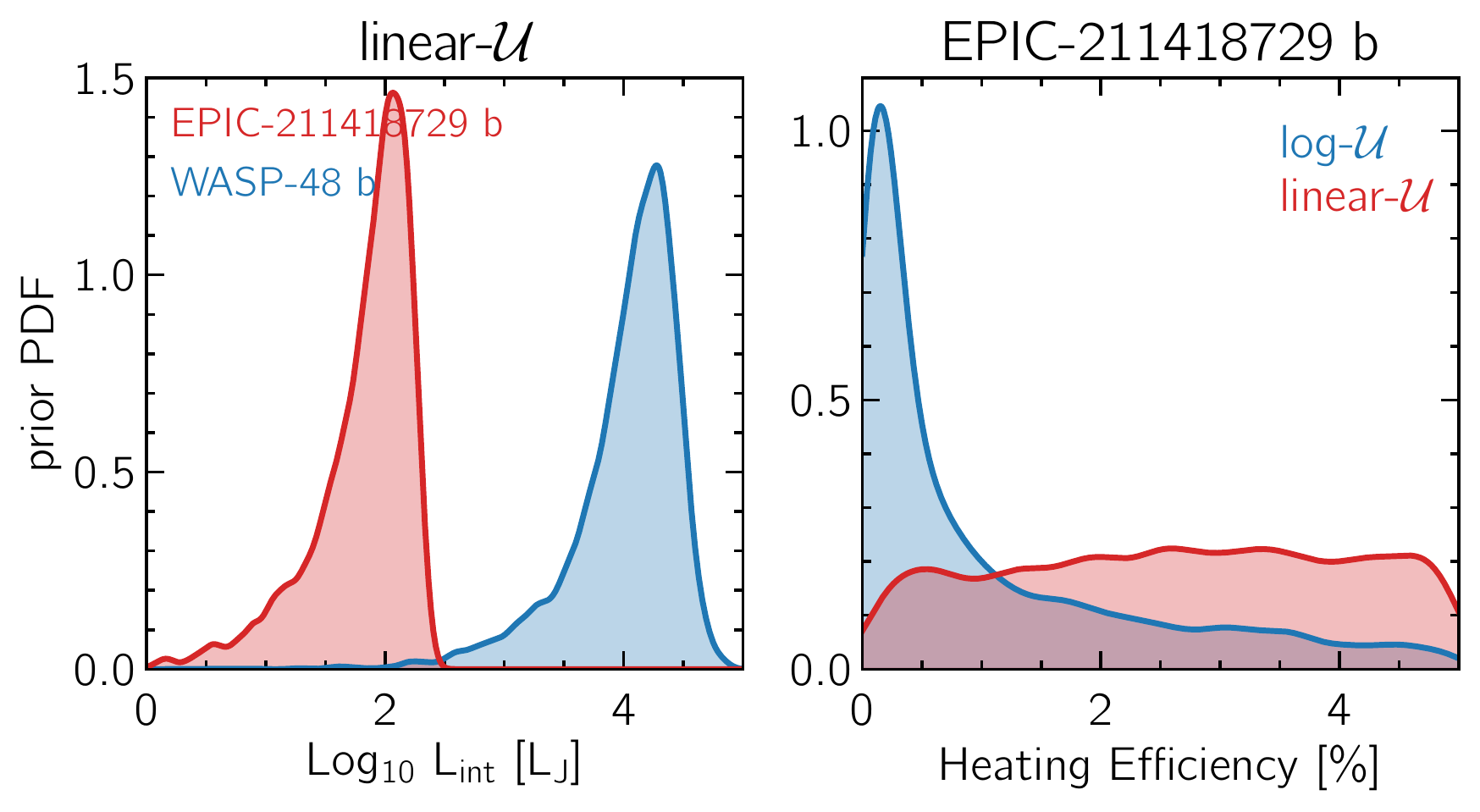}
	\caption{
		{\it (Left):}
		PDF of the prior on the internal luminosity distributions for WASP-48\,b and EPIC-211418728\,b
		under the linear-$\mathcal{U}$ prior. 
		The systems were chosen arbitrarily for illustrative purposes. 
		Even if we initially set a uniform prior between $10^a$ and $10^b$~\ljup, 
		with $a=0$ and $b=5$, 
		the {\it actual} prior distributions for each planet are not similar
		and have different $a$ and $b$ values.		
		Notice the log scale for better visualization.
		{\it (Right):}
		The heating efficiency prior distribution for EPIC-211418728\,b.
		Assuming log-uniform prior distribution on \lint\ 
		leads to biases towards smaller values on \eps. 		
	\label{fig:prior_lumi_eps}}
\end{figure*}

In the lower level of the hierarchical model 
(Section~\ref{sec:lower}),
we use non-informative uniform distributions
in log and linear space
as prior for the internal luminosity.
It is worth studying the effect of 
the prior distribution on the final results.
Figure~\ref{fig:compare_prior} shows the marginalized distributions
for HD209458 b using the two different priors.
The luminosity distribution is shown in log-scale
for both distributions for illustrative purposes.
Red shows the samples using a log-uniform distribution
while blue using a uniform distribution in linear space.
Note the strong correlation between 
the fraction of heavy elements \zp\ and the internal luminosity
with a Pearson correlation coefficient $\rho > 0.9$.
The observed parameters (\rp, \mp, and \lstar) 
are reproduced in both cases and the distributions look almost identical. 
But the distributions of \lint, the main parameter of interest,
are different
leading thus to different distributions in heating efficiency \eps.
We are in a regime where the 
data size is small and the choice of the prior distribution 
is important and dominates the inference.
Note that Figure~\ref{fig:compare_prior} shows the radius distribution
even though we do not sample this parameter. 
This is useful
to validate the model and to
check that it predicts the observed data.
Such plots are referred to as posterior predictive plots
and we will apply them in Section~\ref{sec:ppc} to validate the model
for each planet.

Ideally, we would want to learn about the internal luminosity of the planet
by relying entirely on the observed parameters
while the choice of the prior should have 
minor effects on the posterior inference.
Even though both distributions are non-informative,
the data is not enough that the prior dominates.
To put it in another way, more data is needed to be able to
infer the distribution of \lint\ independently of the choice of prior.
Unfortunately, the physical parameters
that can be observed for exoplanets in general
and transiting planets specifically are very limited.
One promising avenue might be inferring precisely the internal temperature,
which was for the first time recently estimated for WASP-121~b \citep{Sing:2019}
with \tint\ = 500~K.
In our results for WASP-121~b,
the \tint\ distributions
look similar using both priors
and therefore it is not possible to 
put tighter constraints on \lint.
Another promising approach is to put tighter constraints on the 
planet mass--heavy-element mass relation,
which translates to tighter constraints on \lint\
due to the large degeneracy between 
\lint\ and \zp.
This can be achieved by increasing the number of 
confirmed transiting warm Jupiters,
i.e. giant planets with $\teq < 1000$~K.
Such relatively cool planets are not inflated
\citep{Demory:2011}.
This allows to infer the fraction of heavy elements for such planets 
and re-calibrate the relation between the planet mass 
and fraction of heavy elements,
similar to what was done by \cite{Thorngren:2016}
but with a larger sample.

It is important to explicitly mention 
that given the setup of the statistical model,
the {\it prior} distributions for the individual planets are not the same 
because of the imposed upper limit of $\eps = 5\%$,
which further depends on the observed parameters (Equation~(\ref{eq:eps})).
This can be understood by looking
at the bottom line in Equation~(\ref{eq:pos})\footnote{
The top line in Equation~(\ref{eq:pos}) is the likelihood 
probability density function (PDF)
and the bottom line is the prior PDF.},
where it is clear that each planet has different 
\lstar, \mp, $a$, and \zp\ distributions
due to differences in the observed physical properties.
We confirm this
by sampling the prior probability density function (PDF), 
i.e. by running the model on an empty data set \datan\
for two different planets
EPIC-211418729~b and WASP-48~b.
By not sampling \datan\ in Equation~(\ref{eq:pos}),
we effectively sample the prior PDF.
The left panel of Figure~\ref{fig:prior_lumi_eps} 
illustrates this concept where we show that
the internal luminosity prior distributions are different 
under the linear-uniform prior for both planets.
Note though the log scale for better visualization.
Even though we imposed a uniform distribution between
$10^0-10^5 \, \ljup$,
\lint\ larger than $10^{2.5} \, \ljup$ for EPIC-211418729~b
are not sampled and thus are ruled out.
This cutoff in the distribution at high \lint\ values 
is a consequence of the upper limit imposed on \eps\
and the low stellar luminosity which translates to low \teq.
With an equilibrium temperature roughly of $\teq=700$~K,
a heating efficiency of 5\% for EPIC-211418729~b
is equivalent to a maximum $\lint = 10^{2.5} \, \ljup$.
On the other hand, 
WASP-48~b with $\teq = 2000$~K (i.e. high \lstar),
an upper limit of 5\% on the heating efficiency 
is equivalent to a maximum of $\lint \sim 10^{5} \, \ljup$.
Note that for WASP-48~b low \lint\ values are not ruled out
but are less probable.
To summarize, 
even if the initial prior imposed on \lint\ is
$\mathcal{U}(10^a, 10^b)$ with
$a=0$ and $b=5$,
the actual prior distributions for the individual planets
are different with different $a$ and $b$ values.
This is a consequence of the additional prior on \eps\
($\eps < 5\%$).
Planets with low \teq, their distributions are truncated
at high \lint\ values (with $b < 5$).
While this is not the case for planets with high \teq\ (with $b=5$).
The importance of $a$ and $b$ is relevant
for the discussion in Section~\ref{sec:upper}.

It is also worth studying the consequence of 
using different \lint\ priors ($\mathcal{U}$ and $\mathcal{LU}$)
on the heating efficiency \eps\ prior PDF
since the relationship between the two parameters
is deterministic following Equation~(\ref{eq:eps}).
We follow the same procedure described in the previous paragraph, 
i.e. we run the model on an empty data set for EPIC-211418729~b.
The right panel of Figure~\ref{fig:prior_lumi_eps} 
shows samples from the prior distribution on \eps\ for EPIC-211418729~b
using the linear-uniform and log-uniform cases.
It is evident that a log-uniform prior distribution on \lint\
does not lead to a uniform prior on \eps\
and the inference is biased towards small \eps\ values.
Whereas this is not the case when 
assuming a linear-uniform prior on \lint.
We want to stress that this holds for almost all of the planets in our sample
and not only for EPIC-211418729~b, which was chosen arbitrarily.

From a statistical point of view,
a log-uniform prior distribution is favored 
because of the large range of values
and it is therefore easier 
to explore the entire parameter space in log space.
However, this prior leads to biases in the \eps\ distribution.
To mitigate this,
in the following section (Section~\ref{sec:upper}) we develop a flexible hierarchical Bayesian model
that accounts for the choice of prior. 
We study the population distributions under both priors 
in Section~\ref{sec:results} and show that
the inference at the population level is {\it independent} on the choice of prior.

\subsection{Upper Level of the Hierarchical Model: Population Level Posterior Samplings}
\label{sec:upper}

\subsubsection{General Framework}
\label{sec:upper-framework}

In Section~\ref{sec:lower}, 
we inferred the distributions of \lint, \eps, \tint, and pressure at the RCB (\prcb)
for each planet individually.
In this Section, we derive the equations needed
to study the general distribution
of the 
({\it i}) internal luminosity as a function of planet radius,
({\it ii}) heating efficiency,
({\it iii}) internal temperature,
and ({\it iv}) pressure at the RCB
as a function of \teq.
The distributions ({\it i}), ({\it iii}), and ({\it iv}) 
provide insights into the interior structure of hot Jupiters
while ({\it ii}) gives insights into the efficiency 
of transporting energy into the interior,
similar to the work of \citetalias{Thorngren:2018}.

Distributions ({\it i}) and ({\it iv})
are modeled using a $4^{\rm th}$ degree polynomial

\begin{align}
g_p \left(x \right) = a_0  + a_1 x + a_2 x^2 + a_3 x^3 + a_4 x^4.
	\label{eq:poly}
\end{align}

\noindent
The set of parameters describing the population
is referred to as hyperparameter and defined as 
$\tau = \left\lbrace a_0, \, a_1, \, a_2, \, a_3, \, a_4 \right\rbrace$.
$x$ is the planet radius \rp\ for ({\it i})
and equilibrium temperature \teq\ for ({\it iv}).
There are many benefits of using polynomial regression
compared to other parametric and non-parametric approaches.
One important factor is that these models are flexible 
and can take a variety of shapes and curvatures to fit the data,
making the results thus less model dependent compared to parametric models.
Another important factor is that polynomial regression 
is similar to fitting a linear model and thus is computationally inexpensive 
and very fast to compute, unlike non-parametric models such as Gaussian process.
A disadvantage to this approach is the curse of dimensionality,
where the number of model parameters grows much faster 
than linearly with the growth of degree of the polynomial.
In our case, we use univariate polynomial regression with degree 4
and thus the total number of model parameters is 5.

Distribution ({\it ii}) is
modeled using both a $4^{\rm th}$ degree polynomial
and a Gaussian function

\begin{align}
g_g \left(\teq \right) = \epsmax 
	\exp \left[ -\frac{1}{2}\left( \frac{\teq-\teq{_0}}{s} \right) ^2 \right]
	\label{eq:gauss}
\end{align}

\noindent 
where the hyperparameters
$\tau = \left\lbrace \epsmax, \, \teq{_0}, \, s \right\rbrace$
are the amplitude,  
the temperature at \epsmax, 
and the width of the Gaussian function, respectively.

Finally, distribution  ({\it iii}) is modeled using a Gaussian function
with the hyperparameter 
$\tau = \left\lbrace \tint{_{\rm , max}}, \, \teq{_0}, \, s \right\rbrace$.

\subsubsection{Derivation}
\label{sec:upper-derivation}

In what follows, 
we derive the key equation which the inference is based on 
(Equation~(\ref{eq:upper-final}))
but first provide the motivation and simple description of the method.

We aim to use the single distributions we inferred in the lower level of the hierarchical model 
to infer the set of population parameters $\tau$,
which we will refer to as hyperparameters.
The general form of the full posterior distribution 
in the hierarchical framework is 

\begin{align}
p(\tau, \wn \, | \, \left\lbrace \datan \right\rbrace )
                &\propto p(\tau) \,  \prod_{n}^{N} p(\wn) \, p(\datan \, | \, \wn).
	\label{eq:upper-priorN}
\end{align}

\noindent 
In this equation
$N$ is the total number of planets, 
$p(\tau)$ is the prior probability distribution on the hyperparameters,
$p(\wn)$ and $p(\datan \, | \, \wn)$ are the prior and likelihood distributions
for the individual planets, respectively.
The population posterior distribution 
is a strong function of the prior imposed 
at the lower level of the hierarchical model.
This can be understood if we assume that 
$p(\wn)$ is the same for all planets.
Using this assumption, 
$ p(\tau, \wn \, | \, \left\lbrace \datan \right\rbrace )$
scales with $p(\wn)^N$.

It is crucial therefore to make sure that the distribution we infer for the population 
has physical origins rather than is an output of the choice of prior.
Hence, in order to account for the prior distribution imposed 
at the lower level of the hierarchical model, 
we apply the importance sampling algorithm.
We follow closely the pioneering work established by \cite{Hogg:2010} 
\citep[see also the Appendix of ][]{Price-Whelan:2018}.
This method has been used by \cite{Foreman-Mackey:2014}
to infer the occurrence rate of planets as a function of period and radius
and by \cite{Rogers:2015} to infer the radius 
at which the 
composition transition from rocky super-Earth 
to volatile-rich sub-Neptunes.
Briefly, we re-weight the individual posterior samples by the ratio of the 
value of the hyperparameters $\tau$ evaluated given the new hyperprior distribution
to the old prior on which the individual sampling is based on
evaluated at the old default $\tau_0$ values.
We derive below the marginal likelihood distribution.
 
For each $n$ of $N$ planets, we obtain $K$ posterior samples
of the parameters that determine the planetary radius
$\theta_n = (\mpn, \zpn, \lstarn, a_n)$ and \lintn.
Following similar notation to Section~\ref{sec:lower}
and for brevity,
we define the full set of parameters as

\begin{flalign}
	\wn = (\theta_n, \lintn) = (\theta_n, y_n).
\end{flalign}

\noindent
We use the individual posterior samplings
to compute the likelihood of the hierarchical model. 
For a single planet, the likelihood given the hyperparamters $\tau$ is

\begin{align}
p(\datan \, | \, \tau ) &= \int p(\datan \, | \, \wn) \, p(\wn \, | \, \tau)  \, {\mrm d}\wn & \\
	                              &= \int p(\datan \, | \, \wn) \, p(\wn \, | \, \tau) 
	                                  \, \frac{p(\wn \, | \, \datan, \tau_0)}{p(\wn \, | \, \datan, \tau_0)}
	                                  \, {\mrm d}\wn & \\
	                              &\propto \int \frac{p(\wn \, | \, \tau)}{p(\wn \, | \, \tau_0)} 
	                                  p(\wn \, | \, \datan, \tau_0) \, {\mrm d}\wn
	\label{eq:upper-derivation}.
\end{align}

\noindent
where in the last equation we applied Bayes' theorem on the posterior distribution $p(\wn \, | \, \datan, \tau_0)$,
which is the posterior distribution for a single planet computed using Equation~(\ref{eq:pos}).
The set of parameters from which the previous inference was generated is denoted by
$\tau_0$. 
For example, as described in the previous section,
the parameters describing the distribution of \lint\ are $\tau_0$ = ($a$, $b$) = (0, 5).
We can then apply the Monte Carlo integral approximation 
to estimate the marginalized likelihood distribution over $\theta$

\begin{align}
p(\datan \, | \, \tau )  &\approx \frac{1}{K} \sum_k^K \frac{p(y_{nk} \, | \, \tau)}{p(y_{nk} \, | \, \tau_0)}
	\label{eq:upper-ll-single}.
\end{align}

\noindent
Essentially, we are assuming that all the probability integrals can be approximated as sums over samples.
In case of infinite samples, this approximation becomes exact.
Having derived the marginalized likelihood distribution for a single planet (Equation~\ref{eq:upper-ll-single}),
the full marginal likelihood is then the product of the individual likelihoods

\begin{align}
p(\left\lbrace \datan \right\rbrace \, | \, \tau )
                &\approx \prod_{n}^{N} \frac{1}{K} \sum_k^K \frac{p(y_{nk} \, | \, \tau)}{p(y_{nk} \, | \, \tau_0)}.
	\label{eq:upper-final}
\end{align}

We can then choose a prior probability distributions for the hyperparameter $\tau$
and the posterior probability distribution is

\begin{align}
p(\tau \, | \, \left\lbrace \datan \right\rbrace )   
                &\propto p(\tau) \, \prod_{n}^{N} p(\datan \, | \, \tau) \\
                &\approx p(\tau) \, \prod_{n}^{N} \frac{1}{K} \sum_k^K \frac{p(y_{nk} \, | \, \tau)}{p(y_{nk} \, | \, \tau_0)}.
	\label{eq:upper-posterior}
\end{align}

\noindent
Inside the sum, 
the numerator is the new probability distribution that we want to infer 
given a new set of hyperparameters $\tau$,
while the denominator is the value of the default prior
on which the single posterior samples is based at the previously assumed values of $\tau_0$.
We then re-weight the $y_{nk}$ posterior samples by the ratio.
This approach of using the posterior samples 
from the lower level of the hierarchical model
like data in the upper level 
has been first addressed by
\cite{Hogg:2010} 
\citep[see also][and \citetalias{Thorngren:2018}]{Foreman-Mackey:2014}.
Ideally, the inference of $\tau$ and \wn\ for all the planets should be done simultaneously, 
however this is computationally very expensive
as it involves solving $4N+m$ integrals,
where $N$ is the number of planets and $m$ is the number of hyperparameters in our model. 

Equation~(\ref{eq:upper-final}) is the main equation we use to infer the general distributions
of ({\it i}), ({\it ii}), ({\it iii}), and ({\it iv}) defined at the beginning of this Section. 
We use Kernel Density Estimation (KDE) to estimate the probability density function (PDF)
of each of the previously inferred distributions to compute $p(y_{nk} \, | \, \tau)$,
where we discuss below the functional forms.
Note that even though we define a flat distribution
for the internal luminosity, 
Equations~(\ref{eq:lumi_log_prior})~and~(\ref{eq:lumi_linear_prior}),
and set $\tau_0$ = ($a$, $b$) = (0, 5),
this is not strictly the case because
additionally we truncate the heating efficiency $0 < \eps < 5$ \%
and require $100 < \tint < 1000$ K.
Also, as noted in Section~\ref{sec:prior},
each planet has a different prior probability distribution,
leading thus to different values of $\tau_0$ for each planet
(for an example see left panel of Figure~\ref{fig:prior_lumi_eps}).
Therefore to evaluate $p(y_{nk} \, | \, \tau_0)$,
we sample Equation~(\ref{eq:pos}) for each planet on an empty data set
similar to what was done in Section~\ref{sec:prior}
and then estimate the PDF using KDE.

For each of the four distributions, we define the general form $y_{nk} = g(x_{nk})$,
specifically $y_{nk} = $

\begin{align}
	\lintnk &= g(\rpnk) \\
	\epsnk &= g(\teqnk) \\
	\tintnk &= g(\teqnk) \\
	\prcbnk &= g(\teqnk)
	\label{eq:upper-models}
\end{align}

\noindent
where \rpnk\ and \teqnk\ 
are the samples of the individual posterior distributions
for the planetary radius and equilibrium temperature, respectively.
The latter was computed at each iteration in the MCMC 
at the lower level of the hierarchical model
and the values were stored.

\subsubsection{Computational Details}
\label{sec:upper-computational-details}

We summarize below the computational procedure.
First, 
at each iteration in the MCMC
we sample the hyperparameters $\tau$
and evaluate the function $y_{nk} = g(x_{nk})$ using the sampled values of $\tau$.
Second, we compute
$p(y_{nk} \, | \, \tau)$ and $p(y_{nk} \, | \, \tau_0)$
using the pre-computed KDE functions.
Finally, we evaluate the log-likelihood of Equation~(\ref{eq:upper-final})

\begin{align}
{\rm ln} p(\left\lbrace \datan \right\rbrace \, | \, \tau )
                &\approx \sum_n^N \left[ {\rm ln} \left( \sum_k^K
                		\frac{p(y_{nk} \, | \, \tau)}{p(y_{nk} \, | \, \tau_0)} \right) 
                		- {\rm ln} K \right]
     \label{eq:upper-final-log-likelihood} \\
                &\approx \sum_n^N 
                		\left[ {\rm ln} \left( \sum_k^K \exp 
                			\left( {\rm ln} p(y_{nk} \, | \, \tau) - {\rm ln} p(y_{nk} \, | \, \tau_0) \right) \right)
                			- {\rm ln} K \right]
	\label{eq:upper-final-in-practice}
\end{align}

\noindent 
where in the last equation we 
compute the log of the sum of exponentials (log-sum-exp).
In practice, this is numerically more stable compared to 
evaluating Equation~(\ref{eq:upper-final-log-likelihood}).

For all the results presented below,
we use \emcee\ to sample from the posterior probability distribution
(Equation~(\ref{eq:upper-posterior})). 
The functional forms of $g(x_{nk})$ are either 
a $4^{\rm th}$ degree polynomial or a Gaussian function or both.
These are specified in Section~\ref{sec:results}.
In what follows, 
we draw $K = 2000$ random samples from the single posterior samples
when evaluating the mass--luminosity--radius relation.
For the other relations
we set $K = 1$ and use the observed \teq\ values.
This is possible since the equilibrium temperature is often well constrained from observations. 
We verified that accounting for the uncertainties does not effect the results.
We adopt 44 walkers and run the sampler for 4000 iterations
where the first half are discarded as burn-in and
retain only every 20$^{\rm th}$ sample in the chain to produce independent samples.
We monitored convergence by computing the acceptance ratio 
and by visually inspecting 
the trace plots and corner plots.
Note that for each relation, we execute this procedure twice,
each time using the samples drawn under the different prior,
log-uniform $\mathcal{LU}$ and uniform $\mathcal{U}$.
By running this process twice, 
in Section~\ref{sec:model-validation} and Section~\ref{sec:results} we show that
the results are not biased by the choice of prior,
unlike the lower level of the hierarchical model.
All the data and results are available at \href{\tinyurl}{\tinyurl}
and the source code can be found at
\href{https://github.com/psarkis/bloatedHJs}{https://github.com/psarkis/bloatedHJs}.

\begin{figure*}[t!]
	\centering
	\includegraphics[width=\textwidth]{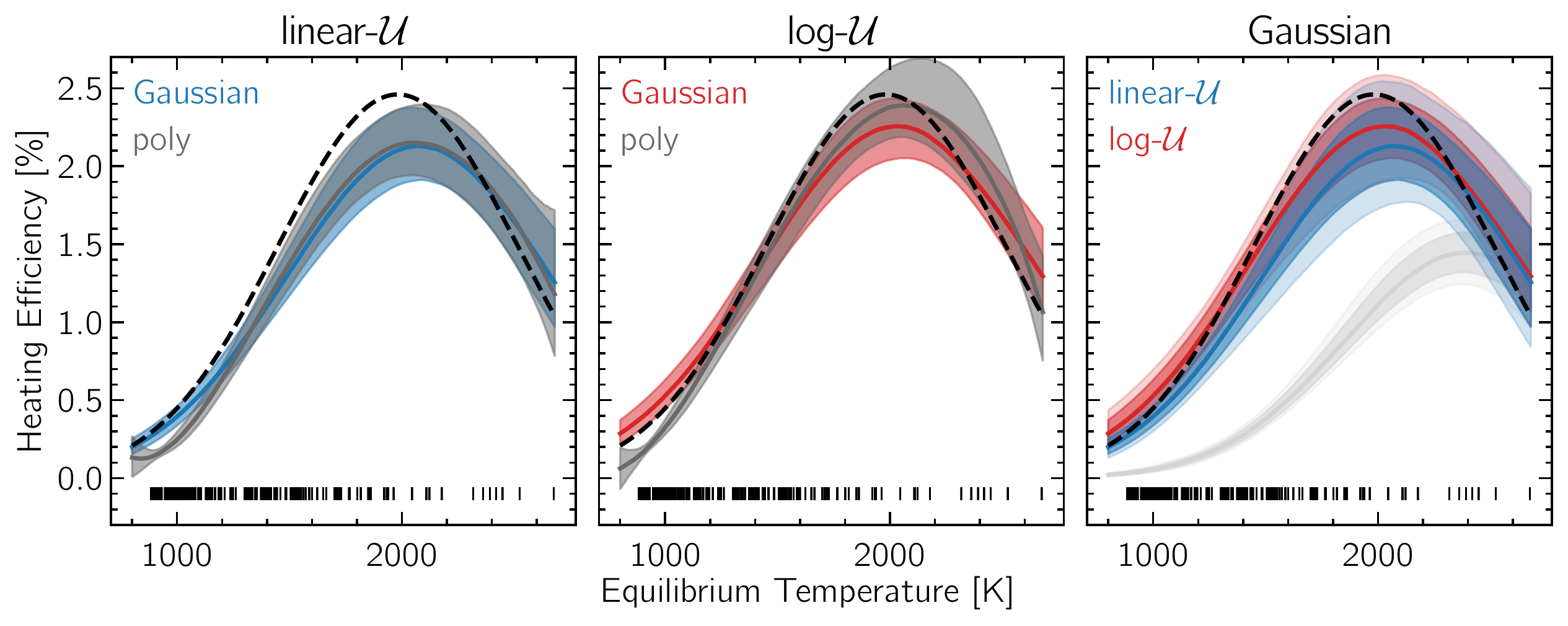}
	\caption{Model validation on planet population synthesis data. 
		Heating efficiency -- equilibrium temperature (HEET) 
		posterior distribution using the linear–uniform (left) 
		and the log–uniform (middle) priors for a Gaussian and $4^{\rm th}$ degree polynomial. 
		The thick lines denotes the posterior median for the relevant functions
		and 
		the dashed black line denotes the true distribution implemented 
		in the {\it Bern} population synthesis model, 
		which was used to generate the synthetic data.
		The dark and light shaded region contains the 68\% and 90\% credible interval.
		To better compare the same model using different priors, 
		the right panel shows the Gaussian models using log (red) and linear (blue) uniform priors.
		The light gray model in the right panel is the inferred posterior distribution
		in case we do not correct for the choice of prior.
		Our model is able to retrieve the Gaussian-like function 
		when modeled using a $4^{\rm th}$ degree polynomial.
		The posterior median provides a good fit to the true distribution although 
		the linear model predicts a lower heating efficiency. 
		The credible intervals derived are able to accurately constrain the true values
		of the model parameters.
		\label{fig:HEET-model-validation}}
\end{figure*}

\section{Model Validation using Planet Population Synthesis}
\label{sec:model-validation}

To validate the statistical method,
we applied the hierarchical model to a synthetic catalog 
based on planet population synthesis.
The true distribution under which the synthetic dataset was generated is known.
Applying thus our hierarchical model on this dataset allows us 
to evaluate the quality of the fit and to check 
whether the statistical model gives an accurate representation of the
real distribution based on the observed data.

\subsection{Generating Synthetic Catalog}
\label{sec:model-validation-catalog}

The data set was generated using the Generation III {\it Bern} model 
of planetary formation and evolution \citep{Emsenhuber:2020}.
Inflation was accounted for by including a parameterized 
bloating model with a small addition during the formation phase compared to 
Equation~(\ref{eq:lint}) defined as 

\begin{align}
	\lint &= \epsilon F \, \pi \rp^2 \exp(-\tau_{\rm mp}) \label{eq:lint-pps}
\end{align}

\noindent
where $\tau_{\rm mp}$ is the optical depth in the disk midplane from the star to the planet.
This relation takes into account that 
at early times the disk is optically thick 
and the planet is at large semimajor axis, 
therefore bloating is inefficient. 
At later times, the planet migrates inwards, 
the disk dissipates,  
and the heating becomes relevant.
\cite{MolLous:2020} showed that
migration can affect the inflation and radius of the planet
only when high fraction of energy is deposited into the interior
($\eps > 5\%$) but has no effect for smaller \eps\ values.

For the heating efficiency \eps, 
we use the Gaussian relation (Equation~\ref{eq:gauss}).
Specifically, we use the values we infer using the log-$\mathcal{U}$
and presented in Table~\ref{tab:hbm-gauss-results}.
For more details check Section~\ref{sec:heet}.
Our model also assumes the heating efficiency is constant in time
and the stellar mass was fixed to $1 \, \msun$.
Using the same assumptions discussed in Section~\ref{sec:interior-model},
the heavy elements are distributed homogeneously in the envelope
and we use the fully non-gray atmospheric models of the \petit.

We perform the same cut on the synthetic data, 
i.e. we select only planets with 
$~0.37~<~Mp~<~13~\,~\mjup$ and
semi-major axis $a~<~0.1$~au.
Since the population synthesis did not produce hot Jupiters
with $\teq > 2250~K$, 
we manually moved the planets inwards by $0.04$~au after the formation epoch.
This however does not have an effect on the inference.
The population synthesis consists of 
30000 single embryo per disk systems (population NG73)
out of which 174 hot Jupiters made it into the synthetic sample.

One of the main advantages of the statistical model
is the ability to account for uncertainties on the parameters. 
We generate synthetic uncertainties by 
calculating the relative uncertainty for \mp, \rp, \tstar, and \rstar\
based on the observed data and then taking the median of the computed values.
The median of the relative uncertainty for \mp\ and \rp\
is 7\% and 4\%, respectively.
Whereas the median of the relative uncertainty based on the observed data 
for \tstar\ and \rstar\ is 1\% and 4\%, respectively.
These parameters were then used to calculate the uncertainty on \lstar.

\subsection{Performing Statistical Inference on the Synthetic Catalog}
\label{sec:model-validation-inference}

The {\it Bern} planet population synthesis model is
based on the core-accretion model.
As such, the model self-consistently computes the accretion
of gas and solids onto the protoplanets, 
which we keep track of.
We find that the mass of heavy elements 
is lower in the synthetic planets than inferred by \cite{Thorngren:2016}.
We therefore refrain from using this relation in the lower level of the hierarchical model 
and replace Equation~(\ref{eq:smodel-zhomo-dist}) by

\begin{align}
\zpn &\sim \mathcal{N}(\zpps, 0.05)
\end{align}

\noindent
where \zpps\ is the value from the population synthesis 
with a standard deviation of 0.05, 
which is equivalent to a relative uncertainty of 5\%.

With this only modification to the original lower level of the hierarchical model
described in Section~\ref{sec:lower},
we apply the method
to infer the distribution of \lint\ and \eps\
for each of the synthetic planets.
We compare the marginalized posterior distributions 
of the parameters to the simulated values 
and confirm that we were able to reproduce
\mp, \rp, \lstar, and thus \teq\ for all the synthetic planets.
We repeated this procedure twice each time
assuming \lint\ follows a log-uniform $\mathcal{LU}$ distribution 
or a linear-uniform $\mathcal{U}$ distribution. 
The individual posterior distribution for most of the synthetic planets
are flat, which highlights the need for a hierarchical model 
that combines the individual distributions 
to extract useful information at the population level.
This is one of the main advantages of using hierarchical Bayesian model 
\citep{Loredo:2019}.

We therefore use the marginalized \eps\ distribution for each planet
to infer the 
heating efficiency -- equilibrium temperature (HEET)
for the synthetic population
following the method described in Section~\ref{sec:upper}.
We model the HEET distribution with both a Gaussian function
and a $4^{\rm th}$ degree polynomial.
The former function is used
to test the ability of our hierarchical model to retrieve
the input parameters of the Gaussian function 
and the latter function 
to test whether our model can indeed predict
a Gaussian-like pattern. 

\subsection{Results Using Synthetic Data}

With this procedure, we end up with four posterior distributions,
which are shown in Figure~\ref{fig:HEET-model-validation}.
The left and middle panel show the inference done assuming
\lint\ follows a linear-uniform and log-uniform distributions, respectively,
for the Gaussian function and $4^{\rm th}$ degree polynomial.
The right panel compares the Gaussian functions
shown in the left and middle panel under both prior distributions.
The black dashed line is the {\it true} distribution as implemented 
in the {\it Bern} population synthesis model. 
The dark and light shaded region shows the 68\% and 90\% credible interval.
The Gaussian-like pattern is retrieved when using a $4^{\rm th}$ degree polynomial
and also in agreement with the inferred Gaussian distribution.
The median posterior using the linear-uniform prior 
underestimates slightly the heating efficiency at the 68\% ($1\sigma$) level
but the true model is contained within the 90\% ($2\sigma$) credible interval.
This test shows that the statistical framework is able to retrieve the 
true distribution.

The light gray distribution in the right panel 
is the inference done assuming log-uniform distribution
without correcting for the choice of prior at the lower level. 
This shows the importance of understanding the prior 
at the lower level and highlights the need to re-weight the distributions. 

\begin{figure*}[ht!]
	\centering
	\includegraphics[width=1\textwidth]{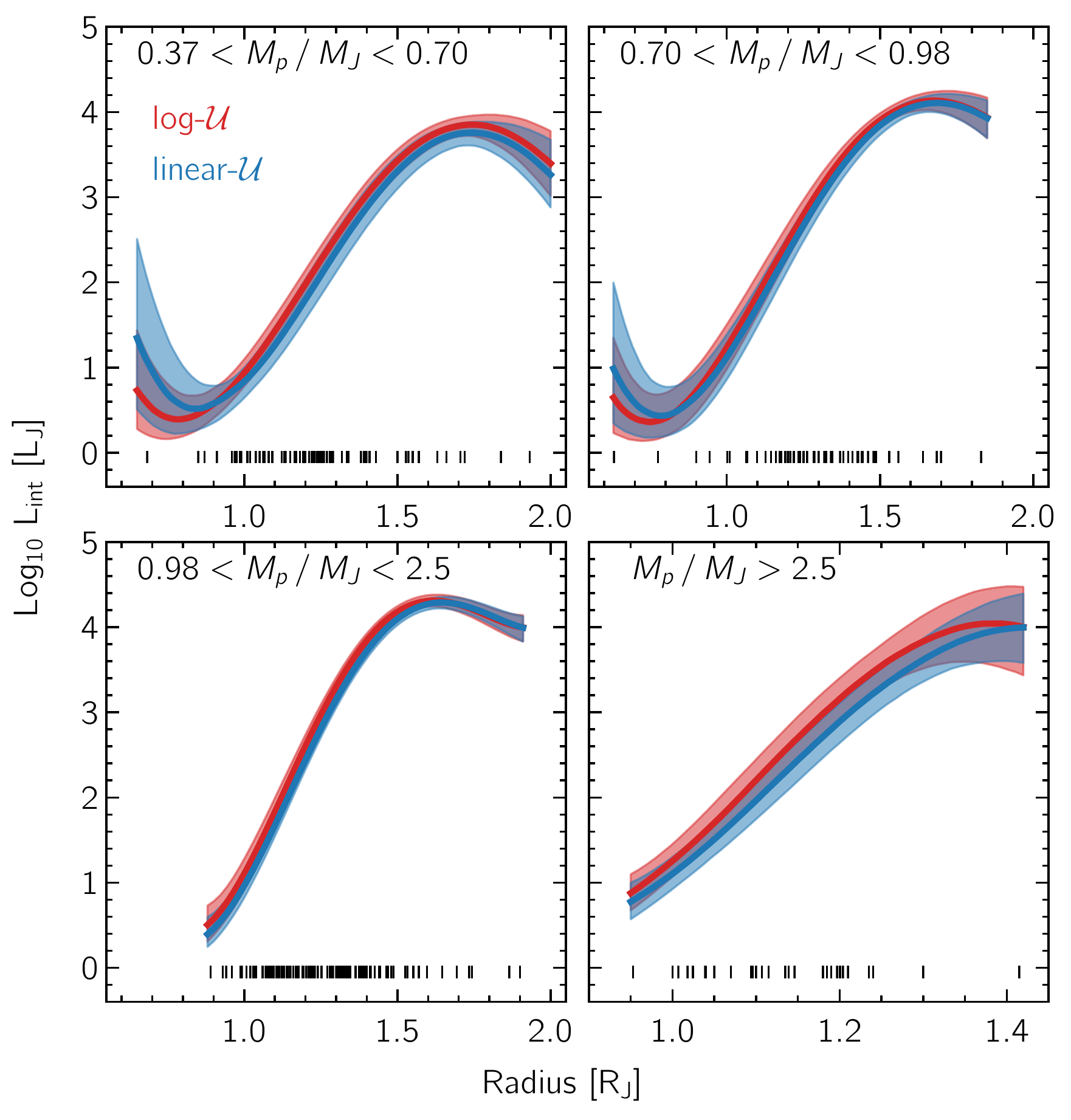}
	\caption{Mass--luminosity--radius (MLR)
	    posterior distribution for four different mass bins 
		showing the median (thick line) and 68\% credible interval (shaded area)
		assuming a uniform prior in log (blue) and linear (red) space.
		Using either prior leads to almost identical results.
		The internal luminosity is high
		with the largest planets having a luminosity $\sim$ four orders of magnitude
		larger than Jupiter.
	\label{fig:MLR_results}}
\end{figure*}

\section{Results Using Real Data}
\label{sec:results}

We now apply the model described in Section~\ref{sec:lower},
i.e. the lower level of the hierarchical model,
to infer the distribution of \lint, \eps, \tint, and \prcb\
for each of the detected planets.
In Section~\ref{sec:ppc}, 
we present diagnostic tools to validate the
lower level of the hierarchical model.
We then use the inferred posterior distributions
to study the 
mass--luminosity--radius (MLR),
\tint\ -- \teq,
\prcb\ -- \teq,
and heating efficiency -- equilibrium temperature (HEET)
distributions for the population of hot Jupiters
following the model introduced in Section~\ref{sec:upper}.
In Section~\ref{sec:mlr},
we show that by properly correcting for the choice of prior,
the MLR distribution at the population level is prior independent.
We hence present the rest of the results under the uniform in linear space prior
in Sections~\ref{sec:heet}~--~\ref{sec:interior-structure}.
For completeness, we show the results using both priors 
in Appendix~\ref{sec:appendix-info}.

\subsection{Posterior Predictive Checks}
\label{sec:ppc}
For each system, we infer the distribution of the internal luminosity 
that reproduces the observed 
radius, mass, and stellar luminosity 
while fixing the semi-major axis to the observed value.
We visually inspect each system to double check 
that the marginalized posterior distributions of the
observed parameters, \mp, \rp, \lstar, and thus \teq, are reproduced. 
Such plots are important to check that the model is a good fit 
and is thus capable of generating data that resemble the observed data.
There are in total 17 systems
where the observed mass and/or radius was not reproduced
and thus we exclude these systems from the data set 
and do not include them in the analysis presented below.
For most of the planets the radii are not possible from theoretical models
as they are at the edge of the computed grid
for a given planet mass, stellar luminosity, and semi-major axis.
The observed radii tend to be larger than what is possible from the theoretical grid
and most of these planets have masses $\mp > 2.5 \, \mjup$.
Note that for three systems the stellar luminosity
and therefore the equilibrium temperature was not reproduced
(HAT-P-20, Qatar-2, and WASP-43).
We decide however to keep these systems since
the difference in the equilibrium temperature is on the order of $\sim 30$ K
and hence the change in the internal luminosity is almost insignificant.

\subsection{Mass--Luminosity--Radius (MLR) distribution}
\label{sec:mlr}

We divide the samples into four mass ranges,
similar to the mass bins estimated by \cite{Sestovic:2018}
but further divide their second mass bin into two:
the sub-Jupiter planets ($0.37 - 0.7 \, \mjup$ and $0.7 - 0.98 \, \mjup$)
and the massive-Jupiter planets ($0.98 - 2.5 \, \mjup$ and $> 2.5 \, \mjup$).
The number of planets in each group is
86, 59, 119, and 33 planets, respectively.
To infer the MLR distribution, 
we run the model 
(Equation~(\ref{eq:upper-final}) or equivalently
Equation~(\ref{eq:upper-final-in-practice}))
for each mass bin by specifying the functional form of $g_p(x)$ 
as a $4^{\rm th}$ degree polynomial using Equation~(\ref{eq:poly}).
As such, 
$x$ is the planet radius \rp\ in Equation~(\ref{eq:poly})
and the hyperparameter 
$\tau = \left\lbrace a_0, \, a_1, \, a_2, \, a_3, \, a_4 \right\rbrace$.

At each iteration in the MCMC, 
we compute \eps\ following Equation~(\ref{eq:eps}),
where the semi-major axis is fixed to the observed value
and \lstar\ and \rp\ are drawn from the individual marginalized posterior distributions.
We further impose an additional log-normal prior on $\eps \sim \mathcal{LN}(-1, 1) $
for the planets with
an equilibrium temperature less than~1000~K.
This reflects our beliefs that planets with low equilibrium temperatures 
are not inflated \citep{Demory:2011}, and thus \eps\ should be small. 
We tested several prior probability distributions on \eps\ and 
verify that our results are not affected by the choice prior.
We repeat the full procedure twice
each time drawing samples
from the lower level of the hierarchical model
under the different priors at the lower level
($\mathcal{LU}$ and $\mathcal{U}$)
and assign uniform uninformative priors on the hyperparameters.
In Table~\ref{tab:results-mlr-linear} and 
Table~\ref{tab:results-mlr-log}
in Appendix~\ref{sec:appendix-info}
we give the 68\% credible interval values assuming 
linear-uniform and log-uniform priors
and
provide the chains online\footnote{
\href{\tinyurl}{\tinyurl}
}.

Figure~\ref{fig:MLR_results} shows the 
posterior distribution inferred for all mass bins 
under the two priors, uniform in log (red) and linear (blue) space.
Notice that the lower right panel has a different scale 
to better visualize the results.
Each data point is represented by a small line at the bottom of the plot 
at the corresponding radius.
Such plots are called rug plots and 
are used to visualize the distribution of the data.
The posterior distributions under both priors
are almost identical and indistinguishable
inline with the conclusion reached in Section~\ref{sec:model-validation}
by validating the hierarchical model on synthetic data.
There are few differences between both models,
such as at small radii for the least massive planets
and at large radii for the most massive ones. 
These differences are mainly dominated by the small number of planets 
in these regions.
This highlights the importance of re-weighting the samples 
by dividing by the prior used to do the sampling 
at the lower level of the hierarchical model.
For the rest of the paper, we show the results under the prior uniform in linear space,
but confirm that the choice of prior at the lower level of the hierarchical model
does not affect the main results and conclusions.

The basic shape of the MLR relation is 
similar across all mass bins,
where as expected larger planets have higher internal luminosity
with a plateau around 1.6 \rjup\
beyond which the luminosity is almost constant.
The small drop towards high radii 
has little statistical significance 
and likely reflects the choice of a fourth-order polynomial.
The inferred internal luminosity
for most of the planets is several orders of magnitude
larger than Jupiter, reaching even up to four orders of magnitude.
We also find that the internal luminosity is mass dependent,
with the most massive planets having the highest internal luminosity. 

A noticeable feature is that the sub-Jupiter planets
with masses $0.37 - 0.98 \, \mjup$
and radii less than 1 \rjup\ 
have an internal luminosity larger than Jupiter.
At first glance,
one might expect such planets to have 
an internal luminosity smaller than Jupiter's.
We note however that the planets that have an equilibrium temperature less than 1000 K,
indeed tend to have $\lint \sim 3 \, \ljup$ and not more.
A higher luminosity is expected because, even with $\teq < 1000$~K,
these planets are still much closer than Jupiter,
which reduces the cooling rate
and thus leads to higher internal luminosity.
As for the planets that have equilibrium temperature larger than 1000 K, 
they tend to have higher fraction of heavy elements distributed in the envelope.
There are only two sub-Jupiter planets in our sample that have radii 
less than 0.7 \rjup, K2-60 and WASP-86,
both of which require large fraction of heavy elements, 0.64 and 0.8, respectively,
ruling out values less than 0.5. 
The high fraction of heavy elements explains the high luminosity values 
and the small number of planets with radii less than 1 \rjup\
is why the distribution is poorly constrained in this regime.

\begin{figure*}[ht!]
	\centering
	\includegraphics[width=\textwidth]{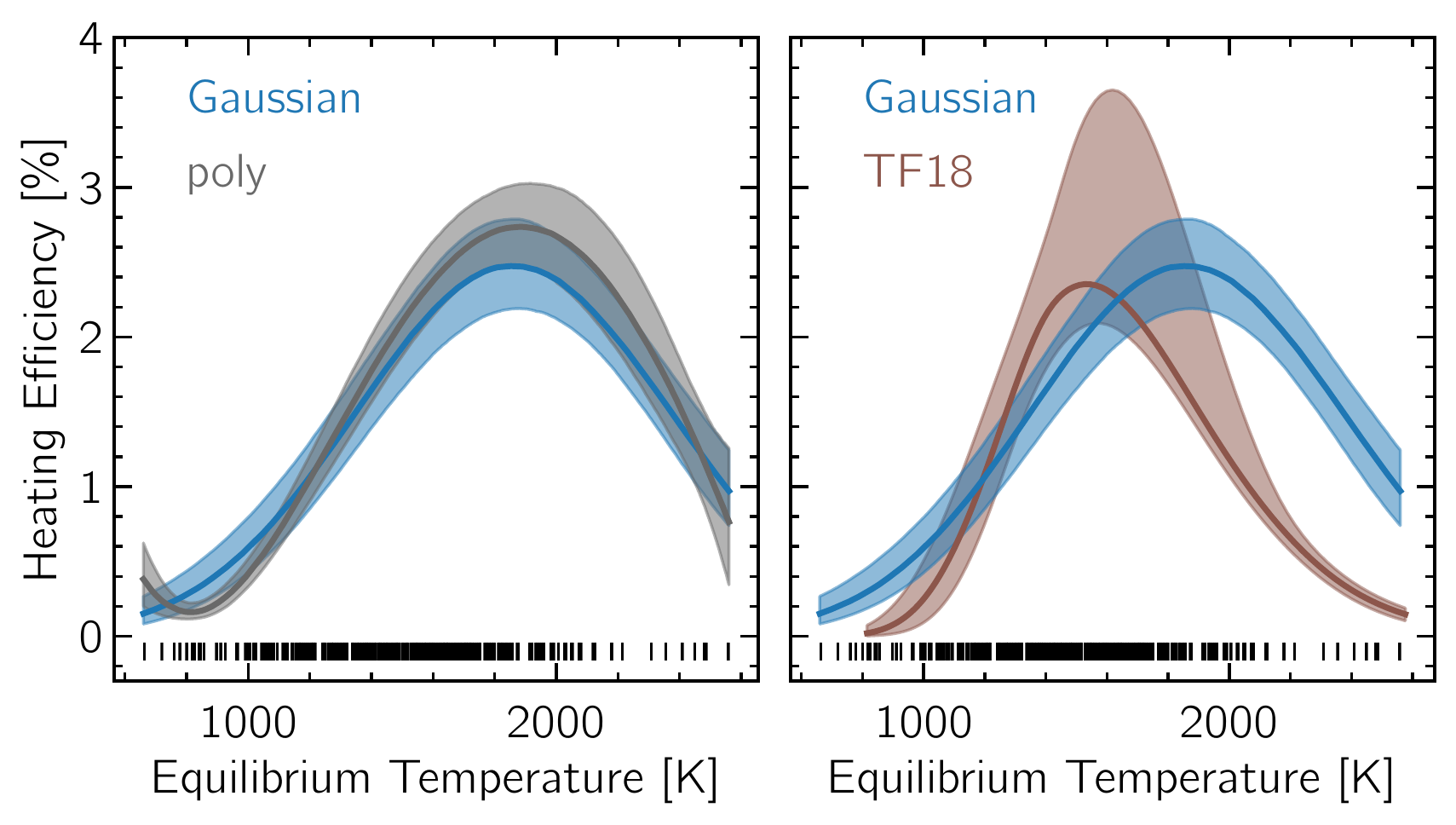}
	\caption{{\it (Left):} 
 	    Heating efficiency -- equilibrium temperature (HEET) 
	   posterior distribution under the linear-uniform prior
		using a Gaussian function and a $4^{\rm th}$ degree polynomial.
		{\it (Right):} the Gaussian function shown on the left side in comparison 
		to the HEET posterior distribution inferred by \citetalias{Thorngren:2018}. 
		The shaded region show the 68\% credible interval.
		There is a good agreement between the Gaussian and poly models,
		which shows that indeed the HEET distribution follows a Gaussian function.
		Our results are in agreement with the findings of \citetalias{Thorngren:2018}
		although the peak in our models is shifted to higher equilibrium temperatures.
	\label{fig:HEET_results}}
\end{figure*}

\subsection{Heating Efficiency Equilibrium Temperature (HEET) distribution}
\label{sec:heet}

Similar to the previous section, 
we also apply the model defined in Section~\ref{sec:upper}
to study the HEET relation using both function forms:
$g_p$ a $4^{\rm th}$ degree polynomial (Equation~(\ref{eq:poly}))
and
$g_g$ a Gaussian function (Equation~(\ref{eq:gauss}))
with $\taw~=~\left\lbrace \epsmax, \, \teq{_0}, \,  s \right\rbrace$.
The former is a flexible function 
that allows us to constrain the general shape of the relation 
by relying entirely on the data as motivated in the previous section,
while the latter allows us to compare our results to 
\citetalias{Thorngren:2018} and to theoretical predictions.
Following the same methodology applied to the MLR relation, 
we further impose for the $g_p$ model the $\mathcal{LN}(-1, 1) $ 
prior on the heating efficiency
for planets with equilibrium temperatures less than~1000~K.
Note that the individual distributions are flat, similar to the 
distributions of the synthetic planets and 
useful information can only be extracted by combining
the individual distributions. 

In Table~\ref{tab:results-heet-poly} in Appendix~\ref{sec:appendix-info} we give the 68\% 
credible interval values assuming $\mathcal{LU}$ and $\mathcal{U}$ priors 
using the polynomial model.
The Gaussian models are shown in Table~\ref{tab:hbm-gauss-results}
and the MCMC chains are available online\footnote{
\href{\tinyurl}{\tinyurl}
}.
The true distribution 
that was used to generate the synthetic data 
in Section~\ref{sec:model-validation}
are the values we obtained using the log-$\mathcal{U}$ prior
and shown in Table~\ref{tab:hbm-gauss-results}.

The left panel of Figure~\ref{fig:HEET_results} shows that the posterior distributions
are similar under both functional forms,
with the polynomial function leading slightly to higher efficiencies.
Using an independent interior structure model 
and a larger sample focused on FGK main-sequence stars,
our results are qualitatively consistent with \citetalias{Thorngren:2018}.
We  confirm the Gaussian pattern
holds independent of the choice of prior 
(see Figure~\ref{fig:HEET_appendix} in Appendix~\ref{sec:appendix-info}).
This pattern was predicted by ohmic dissipation
first based on simulations \citep[e.g.][]{Menou:2012}
and then later supported by \citetalias{Thorngren:2018}.
Our analysis provides further evidence of the Gaussian-like distribution.

\begin{table}[t!]
\caption{%
Comparison of the Gaussian function using the log and linear uniform prior
along with comparison to \citetalias{Thorngren:2018} results.
\label{tab:hbm-gauss-results}}
\begin{center}
\begin{tabular}{cccc}
\hline
\hline
\taw    &  log-$\mathcal{U}$  & linear-$\mathcal{U}$ & \citetalias{Thorngren:2018} \\
\hline
 & & & \\

\epsmax [\%] & $2.46_{-0.24}^{+0.29}$ & $2.49_{-0.28}^{+0.31}$ & $2.37^{+1.30}_{-0.26}$ \\
 & & & \\
$\teq{_0}$ [K]   & $1982_{-58}^{+83}$ & $1862_{-61}^{+67}$ & $1566_{-61}^{+55}$ \\
 & & & \\
$s$ [K]       & $532_{-73}^{+110}$ & $508_{-48}^{+66}$ & $327^{+25}_{-43}$ \\

 & & & \\
\hline
\end{tabular}
\end{center}
\end{table}

\subsubsection{Comparison to \citetalias{Thorngren:2018}}

To compare our results to \citetalias{Thorngren:2018},
we report the median and the 68\% credible interval of \citetalias{Thorngren:2018}
in Table~\ref{tab:hbm-gauss-results}.
We also show the posterior distributions 
in the right panel of Figure~\ref{fig:HEET_results}.
The heating efficiency increases 
until a  maximum is reached at $\teq{_0}$,
beyond which the efficiency decreases.
Our result regarding the maximum heating efficiency 
agrees well within $1\sigma$ with \citetalias{Thorngren:2018},
where we determine $\epsmax \sim 2.50 \, \%$,
compared to $\sim 2.37 \, \%$.
In our model, the peak occurs at $\sim 1860$~K,
while \citetalias{Thorngren:2018} estimate the transition at $\sim 1566$~K.
This discrepancy can be attributed either to differences 
in the statistical framework or in the interior structure model.
We will address both next.

While \citetalias{Thorngren:2018} used a non-parametric 
Gaussian Process (GP) approach to model the HEET distribution, 
they found consistent results with the Gaussian function.
In our study, instead of modeling the HEET distribution with a 
non-parametric GP model, 
we use a flexible $4^{\rm th}$ degree polynomial
that we stress is very fast to compute\footnote{
It takes around 5 minutes on a modern laptop
to evaluate the upper model for $K=1$, 
i.e. without accounting for uncertainties on the x-axis.
For $K=2000$, it takes around 2 CPU hours 
on a server using 20 cores.}
and find consistent results with the Gaussian function. 
To test whether this discrepancy could be due to the statistical framework,
we ran our statistical model using the individual distributions 
inferred by the analysis of 
\citetalias{Thorngren:2018}, which were shared with us.
Note that using their data, there is no need to re-weight
the distributions. 
See Section~\ref{sec:dt-results-not-prior-dependent} for a detailed explanation.
We confirm we were able to recover their posterior distribution
using both a Gaussian function and a $4^{\rm th}$ degree polynomial.
There is a very good agreement at the $1\sigma$ level, 
except for $ \teq < 1000$~K where the results are slightly different.
The amplitudes are in agreement at the $1\sigma$ level 
even though we find tighter credible intervals
at the $1\sigma$ level but very good agreement at $2\sigma$.
With this we conclude that the differences are not due to the statistical framework.

We now study the differences in the interior structure model
by comparing
the solid black and dashed red models in the left panel of Figure~\ref{fig:sample}
computed using our model \completo\ 
and by \citetalias{Thorngren:2018}\footnote{
This is the same red dashed model 
shown in Figure~1 in the \citetalias{Thorngren:2018} paper.}, 
respectively.
Both of these models are for a 1 \mjup\ planet with a pure H/He envelope
without accounting for inflation. 
In our case the planets are 5~Gyr old.
Using our structure model,
\rp\ ranges between $1.12 - 1.22$~\rjup\ 
for \teq\ between $800-2500$~K.
In comparison, 
\rp\ is between $1.11 - 1.31$~\rjup\
using the \citetalias{Thorngren:2018} models
for the same \teq\ interval.
While the the radii at low \teq\ are almost identical, 
the differences at high \teq\ are up to $\sim 0.1 \, \rjup$.
We notice that both models lead to different radii 
starting at $\teq > 1500$~K.
This difference could explain the higher heating efficiency we infer 
at $\teq > 2000$~K.  
Since the planets in our model have smaller radii starting at 1500~K,
then more energy needs to be transported into the interior 
to reproduce the observed radius, 
leading to higher \eps\ values
compared to \citetalias{Thorngren:2018}.
Note that this is a simple case scenario where the models
are for planets made entirely of pure H/He.
While this scenario explains the trend,
more tests are needed to compare the radii at different \teq\
for different fraction of heavy elements
since the details of the EOS for the heavy element 
could in principle also be a source of discrepancy between the models.

The discrepancy in the radii could be caused by differences
in the atmospheric modelling. 
We next compare the atmospheric models of \petit\
and \cite{Fortney:2007},
which was used by 
\citetalias{Thorngren:2018} and did not include TiO and VO.
The previously computed atmospheric grid using \petit\ 
include TiO and VO (see Section~\ref{sec:atmo-model}).
We therefore calculate the PT structure for 
a typical hot Jupiter at solar compostion, 
$\tint=100$~K, $\teq=2000$~K, 
and $\logg=3.27$ without accounting for these absorbers.
We find that in the absence of TiO and VO
no inversion was formed with similar profiles using both atmospheric models.
We then calculate the entropy of both structures using our EOS.
We find that the entropy in the convective layers at pressure of $10^4$~bar
is $7.55$~kB/baryon using \petit\ 
compared to an entropy of $ < 7.65$~kB/baryon 
at pressure of 1000~bar
using the models of \cite{Fortney:2007}.
Note that at this pressure the structure is still not convective 
and thus the entropy is smaller than $7.65$~kB/baryon
in the convective layers
and most likely the difference is $ < 0.1$~kB/baryon between both models.
Note that when including TiO/VO the entropy is 
$\sim 7.5$~kB/baryon.
A higher entropy leads to larger radii
\citep[e.g.][]{Spiegel:2013,Marleau:2014}
and as such we conclude that the difference between both models
presented in Figure~\ref{fig:sample}
could be due to differences in the opacities 
high up in the atmosphere, namely TiO/VO,
which then have a larger effect on the deep atmosphere
due to the (anti) greenhouse effect.

A more systematic comparison between both atmospheric models
for different \teq\ and \logg\
is required to further quantify the discrepancies,
which is beyond the scope of this paper.
We note that both studies do not account for systematic differences 
in the structure models.
Such comparisons will therefore allow similar future studies
to account for the systematic differences
and thus infer more reliable credible intervals.

\subsubsection{Are the Results of \citetalias{Thorngren:2018} Prior Dependent?}
\label{sec:dt-results-not-prior-dependent}

In short, no.

In our study, we sample \lint\ and then compute \eps\
using Equation~(\ref{eq:eps}).
We imposed two different prior distributions on \lint\ 
because we do not have a priori knowledge which distribution best 
represent the population. 
Within a statistical framework, a log-uniform distribution 
is preferred in order to explore the entire parameter space.
However, as we showed in the right panel of Figure~\ref{fig:prior_lumi_eps},
this leads to biases giving more weight to lower heating efficiency. 
Whereas, a linear-uniform prior distribution on \lint\
leads to approximately a uniform prior distribution on \eps. 

As discussed in Section~\ref{sec:upper-derivation},
the choice of prior distribution is important as
the posterior distribution scales to the number of planets $N$. 
Thus the need to re-weight the distributions at the upper level.
Another way to approach this study is to perform a full 
hierarchical Bayesian modeling where 
the inference on both the individual planets and population is
made simultaneously \citep{Wolfgang:2015, Wolfgang:2016}.

In the study of \citetalias{Thorngren:2018}, the setup is different. 
They sample \eps\ and impose a uniform prior between $0-5\%$.
There are no additional conditions that truncate the \eps\ 
distribution, which itself is flat non-informative prior.
Therefore, there is no need to re-weight the distributions.

In general, it is always a good practice 
to {\it sample the prior PDF distribution} \citep{Hogg:2018}.
This step is important to 
check whether MCMC samples correctly
the specified prior distributions.

\subsubsection{Is the Decrease in Efficiency at High \teq\ Real?}
\label{sec:heet-significant-decrease-high-teq}

\begin{figure}[t!]
	\includegraphics[width=0.5\textwidth]{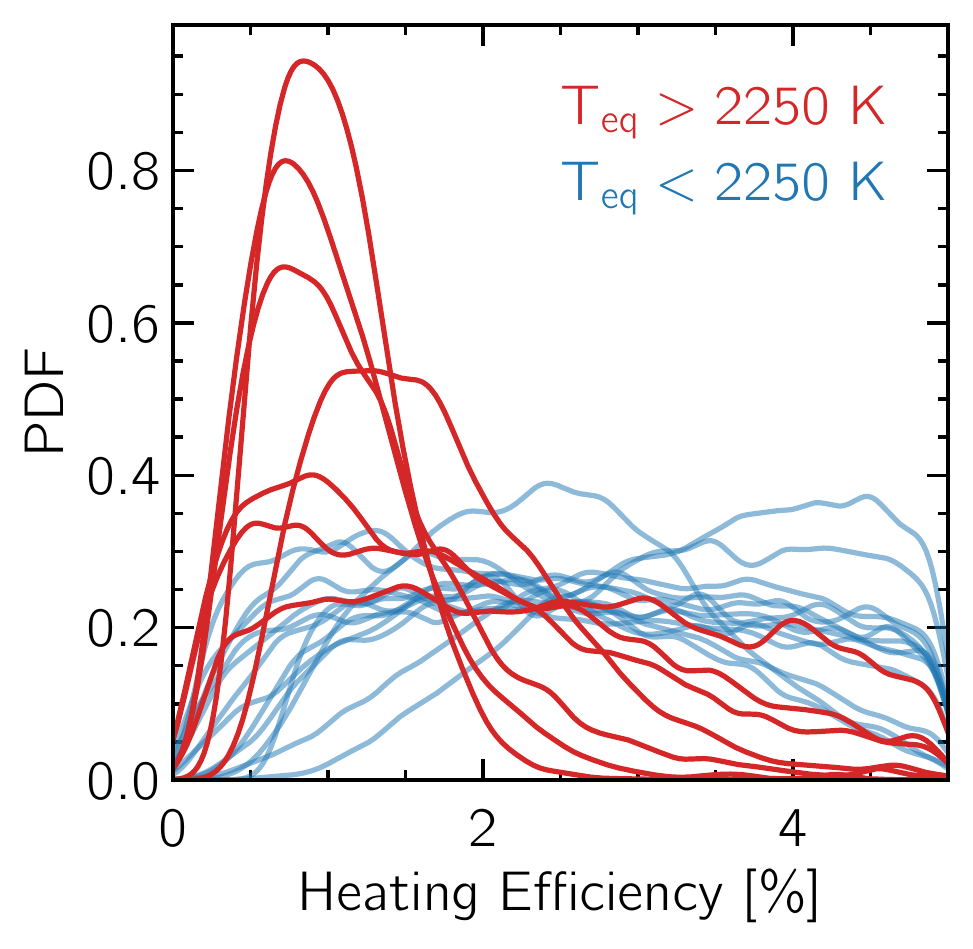}
	\caption{Posterior distributions of the heating efficiency \eps\ 
	for all the planets with $\teq>2000$~K. 
	The colors indicate planets with $\teq<2250$~K and $\teq> 2250$~K
	in blue and red, respectively.
	Six out of the seven planets shown in red favor small heating efficiency values
	with the most probable value close to $\eps \sim 1\%$.
	This provides evidence that the interior structure model disfavor high \eps\ values
	and thus the decrease seen in the HEET distribution is real given our structure model.
	\label{fig:heet-decrease-real}}
\end{figure}

In order to check whether the decrease in the heating efficiency 
at high \teq\ is real or not, 
we re-ran the lower-level model for all the planets with $\teq > 2000~K$.
We assumed a linear-uniform prior for \lint\ to ensure
the results are not biased towards small \eps\ values.
Additionally, we did not put constraints on \tint.
This is important for the highly irradiated planets
where $\tint \sim 1000$~K translates to $\eps < 5\%$.

The main goal of this exercise is to check whether 
the interior structure model 
allows for high \eps\ values for all the planets with $\teq > 2000$~K.
The cutoff was chosen to be close to the peak of the Gaussian 
function as inferred previously in Section~\ref{sec:heet}
(see also Table~\ref{tab:hbm-gauss-results}).
This allows us to compare the \eps\ distribution 
for planets with \teq\ close to 2000~K
to the highly irradiated ones. 
If the structure model allows for high \eps\ values
for the mostly irradiated planets, 
then we do not have enough evidence that the decrease in 
the heating efficiency is real. 
Otherwise, there is evidence that the decrease is real. 

Figure~\ref{fig:heet-decrease-real} shows the posterior distributions
of the heating efficiency \eps.
Planets with $\teq < 2250$~K are in blue
and planets with $\teq > 2250$~K are shown in red.
As can be seen, all but one of the planets with 
$\teq > 2250$~K
disfavor high \eps\ values with the most probable value 
around $\sim 1\%$.
The only planet where high \eps\ values are likely
is the massive hot Jupiter WASP-18\,b \citep[10.52 \, \mjup;][]{Maxted:2013}.
At this mass, the radius is a weak function of \lint\ and \eps\
as it is difficult to inflate massive planets \citep{Sestovic:2018}.
We therefore consider WASP-18\,b an exceptional case, 
especially that all the planets with 
$\teq > 2000$~K have $\mp < 2.4 \, \mjup$.

This analysis illustrates that 
hot Jupiters with $\teq > 2250$~K
require low heating efficiencies to reproduce their radii
using our interior structure model,
which supports the Gaussian-like pattern and the decrease 
at 2000~K.
With 20 planets having $\teq > 2000$~K 
out of which only 7 planets have $\teq > 2250$~K,
future ultra-short hot Jupiters discoveries are essential 
to further confirm or refute this trend.

\subsection{Distributions of Internal Temperature and Pressure at the RCB}
\label{sec:interior-structure}

\begin{figure*}[t!]
	\includegraphics[width=\textwidth]{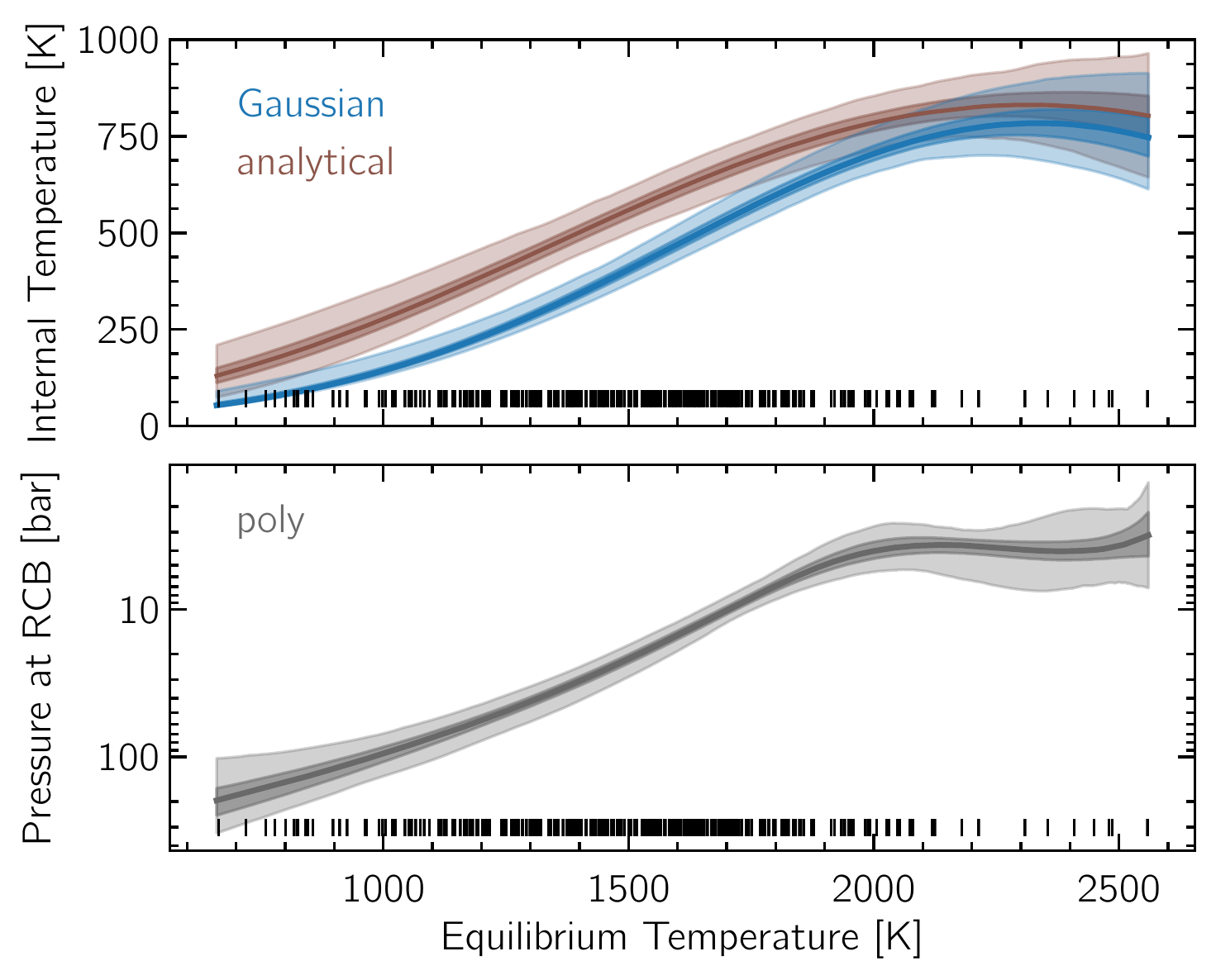}
	\caption{\tint--\teq\ and \prcb--\teq\ diagrams in the upper and lower panel, respectively.
	The dark and light shaded regions present the 68\% and 95\% credible intervals.
	Although the analytical approach overestimates the internal temperature at \teq\ between $1000-1800$~K,
	there is a good agreement at $\teq<1000$~K and $\teq>1800$~K.
	Due to the increase in the internal temperature with equilibrium temperature, 
	the \prcb\ moves to lower pressures with increasing \teq, reaching up to $\sim 3$ bar for the 
	most irradiated planets.
	\label{fig:TT_PT_results}}
\end{figure*}

Having inferred the population level distributions of
the internal luminosity distribution and the heating efficiency,
it is interesting to study the effect of energy dissipation
on the interior structure of the planet.
In particular, we show that 
as a consequence of transporting energy into the interior,
hot Jupiters have very hot interiors 
which in turn pushes the RCB to low pressures.
Our findings are in agreement with \cite{Thorngren:2019} (hereafter \citetalias{Thorngren:2019}),
where they used the HEET relation presented in \citetalias{Thorngren:2018}
to compute \tint\ and then generate PT atmospheric models 
for a range of \teq\ and surface gravities
to locate the \prcb.

As mentioned in Section~\ref{sec:lower},
we keep track of the PT profiles,
and thus we can infer the distribution of the 
internal temperature and the pressure of the RCB
for each planet.
We again apply the model defined in Section~\ref{sec:upper}
to study the distributions of \tint\ and \prcb\ 
as a function of \teq.
We model the distributions,
\tint~--~\teq\ and \prcb~--~\teq,
as a Gaussian function and $4^{\rm th}$ degree polynomial, respectively.
At steady state, 

\begin{align}
\tint  &= \eps^{1/4} \teq \label{eq:tint-teq-relation} \\
         &= g_g \left(\teq \right)^{1/4} \teq \label{eq:tint-epsilon-teq-relation}
\end{align}

\noindent
where the last equation was obtained by replacing
$\lint = 4 \pi \rp^2 \sigma \tint^4$
in Equation~(\ref{eq:lint})
and combining Equations~(\ref{eq:teq}) and (\ref{eq:incident-flux}).
We use the samples from our previous analysis using the Gaussian model 
(see Section~\ref{sec:heet})
to compute \tint\ using Equation~(\ref{eq:tint-epsilon-teq-relation})
and compare the results to the hierarchical Bayesian approach.
We refer to the former method as the analytical approach.
For all the models, we assign uniform distributions on all the hyperparameters.

Figure~\ref{fig:TT_PT_results} shows the inferred posterior distribution
for the internal temperature (upper panel) and pressure at the RCB (lower panel)
as a function of the equilibrium temperature.
The analytical approach leads similar results to the Bayesian approach 
at the lowest and highest equilibrium temperatures.
However, \tint\ is overestimated at the 2$\sigma$ level
for \teq\ between 1000 and 1800~K.
This difference could be because we did not account for intrinsic scatter
in the model, which we leave for future work.

For both models, almost all hot Jupiters have \tint\ larger than 200~K,
while, for comparison, the internal temperature of Jupiter is 100~K \citep{Li:2012, Guillot:2014}.
This is expected given the observed inflated radii.
WASP-121\,b is the only exoplanet to date
whose internal temperature was constrained 
from observations of Mg and Fe in the transmission spectrum,
with a reported value of 500~K \citep{Sing:2019}.
With an equilibrium temperature of $\teq = 2358 \pm 52$~K 
\citep{Delrez:2016},
we infer $\tint \sim 800$~K and 
by inspecting the individual posterior distribution of WASP-121\,b, 
we rule out values below 500~K.
This is the first hint from observations that hot Jupiters possess hot interiors,
which is associated with a high internal entropy.

Another notable parameter to study is the pressure of the RCB 
as this partly controls the planetary cooling rate
\citep{Arras:2006, Spiegel:2013}.
It is known that 
high equilibrium temperature pushes the RCB deeper into the planet 
\citep[e.g.][]{Fortney:2007},
however high internal temperature pushes the RCB to lower pressures.
Therefore, the location of the RCB is not known beforehand
for planets with high equilibrium and internal temperatures.
The lower panel of Figure~\ref{fig:TT_PT_results} 
shows that the RCB is situated at low pressures 
or at shallow depths for high \teq.
The effect of the high internal temperature is thus dominant.
The planets receiving high stellar irradiation tend to have hot interiors, 
typically around $\sim 800$ K,
which pushes the RCB to low pressures, 
reaching $\sim 3$ bar for the most extreme cases.

Our results agree well with \citetalias{Thorngren:2019}.
While we report a maximum \tint\ of 800~K at $\teq \sim  2500$~K,
\citetalias{Thorngren:2019} finds the maximum \tint\ of 700~K at $\teq \sim  1800$~K.
The difference is mainly due to the differences in the \eps(\teq) distribution 
(see Section~\ref{sec:heet}).
We estimate the RCB to be at 100~bar and 4~bar
for $\teq = 1000$~K and 2000~K, respectively, 
in agreement with the findings of \citetalias{Thorngren:2019}.
Qualitatively, both models show the same pattern
where the hot interior of hot Jupiters 
drive the RCB to lower pressures.

We provide the 68\%  credible interval values for the 
Gaussian model under both priors for the \tint--\teq\ distribution 
in Table~\ref{tab:results-tt-gauss}.
The values for the \prcb--\teq\ distribution
are shown in Table~\ref{tab:results-pt-poly}, also under both priors.
For both distributions the chains are available online\footnote{
\href{\tinyurl}{\tinyurl}}.

\section{Discussion}
\label{sec:discussion}

Building on the assumption that hot Jupiters 
are inflated because of a process leading to high internal luminosity,
we infer for each planet the internal luminosity distribution 
that reproduces the radius given the 
planet mass and equilibrium temperature from observations
and using the mass--heavy-element relation \citep{Thorngren:2016} 
as a prior for the fraction of heavy elements.
We then combine the individual distributions to constrain 
the population mass--luminosity--radius (MLR) distribution.
Assuming that the source of extra heat in the interior 
is the irradiation by the host star
(e.g. tides or magnetic fields), 
we then compute the fraction of the incident flux \eps\ 
deposited in the interior and study the 
heating-efficiency--equilibrium-temperature (HEET) distribution for the full population.
Finally, as a by-product of our structure model,
we can also gain insights into the interior structure of the planets
by inferring the distributions of the internal temperature
and the pressure at the RCB.

In what follows,
in Section~\ref{sec:discussion-interior} we
discuss the consequences of the hot interior hot Jupiters possess
on the internal structure.
Then we discuss our results within the context of the 
competing heating mechanisms,
mainly ohmic dissipation in 
Section~\ref{sec:discussion-ohmic-dissipation} and 
advection of potential temperature in Section~\ref{sec:discussion-advection}.
In Section~\ref{sec:discussion-general-comparison},
we give a general comparison with analytical relations
and discuss the limitations and caveats of our results 
in Section~\ref{sec:limitations}.

\subsection{Insights into the Interior Structure of Hot Jupiters}
\label{sec:discussion-interior}

We have shown that hot Jupiters have hot interiors, 
with an internal temperature as high as 800~K.
This has important consequences 
on the location of the RCB,
which in turn is important for the heating mechanism.
\cite{Komacek:2017b} showed that heat dissipated 
in the convective layers suppresses cooling
and thus enables the planet to maintain a large radius.
Heat deposited in the radiative layer, however,
does not significantly inhibit cooling.
Most it is re-radiated away
leading therefore to small radii.
The location of the RCB is hence important to 
constrain the minimum depth at which the heat should be deposited
and thus the efficiency of the heating mechanism.
We find that the RCB is around 100~bar for planets
with equilibrium temperatures of about 1000~K,
and can reach 3~bar for the highly irradiated planets,
which is significantly lower than previous estimates of 1000~bar
without accounting for a bloating mechanism \citep{Fortney:2007}.
Our results are in agreement with \citetalias{Thorngren:2019}
based on coupling the heating efficiency relation \citepalias{Thorngren:2018}
to a planetary interior structure model.

Mechanisms based on transporting heat into the deep interior, 
such as atmospheric circulation \citep{Showman:2002},
ohmic dissipation \citep{Batygin:2010}, or 
advection of potential temperature \citep{Tremblin:2017} 
rely on the existence of winds in the interior.
While the extra heat must be deposited in the convective layer 
in order to inflate the planet
\citep{Komacek:2017b},
the actual wind speeds are not constrained 
from Global Circulation Models (GCMs) 
due to inaccurate coupling between the atmosphere and deep interior.
Recently, \cite{Carone:2019} showed that through a better treatment
of the lower boundary condition, i.e. by accounting for a hot interior,
shallow zonal winds are present at 100 bar.
With new estimates and better understanding 
of the internal temperature and pressure at the RCB,
the depth of the wind zone and wind speeds
can be constrained from GCM models,
which in turn will be key inputs 
to further study the efficiency of the proposed mechanisms.

\subsection{Comparison to Ohmic Dissipation}
\label{sec:discussion-ohmic-dissipation}

The general idea of ohmic dissipation is that 
equilibrium temperatures larger than 1000 K
lead to thermally ionized atmospheres
that couples to the magnetic field
and in the presence of strong winds produces currents,
which then dissipate thermally in the deep interior 
\citep{Batygin:2010, Batygin:2011}.
However, in the high equilibrium temperature regime
and therefore high atmospheric ionization fraction,
ions slow down the winds due to Lorentz force, 
which in turn decrease the efficiency of ohmic dissipation 
\citep{Perna:2010a, Perna:2010b}.
Scaling law relations based on ohmic dissipation showed
that indeed the heating efficiency increases with equilibrium temperature
until a maximum is reached beyond which the efficiency decreases
\citep{Menou:2012},
which was also confirmed by \citetalias{Thorngren:2018} 
and now in our study.
The scaling laws also suggest that the location of the peak 
depends on the strength of the magnetic field.
Therefore, studying the functional form of the HEET distribution 
provides insights within the context of ohmic dissipation.

Based on our analysis,
we find that the HEET distribution can be modeled by a Gaussian function,
in agreement with \citetalias{Thorngren:2018}
and with the theoretical predictions.
We find however that the location of the peak is at 1860~K,
which is higher compared to the work of \citetalias{Thorngren:2018}
that reported the peak around 1566~K
(see Table~\ref{tab:hbm-gauss-results}).
\cite{Menou:2012} showed that the transition is 
a function of the strength of the magnetic field
(see his Figure~4)
where stronger magnetic fields push the peak to higher equilibrium temperatures
(the peak is at $\sim 1800$~K for a 30 G field). 
\cite{Ginzburg:2016} estimate the transition around $\sim 1500$~K
based on analytical models
and \cite{Rogers:2014} at $\sim 1500-1600$~K based on 
magnetohydrodynamic simulations.
\cite{Yadav:2017} estimate the surface magnetic field strength of hot Jupiters
using the energy flux scaling law from \cite{Christensen:2009} 
and account for the extra heat injected using 
the heating efficiency relation presented by \citetalias{Thorngren:2018}.
They found magnetic field strengths around $50-100$~G 
for the most inflated hot Jupiters.
There are no theoretical atmospheric circulation models 
with such strong magnetic fields,
which might hence change the location of the peak.
The transition is still not well constrained and
might depend on the field strength
but the Gaussian distribution is robust and most importantly is prior independent.
Future observations of magnetic field strengths could potentially provide a better overview
but for now they remain unconstrained from an observational point of view
\citep[for a current review see][]{Griessmeier:2017, Lazio:2018}.

\subsection{Comparison to Advection of Potential Temperature}
\label{sec:discussion-advection}

\begin{figure}[t!]
	\includegraphics[width=0.5\textwidth]{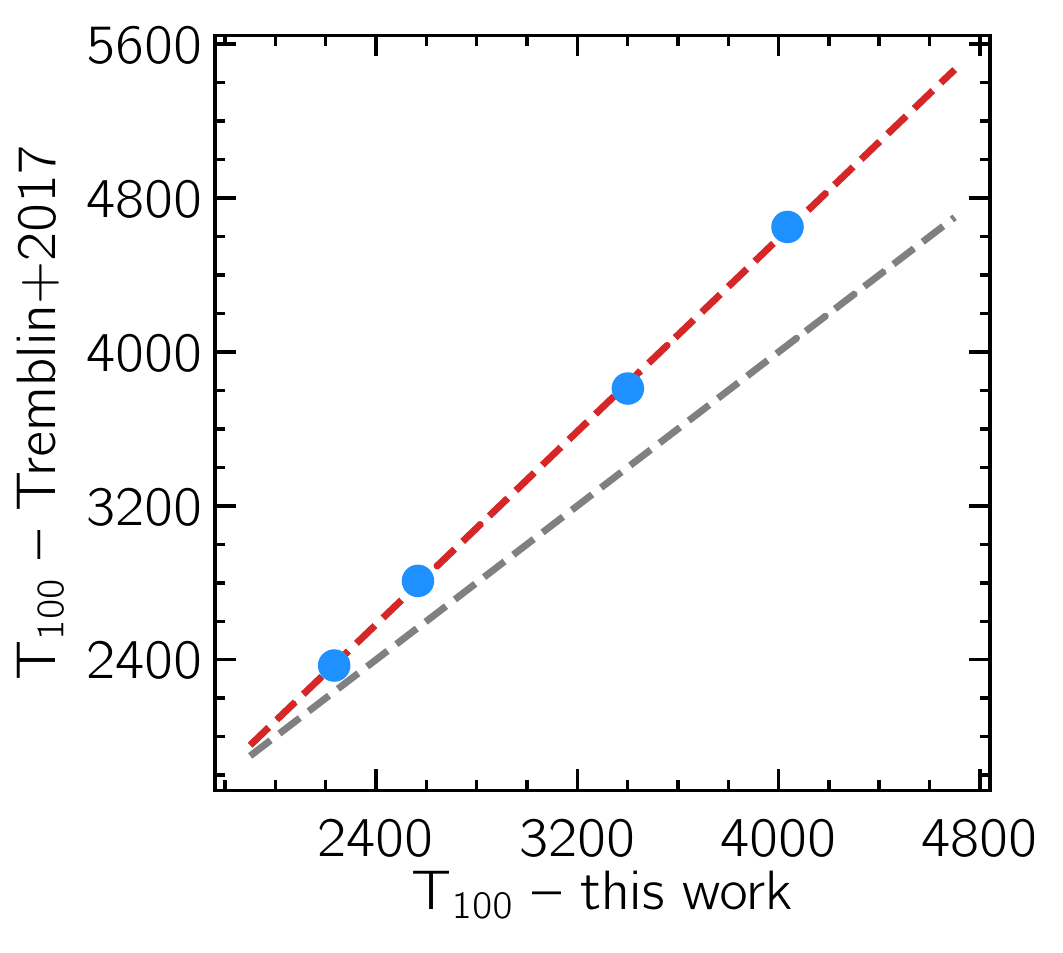}
	\caption{Temperature at 100~bar derived from our PT structures
	compared to the values from the average PT profiles
	using 2D circulation models presented by \cite{Tremblin:2017}
	resulting from the advection of potential temperature.
	All the models correspond to a planet with $\logg = 2.97 \pm 0.15$
	and increasing stellar incident flux. 
	The gray dashed line shows the 1:1 relation 
	while the red dashed line shows the fit to the data.
	\label{fig:tremblin_comparison}}
\end{figure}

Another source of heat could be the movement of high-entropy fluid parcels
deeper into the atmosphere, a process known as advection of potential temperature.
Within this context, \cite{Tremblin:2017} suggested, 
using two-dimensional (2D) circulation model,
that this process leads to a hot interior
that can naturally explain the radius anomaly of hot Jupiters.
This was further supported recently by 3D GCM simulations
\citep{Sainsbury-Martinez:2019}.
The 2D models show that a stronger stellar incident flux
leads to hotter interior adiabat (see their Figure~5).
We compare our results based on the 1D model 
to the 2D models by selecting 
four planets from our sample that matches their simulation parameters,
i.e. $\logg = 2.97 \pm 0.15$ with the corresponding equilibrium temperatures.
We do not include the model with the lowest equilibrium temperature
($\sim 500$~K) as it does not match 
any of the selected systems in our sample.
The planets we selected as a function of increasing stellar incident flux are
HAT-P-17\,b, Corot-4\,b, HD209458\,b, and HATS-35\,b.
We then compare the temperatures at 100~bar (T$_{100}$)
using the PT profiles based on the 2D models 
to the ones based on our 1D model 
presented in Section~\ref{sec:interior-model}.
The results are illustrated in Figure~\ref{fig:tremblin_comparison},
where the derived temperatures at 100~bar
are shown in blue circles
and the red dashed line shows the fit to the data.
The gray dashed line shows the 1:1 relation
on which the points would lie if their model 
and our data derived from observations would predict identical temperatures.
We find that roughly the results agree well with a slope of 1.25,
deviating from the 1:1 relation.
We note however that these values are model dependent
and any change in the treatment of the atmospheric model,
e.g. including clouds and new opacity sources, will change these values.
The temperatures estimated from the average PT profiles
using the 2D circulation models are larger than 
the values predicted by our model, varying from 6\% up to 15\%
for the most irradiated planets.
This is expected since the 2D models tend to overestimate
the radii compared to the observed ones \citep{Tremblin:2017}.
Our results concerning the adiabatic profile are also in agreement,
where the 2D and 3D atmospheric circulation models
suggest a hot adiabat starting at $\sim$10~bar,
significantly at lower pressures compared to standard irradiated models
\citep[e.g.][]{Fortney:2007}.
This is in agreement with our findings and conclusions
that future GCM models should account for the extra heat in the interior
of inflated hot Jupiters
and in-line with the work of \cite{Carone:2019}.
We note however that convection is not included
in the models of \cite{Tremblin:2017}.
In this context, the RCB should not be interpreted as a 
Radiative-Convective-Boundary but rather as a proxy for 
the Radiative-Circulation-Boundary.
As such, the energy flux is downwards and not upwards,
which in turn leads to a hotter adiabat.

\subsection{General Comparison to Previous Studies}
\label{sec:discussion-general-comparison}
 
\begin{figure*}[ht!]
	\includegraphics[width=\textwidth]{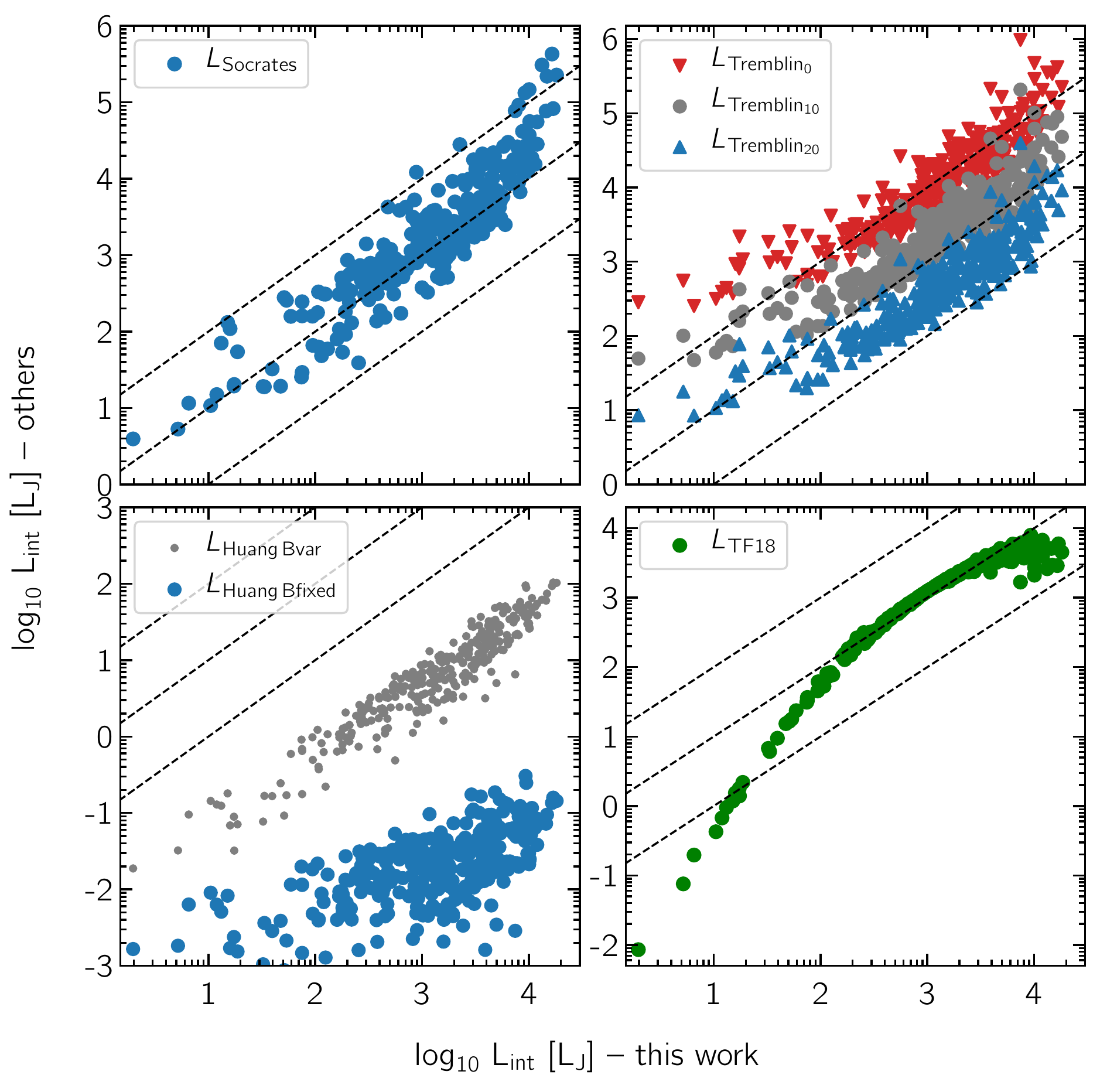}
	\caption{Comparison of the internal luminosity 
		derived in this work from observations and other (theoretical) studies.
		The solid dashed lines are from top to botton the 1:10, 1:1, and 1:0.1 relations.
		The different panels show the results in comparison with analytical relations 
		\citep{Huang:2012, Socrates:2013}, 
		numerical modeling \citep{Tremblin:2017}, 
		and based on a statistical approach similar to ours \citepalias{Thorngren:2018}.
		See text for explanations on the different versions of $L_{\rm Huang}$ and $L_{\rm Tremblin}$.
		Notice the different scales in each panel.		
		The results based on the analytical approximations of \cite{Huang:2012}
		underestimate \lint.
		There is an agreement with \citetalias{Thorngren:2018}, \cite{Tremblin:2017},
		and \cite{Socrates:2013}
		giving thus evidence for advection of potential temperature
		and thermal tides as possible mechanisms to explain the radius inflation conundrum.
	\label{fig:lbloat_comparison}}
\end{figure*}

It is useful and informative to compare the results of our model
with analytical relations.
We consider the analytical approximations
of the internal luminosity based on 
ohmic dissipation \citep[$L_{\rm Huang}$; Equation (14) of][]{Huang:2012}
and thermal tides \citep[$L_{\rm Socrates}$; Equation (8) of][]{Socrates:2013}:

\begin{align}
	L_{\rm Huang} = 3 &\times 10^{22} \, {\rm erg \, s^{-1} } 
			\left( \frac{B_{\phi 0}}{10 \, {\rm G}} \right)^2
			\left(  \frac{\sigma_t}{10^6 \, {\rm s^{-1}}}  \right)^{-1} \nonumber &\\[2ex]
			&\times 
			\left(  \frac{\teq}{1500 \, {\rm K}} \right) 
			\left(  \frac{\rp}{\rjup} \right)^4
			\left(  \frac{\mp}{\mjup} \right)^{-1}
		 \label{eq:lhuang} \\
	L_{\rm Socrates} = 1.5 &\times 10^{28} \, {\rm erg \, s^{-1} } 
			\left( \frac{P}{4 \, {\rm days}} \right)^{-2} \nonumber &\\[2ex]
			&\times
			\left(  \frac{\teq}{2000 \, {\rm K}} \right) ^3
			\left(  \frac{\rp}{10^{10} \, {\rm cm}} \right)^4. \label{eq:lsocrates}
\end{align}

\noindent
In the above equations,
$B_{\phi 0}$ is the toroidal component of the magnetic field
at a reference pressure of 10~bar,
$\sigma_t$ is the electrical conductivity in the dissipation region,
and $P$ is the orbital period.
To compute $L_{\rm Huang}$,
we fix $\sigma_t$ to the nominal value $10^6 \, {\rm s^{-1}}$
and consider two different cases for $B_{\phi 0}$.
In the first case, we fix $B_{\phi 0}$ to 10~G
and in the second case,
we compute the mean magnetic field strength at the surface of the dynamo
based on the scaling law of \cite{Christensen:2009}
in the form given by \cite{Reiners:2010}:

\begin{align}
	B_{\rm dyn} = 4.8 &\times 10^{3} \, {\rm G}
			\times \left(  \frac{ML^2}{R^7} \right) ^{1/6}
		 \label{eq:christensen}
\end{align}

\noindent
where $M$, $L$, and $R$ are the mass, luminosity, and radius of the planet
normalized to solar units. 
We assume $B_{\phi 0} = B_{\rm dyn}$.
Note that using this relation, $B_{\rm dyn}$ ranges roughly between 
30 and 480~G for our sample,
in agreement with the previous estimates of \cite{Yadav:2017}.
We refer to these cases as $L_{\rm Huang, \, Bfixed}$ and $L_{\rm Huang, \, Bvar}$, respectively.
It is straightforward then to calculate
$L_{\rm Socrates}$, $L_{\rm Huang, \, Bfixed}$ and $L_{\rm Huang, \, Bvar}$
for each hot Jupiter in our sample using the relevant physical properties.

We also examine our results within the context
of advection of high-entropy material
based on models of \cite{Tremblin:2017}.
Our aim is to compare the internal luminosity of the planets 
that this mechanism predicts
to the internal luminosities derived in Section~\ref{sec:lower}.
\cite{Tremblin:2017} 
computed 2D PT profiles only for four planets 
with different \teq values.
We thus need to estimate the internal luminosity 
of all the planets based on the model of advection of potential temperature,
for which we follow the procedure described next.
We first compute the entropy 
using the SCvH EOS \citep{Saumon:1995}
and $T_{100}$,
which was derived from the 2D PT profiles 
based on four fiducial planets with different \teq\
(see Section~\ref{sec:discussion-advection} for more details).
Second, we fit a relation between the equilibrium temperatures 
of the four planets and their estimated entropy.
Finally, to convert the entropy into an internal luminosity, 
we use the entropy--mass--luminosity relation
from an updated version of the population synthesis of \cite{Mordasini:2018}.
The second step allows us to 
compute the entropy for 
all the selected hot Jupiters in our sample using the observed \teq.
Having calculated the entropy and knowing \mp\ from observations, 
the last step allows us to compute the internal luminosity of the planets.
With this procedure, we therefore calculate the internal luminosity of the
planets predicted by this mechanism
based on these fits and based on \teq\ and \mp\ from observations.
We consider three cases for comparison by assuming the planets are
composed of H/He and 
setting the fraction of heavy elements to 0\%, 10\%, and 20\%.
We refer to these models as 
$L_{\rm Tremblin_0}$, $L_{\rm Tremblin_{10}}$, and $L_{\rm Tremblin_{20}}$,
respectively.
We point out that the values should be taken with caution
as there are strong approximations involved in this approach.

Finally, to compare our results to \citetalias{Thorngren:2018}, 
we use the analytical \eps(\teq) (Equation~(34) in their paper)
to compute \eps\ and then estimate $L_{\rm TF18}$ using Equation~(\ref{eq:lint}).

Figure~\ref{fig:lbloat_comparison} compares our results to the various studies
where the dashed lines are the 1:10, 1:1, and 1:0.1 relations.
The predicted luminosities based on the analytical solution
of thermal tides as suggested by \cite{Socrates:2013}
and the advection of potential temperature \citep{Tremblin:2017}
are on the same order of magnitude 
as the ones we derive in this work based on observations.
The advection of potential temperature \citep{Tremblin:2017}
predicts high luminosity values for the least luminous planets in our sample.
This is expected since their model tend to overestimate the radii compared to observations,
even for planets with incident flux below the threshold of inflation
(stellar incident flux of $\sim 2 \times 10^8 \, \rm{erg \, s^{-1} \, cm^{-2}}$
or $\teq\ \approx 1000 \, \rm{K}$).

The relation of \cite{Huang:2012} based on ohmic dissipation 
leads to small internal luminosity values.
Note that this relation is an order-of-magnitude estimation of the total ohmic power.
We therefore caution that these results do not provide 
evidence against ohmic dissipation, 
but rather that this relation underestimates the ohmic power.
Based on our results and the work of \citetalias{Thorngren:2018},
there is compelling evidence from the HEET relation
that ohmic dissipation can explain the radii of hot Jupiters.
The ohmic power values estimated 
by \cite{Batygin:2010} and \cite{Menou:2012}
are up to three orders of magnitude higher than the values predicted by \cite{Huang:2012}
and thus on the same order of magnitude estimated in this work.
Moreover, the small internal luminosity values 
using the relation of \cite{Huang:2012}
could also 
explain the findings of \cite{Lopez:2016},
where it was shown that the relation did not lead to re-inflation of hot Jupiters.

For our models with $ \lint < 10^2 \, \ljup$,
the model of \citetalias{Thorngren:2018} predicts smaller values of \lint.
This difference is a direct consequence of the discrepancy 
in \eps\ as shown in the right panel of Figure~\ref{fig:HEET_results},
where as discussed in Section~\ref{sec:heet}
we predict higher heating efficiencies
for the least and the most irradiated planets.

Converting the luminosity values to a heating efficiency
using Equation~(\ref{eq:lint}),
the models of \cite{Socrates:2013} and \cite{Tremblin:2017}
do not lead to a decrease in the heating efficiency at the highest equilibrium temperatures.
The former predicts a continuous increase as was shown by \citetalias{Thorngren:2018}
with values as high as 20--25\%
and the latter seems to increase moderately up to 30\%, 10\%, and 2\%
for \zp = 0, 0.1, and 0.2, respectively.
This is expected given the steeper increase in the luminosity values 
above $10^4 \, \ljup$ for both models.
These are the highly inflated and highly irradiated hot Jupiters
($\rp > 1.4 \, \rjup$ and $\teq > 1900$~K).
Note that the peak in the HEET distribution in our model
occurs close to 1900~K (see Section~\ref{sec:heet} and Table~\ref{tab:hbm-gauss-results}), 
beyond which \eps\ decreases for higher \teq.
This explains why the models of \cite{Socrates:2013} and \cite{Tremblin:2017}
do not predict a Gaussian function,
i.e. why \eps\ does not decrease at high \teq.
We stress that these models can nevertheless explain the observed radii 
of most of the hot Jupiters and can be the dominant mechanisms responsible for inflation
even in the absence of the Gaussian function.
It could be thus that these mechanisms are too efficient 
in inflating hot Jupiters at temperatures above than 1900~K.
Thermal tides have received less attention within the context
of the radius anomaly problem and thus more work is 
needed to understand the physical regime where this mechanism is efficient.

In summary, we provide evidence that thermal tides and 
advection of potential temperature can reproduce 
the large observed radii of most of the hot Jupiters
based on the internal luminosity predicted using these models.
Moreover, the HEET distribution suggests that ohmic dissipation can also 
explain the radii of the close-in giant planets
(see Section~\ref{sec:discussion-ohmic-dissipation}). 
We therefore conclude that all of these three mechanisms
can explain the inflation of hot Jupiters.
This is in line with our main goals where we stress that these mechanisms were
tested on only a handful of exoplanets and a statistical approach is necessary 
to confirm or refute these mechanisms for the entire population.

\subsection{Limitations and Caveats}
\label{sec:limitations}

There are important caveats and limitations 
related to this work that should be explicitly mentioned.

Our results and conclusions are based on a simple 1D interior structure model. 
Hot Jupiters however are tidally locked,
which gives rise to a temperature gradient between the day-side and the night-side.
The RCB at the night-side might thus be at lower pressures
compared to the day-side leading to uneven cooling.
As a consequence of that, 
\cite{Spiegel:2013} showed using a 1+1D model
that the net effect of incorporating night-side cooling
leads to higher cooling rates compared to the default 1D models.
2D circulation models also showed that 
the location of the RCB differs 
from the day-side to the night-side,
which further enhances the cooling rate \citep{Rauscher:2014}
and thus requires even higher efficiency to explain the radii of hot Jupiters.
This is especially important for the highly irradiated planets
as it was shown that the day-side--night-side temperature differences
increases with stellar irradiation \citep{Komacek:2016, Komacek:2017a}.

In addition, we assume that the heat is deposited in the interior of the planet
and we do not account for dissipation in the intermediate layers.
A better treatment would be to deposit the heat 
over a range of depths similar to 
\cite{Ginzburg:2016} or \cite{Komacek:2017b}.
Moreover, even though we showed that the Gaussian profile 
of the HEET distribution is in agreement
with ohmic dissipation there are few shortcomings to this.
A key component for ohmic dissipation is 
the electrical conductivity $\sigma$,
where the ohmic power is proportional to $1/\sigma$ \citep{Batygin:2010}.
The electrical conductivity increases dramatically in the interior
leading to efficient heating only at lower densities and thus at lower pressures.
However, the layers that contribute to the inflation are not at the surface
where the conductivity is maximum 
but rather at deeper layers \citep[between 100 and 1000 bar;][]{Batygin:2011}.
\cite{Wu:2013} confirmed these results by 
showing that heat deposited at 100~bar 
requires significantly less heating efficiency 
in comparison to 10 or 3~bar
(0.3\% compared to 3\% and 200\%, respectively, see their Figure~3).
It is therefore unclear whether the Gaussian functional form
holds for energy dissipated at lower pressures.

The depth of the heating has also direct consequences on the interior structure.
For example, \cite{Huang:2012}
included ohmic heating only in the radiative layers deeper than 10~bar
and showed that as a consequence of that the RCB moves to deeper pressures.
However, their model cannot reproduce the radii of massive planets.
Understanding the location of the RCB is crucial as it regulates the 
planetary cooling rate and thus the contraction rate
\citep{Arras:2006, Marleau:2014}.
Future developments of state-of-the-art GCM models 
that solve the complete equations without approximations
and that couple the upper atmosphere with the deep convective layers 
will provide a complete picture of the underlying physical processes.

Finally, in this work we did not account for observational biases.
A large number of the hot Jupiters discovered to date
are discovered using ground based telescopes,
such as WASP \citep{Pollacco:2006} 
and the HATNet and HATSouth \citep{Bakos:2018}
exoplanet surveys.
There is a lack of hot Jupiters with radii smaller than 
$\sim$~1.4~\rjup\ around early- and mid-F stars.
This is because detecting such planets is still challenging from the ground
as the transit depths are shallow and less than 0.5\%.
\cite{Heng:2012} showed that {\it if} ohmic dissipation
can explain the anomalously large radii of hot Jupiters, 
then this naturally leads to scatter in the radii at a given stellar incident flux
due to variations in the opacity, albedo, cloud/hazes properties,
and the magnetic fields strength.
It is therefore still not quite clear whether 
the lack of \q{medium-inflated} hot Jupiters around F stars
is due to observational biases or variations 
in the efficiency of the heating mechanism.
The NASA {\it Transiting Exoplanet Survey Satellite} mission
\citep[TESS;][]{Ricker:2015} will discover such planets if they exist
and will help to better constrain the efficiency of the heating mechanisms
either by the lack or existence of such planets.
Subsequently high precision follow-up observations with 
the {\it CHaracterising ExOPlanet Satellite} \citep[CHEOPS;][]{Broeg:2013}
will help to get very accurate radii.

\section{Conclusion}
\label{sec:conclusion}

In this work, we developed a flexible and robust hierarchical Bayesian model 
to couple the observed physical parameters of hot Jupiters to an interior structure model.
The model accounts for observational uncertainties
and for the scatter in the relation between 
planet mass and heavy-element fraction. 
We validated the statistical method by applying it
to synthetic planets based on planet population synthesis
and showed that we are able to retrieve the true distribution.   
We then applied this method to quantify the internal luminosity 
needed to explain the radii of a sample of 314 hot Jupiters. 
We tested this model under two different priors
(assuming a log-uniform and a linear-uniform distributions for \lint)
and showed that the population level distributions
are prior independent 
(Figure~\ref{fig:MLR_results}). 
This provides useful and robust constraints 
on the interior structure of hot Jupiters. 
We find that such planets tend to have hotter interiors
compared to previous assumptions,
and as a result, the RCB is located at low pressures,
in agreement with recent work by \cite{Thorngren:2019}
(Figure~\ref{fig:TT_PT_results}).

Assuming the planet has reached steady state
and assuming that the additional source of heat is the stellar irradiation,
we compute the heating efficiency \eps,
defined as the fraction of stellar irradiation deposited into the interior of the planet
that is needed to explain the observed inflated radii.
We find that the heating-efficiency--equilibrium-temperature relation
is described by a Gaussian function (Figure~\ref{fig:HEET_results}),
in agreement with previous work by \citetalias{Thorngren:2018},
however, the peak is not consistent in both studies.
We found that the models of \citetalias{Thorngren:2018}
predict larger radii than our models for $\teq > 1500$~K,
which we attribute due to differences in 
the atmospheric modelling.
The Gaussian-like pattern is more importantly 
in agreement with theoretical predictions
based on the ohmic dissipation model \citep{Menou:2012}.
We also show that thermal tides \citep{Arras:2010, Socrates:2013}
and advection of potential temperature \citep{Tremblin:2017}
can explain the observations of most of the planets in our sample
and thus are possible mechanisms responsible 
for the anomalously large radii of hot Jupiters
(Figure~\ref{fig:lbloat_comparison}).

To conclude, we provide new insights into the interior of hot Jupiters
by coupling observations to theoretical
models within a powerful statistical framework.
With a better understanding of the interior,
we highlight the importance of accounting for 
the extra heat flux in the interior in 3D GCM models,
which will further improve our understanding of wind speeds
and hence on the efficiency of the heating mechanisms.

The future of hot Jupiters is exciting and bright.
Simulations of the exptected TESS yield \citep{Barclay:2018}
predict that TESS will discover more than 250 hot Jupiters suitable for RV follow-up
($\rp > 1 \, \rjup$) with orbital periods $<$ 10 days orbiting bright stars ($V < 14$~mag),
almost doubling the number of hot Jupiters discovered.
The mission already detected few hot Jupiters 
\citep[e.g.][]{Kossakowski:2019, Wang:2019}
with many yet to be discovered.
Furthermore, CHEOPS \citep{Broeg:2013}
is capable of detecting the phase curves of hot Jupiters,
which provide information on the day-night temperature contrast.
CHEOPS will therefore play 
a major role in providing clues into the efficiency of energy transport in hot Jupiter atmospheres
\citep[e.g. HD189733\,b;][]{Knutson:2007}.
With a better understanding of the interior structure of hot Jupiters 
thanks to the development of flexible and computationally efficient statistical tools, 
we will be able to provide further constraints on the radius inflation conundrum.

The source code for the hierarchical model is open 
source and available at 
\href{https://github.com/psarkis/bloatedHJs}{https://github.com/psarkis/bloatedHJs}
under the MIT open source software license.
Part of the code is still being added and available upon request.

The posterior samples at the population level 
are also available online at 
\href{\tinyurl}{\tinyurl}.

{\it Software:} 
	\texttt{corner} \citep{corner},
	\emcee\ \citep{Foreman-Mackey:2013},
	\texttt{Jupyter} \href{https://jupyter.org/}{https://jupyter.org/},
	\texttt{matplotlib} \citep{Hunter:2007}, 
	\texttt{numpy} \citep{Walt:2011}, 
	\petit\ \citep{Molliere:2015, Molliere:2017},
	\texttt{pandas} \citep{pandas:2010},
	\texttt{scikit-learn} \citep{scikit}, 
	\texttt{scipy} \citep{scipy}.

% -----------------------------------------------------------

\begin{acknowledgements}
	P.S., C.M., G.-D.M. acknowledge the support from the Swiss National Science Foundation 
	under grant BSSGI0$\_$155816 ``PlanetsInTime''. 
	Parts of this work have been carried out within the framework of the NCCR PlanetS 
	supported by the Swiss National Science Foundation.
	T.H. and P.M. acknowledge support from the European Research Council 
	under the Horizon 2020 Framework Program via the ERC Advanced Grant Origins 83 24 28.
	G.-D.M. acknowledges the support of the DFG priority program SPP 1992 ``Exploring the Diversity of Extrasolar Planets'' (KU 2849/7-1).
	P.S. would like to thank David Hogg and Morgan Fouesneau 
	for useful discussions related to the importance sampling technique
	and multilevel modeling
	and Daniel Thorngren for sharing the posterior distributions of the individual planets
	and the data used in Figure~\ref{fig:sample}.
	P.S. thanks Saavi and Gabriele for exchanging priority time on the aida server.
\end{acknowledgements}

\bibliographystyle{aa} % style aa.bst 
\bibliography{/Users/sarkis/work/bloatedJupiters/paper/hj.bib}

\begin{thebibliography}{104}
\expandafter\ifx\csname natexlab\endcsname\relax\def\natexlab#1{#1}\fi

\bibitem[{{Arras} \& {Bildsten}(2006)}]{Arras:2006}
{Arras}, P. \& {Bildsten}, L. 2006, \apj, 650, 394

\bibitem[{{Arras} \& {Socrates}(2010)}]{Arras:2010}
{Arras}, P. \& {Socrates}, A. 2010, \apj, 714, 1

\bibitem[{{Bakos}(2018)}]{Bakos:2018}
{Bakos}, G.~{\'A}. 2018, {The HATNet and HATSouth Exoplanet Surveys}, 111

\bibitem[{{Baraffe} {et~al.}(2008){Baraffe}, {Chabrier}, \&
  {Barman}}]{Baraffe:2008}
{Baraffe}, I., {Chabrier}, G., \& {Barman}, T. 2008, \aap, 482, 315

\bibitem[{{Baraffe} {et~al.}(2004){Baraffe}, {Selsis}, {Chabrier}, {Barman},
  {Allard}, {Hauschildt}, \& {Lammer}}]{Baraffe:2004}
{Baraffe}, I., {Selsis}, F., {Chabrier}, G., {et~al.} 2004, \aap, 419, L13

\bibitem[{{Barclay} {et~al.}(2018){Barclay}, {Pepper}, \&
  {Quintana}}]{Barclay:2018}
{Barclay}, T., {Pepper}, J., \& {Quintana}, E.~V. 2018, \apjs, 239, 2

\bibitem[{{Batygin} \& {Stevenson}(2010)}]{Batygin:2010}
{Batygin}, K. \& {Stevenson}, D.~J. 2010, \apjl, 714, L238

\bibitem[{{Batygin} {et~al.}(2011){Batygin}, {Stevenson}, \&
  {Bodenheimer}}]{Batygin:2011}
{Batygin}, K., {Stevenson}, D.~J., \& {Bodenheimer}, P.~H. 2011, \apj, 738, 1

\bibitem[{{Bodenheimer} {et~al.}(2001){Bodenheimer}, {Lin}, \&
  {Mardling}}]{Bodenheimer:2001}
{Bodenheimer}, P., {Lin}, D.~N.~C., \& {Mardling}, R.~A. 2001, \apj, 548, 466

\bibitem[{{Broeg} {et~al.}(2013){Broeg}, {Fortier}, {Ehrenreich}, {Alibert},
  {Baumjohann}, {Benz}, {Deleuil}, {Gillon}, {Ivanov}, {Liseau}, {Meyer},
  {Oloffson}, {Pagano}, {Piotto}, {Pollacco}, {Queloz}, {Ragazzoni}, {Renotte},
  {Steller}, \& {Thomas}}]{Broeg:2013}
{Broeg}, C., {Fortier}, A., {Ehrenreich}, D., {et~al.} 2013, in European
  Physical Journal Web of Conferences, Vol.~47, European Physical Journal Web
  of Conferences, 03005

\bibitem[{{Burrows} {et~al.}(2000){Burrows}, {Guillot}, {Hubbard}, {Marley},
  {Saumon}, {Lunine}, \& {Sudarsky}}]{Burrows:2000}
{Burrows}, A., {Guillot}, T., {Hubbard}, W.~B., {et~al.} 2000, \apj, 534, L97

\bibitem[{{Burrows} {et~al.}(2007){Burrows}, {Hubeny}, {Budaj}, \&
  {Hubbard}}]{Burrows:2007}
{Burrows}, A., {Hubeny}, I., {Budaj}, J., \& {Hubbard}, W.~B. 2007, \apj, 661,
  502

\bibitem[{{Carone} {et~al.}(2019){Carone}, {Baeyens}, {Molli{\`e}re}, {Barth},
  {Vazan}, {Decin}, {Sarkis}, {Venot}, \& {Henning}}]{Carone:2019}
{Carone}, L., {Baeyens}, R., {Molli{\`e}re}, P., {et~al.} 2019, arXiv e-prints,
  arXiv:1904.13334

\bibitem[{{Chabrier} \& {Baraffe}(2007)}]{Chabrier:2007}
{Chabrier}, G. \& {Baraffe}, I. 2007, \apjl, 661, L81

\bibitem[{{Christensen} {et~al.}(2009){Christensen}, {Holzwarth}, \&
  {Reiners}}]{Christensen:2009}
{Christensen}, U.~R., {Holzwarth}, V., \& {Reiners}, A. 2009, \nat, 457, 167

\bibitem[{{Collins} {et~al.}(2017){Collins}, {Kielkopf}, \&
  {Stassun}}]{Collins:2017}
{Collins}, K.~A., {Kielkopf}, J.~F., \& {Stassun}, K.~G. 2017, \aj, 153, 78

\bibitem[{{Delrez} {et~al.}(2016){Delrez}, {Santerne}, {Almenara}, {Anderson},
  {Collier-Cameron}, {D{\'\i}az}, {Gillon}, {Hellier}, {Jehin}, {Lendl},
  {Maxted}, {Neveu-VanMalle}, {Pepe}, {Pollacco}, {Queloz}, {S{\'e}gransan},
  {Smalley}, {Smith}, {Triaud}, {Udry}, {Van Grootel}, \& {West}}]{Delrez:2016}
{Delrez}, L., {Santerne}, A., {Almenara}, J.~M., {et~al.} 2016, \mnras, 458,
  4025

\bibitem[{{Demory} \& {Seager}(2011)}]{Demory:2011}
{Demory}, B.-O. \& {Seager}, S. 2011, \apjs, 197, 12

\bibitem[{{Dorn} {et~al.}(2019){Dorn}, {Harrison}, {Bonsor}, \&
  {Hands}}]{Dorn:2019}
{Dorn}, C., {Harrison}, J.~H.~D., {Bonsor}, A., \& {Hands}, T.~O. 2019, \mnras,
  484, 712

\bibitem[{{Emsenhuber} {et~al.}(2020){Emsenhuber}, {Mordasini}, {Burn},
  {Alibert}, {Benz}, \& {Asphaug}}]{Emsenhuber:2020}
{Emsenhuber}, A., {Mordasini}, C., {Burn}, R., {et~al.} 2020, arXiv e-prints,
  arXiv:2007.05561

\bibitem[{{Enoch} {et~al.}(2012){Enoch}, {Collier Cameron}, \&
  {Horne}}]{Enoch:2012}
{Enoch}, B., {Collier Cameron}, A., \& {Horne}, K. 2012, \aap, 540, A99

\bibitem[{{Folkner} {et~al.}(2017){Folkner}, {Iess}, {Anderson}, {Asmar},
  {Buccino}, {Durante}, {Feldman}, {Gomez Casajus}, {Gregnanin}, {Milani},
  {Parisi}, {Park}, {Serra}, {Tommei}, {Tortora}, {Zannoni}, {Bolton},
  {Connerney}, \& {Levin}}]{Folkner:2017}
{Folkner}, W.~M., {Iess}, L., {Anderson}, J.~D., {et~al.} 2017, Geophysical
  Research Letters, 44, 4694

\bibitem[{Foreman-Mackey(2016)}]{corner}
Foreman-Mackey, D. 2016, The Journal of Open Source Software, 1, 24

\bibitem[{{Foreman-Mackey} {et~al.}(2013){Foreman-Mackey}, {Hogg}, {Lang}, \&
  {Goodman}}]{Foreman-Mackey:2013}
{Foreman-Mackey}, D., {Hogg}, D.~W., {Lang}, D., \& {Goodman}, J. 2013, \pasp,
  125, 306

\bibitem[{{Foreman-Mackey} {et~al.}(2014){Foreman-Mackey}, {Hogg}, \&
  {Morton}}]{Foreman-Mackey:2014}
{Foreman-Mackey}, D., {Hogg}, D.~W., \& {Morton}, T.~D. 2014, The Astrophysical
  Journal, 795, 64

\bibitem[{{Fortney} {et~al.}(2008){Fortney}, {Lodders}, {Marley}, \&
  {Freedman}}]{Fortney:2008}
{Fortney}, J.~J., {Lodders}, K., {Marley}, M.~S., \& {Freedman}, R.~S. 2008,
  \apj, 678, 1419

\bibitem[{{Fortney} {et~al.}(2007){Fortney}, {Marley}, \&
  {Barnes}}]{Fortney:2007}
{Fortney}, J.~J., {Marley}, M.~S., \& {Barnes}, J.~W. 2007, \apj, 659, 1661

\bibitem[{{Ginzburg} \& {Sari}(2016)}]{Ginzburg:2016}
{Ginzburg}, S. \& {Sari}, R. 2016, \apj, 819, 116

\bibitem[{{Griessmeier}(2017)}]{Griessmeier:2017}
{Griessmeier}, J.~M. 2017, in Planetary Radio Emissions VIII, ed. G.~{Fischer},
  G.~{Mann}, M.~{Panchenko}, \& P.~{Zarka}, 285--299

\bibitem[{{Grunblatt} {et~al.}(2017){Grunblatt}, {Huber}, {Gaidos}, {Lopez},
  {Howard}, {Isaacson}, {Sinukoff}, {Vanderburg}, {Nofi}, {Yu}, {North},
  {Chaplin}, {Foreman-Mackey}, {Petigura}, {Ansdell}, {Weiss}, {Fulton}, \&
  {Lin}}]{Grunblatt:2017}
{Grunblatt}, S.~K., {Huber}, D., {Gaidos}, E., {et~al.} 2017, \aj, 154, 254

\bibitem[{{Grunblatt} {et~al.}(2016){Grunblatt}, {Huber}, {Gaidos}, {Lopez},
  {Fulton}, {Vanderburg}, {Barclay}, {Fortney}, {Howard}, {Isaacson}, {Mann},
  {Petigura}, {Silva Aguirre}, \& {Sinukoff}}]{Grunblatt:2016}
{Grunblatt}, S.~K., {Huber}, D., {Gaidos}, E.~J., {et~al.} 2016, \aj, 152, 185

\bibitem[{{Guillot}(2010)}]{Guillot:2010}
{Guillot}, T. 2010, \aap, 520, A27

\bibitem[{{Guillot} \& {Gautier}(2014)}]{Guillot:2014}
{Guillot}, T. \& {Gautier}, D. 2014, arXiv e-prints, arXiv:1405.3752

\bibitem[{{Guillot} \& {Showman}(2002)}]{Guillot:2002}
{Guillot}, T. \& {Showman}, A.~P. 2002, \aap, 385, 156

\bibitem[{{Hartman} {et~al.}(2016){Hartman}, {Bakos}, {Bhatti}, {Penev},
  {Bieryla}, {Latham}, {Kov{\'a}cs}, {Torres}, {Csubry}, {de Val-Borro},
  {Buchhave}, {Kov{\'a}cs}, {Quinn}, {Howard}, {Isaacson}, {Fulton}, {Everett},
  {Esquerdo}, {B{\'e}ky}, {Szklenar}, {Falco}, {Santerne}, {Boisse},
  {H{\'e}brard}, {Burrows}, {L{\'a}z{\'a}r}, {Papp}, \&
  {S{\'a}ri}}]{Hartman:2016}
{Hartman}, J.~D., {Bakos}, G.~{\'A}., {Bhatti}, W., {et~al.} 2016, \aj, 152,
  182

\bibitem[{{Heng}(2012)}]{Heng:2012}
{Heng}, K. 2012, \apjl, 748, L17

\bibitem[{{Hogg} \& {Foreman-Mackey}(2018)}]{Hogg:2018}
{Hogg}, D.~W. \& {Foreman-Mackey}, D. 2018, \apjs, 236, 11

\bibitem[{{Hogg} {et~al.}(2010){Hogg}, {Myers}, \& {Bovy}}]{Hogg:2010}
{Hogg}, D.~W., {Myers}, A.~D., \& {Bovy}, J. 2010, The Astrophysical Journal,
  725, 2166

\bibitem[{{Huang} \& {Cumming}(2012)}]{Huang:2012}
{Huang}, X. \& {Cumming}, A. 2012, \apj, 757, 47

\bibitem[{{Hunter}(2007)}]{Hunter:2007}
{Hunter}, J.~D. 2007, Computing in Science and Engineering, 9, 90

\bibitem[{{Jin} \& {Mordasini}(2018)}]{Jin:2018}
{Jin}, S. \& {Mordasini}, C. 2018, \apj, 853, 163

\bibitem[{{Jin} {et~al.}(2014){Jin}, {Mordasini}, {Parmentier}, {van Boekel},
  {Henning}, \& {Ji}}]{Jin:2014}
{Jin}, S., {Mordasini}, C., {Parmentier}, V., {et~al.} 2014, \apj, 795, 65

\bibitem[{{Knutson} {et~al.}(2007){Knutson}, {Charbonneau}, {Allen}, {Fortney},
  {Agol}, {Cowan}, {Showman}, {Cooper}, \& {Megeath}}]{Knutson:2007}
{Knutson}, H.~A., {Charbonneau}, D., {Allen}, L.~E., {et~al.} 2007, \nat, 447,
  183

\bibitem[{{Komacek} \& {Showman}(2016)}]{Komacek:2016}
{Komacek}, T.~D. \& {Showman}, A.~P. 2016, \apj, 821, 16

\bibitem[{{Komacek} {et~al.}(2017){Komacek}, {Showman}, \&
  {Tan}}]{Komacek:2017a}
{Komacek}, T.~D., {Showman}, A.~P., \& {Tan}, X. 2017, \apj, 835, 198

\bibitem[{{Komacek} \& {Youdin}(2017)}]{Komacek:2017b}
{Komacek}, T.~D. \& {Youdin}, A.~N. 2017, \apj, 844, 94

\bibitem[{{Kossakowski} {et~al.}(2019){Kossakowski}, {Espinoza}, {Brahm},
  {Jord{\'a}n}, {Henning}, {Rojas}, {K{\"u}rster}, {Sarkis}, {Schlecker},
  {Pozuelos}, {Barkaoui}, {Jehin}, {Gillon}, {Matthews}, {Horch}, {Ciardi},
  {Crossfield}, {Gonzales}, {Howell}, {Matson}, {Schlieder}, {Jenkins},
  {Ricker}, {Seager}, {Winn}, {Li}, {Rose}, {Smith}, {Dynes}, {Morgan},
  {Villasenor}, {Charbonneau}, {Jaffe}, {Yu}, {Bakos}, {Bhatti}, {Bouchy},
  {Collins}, {Collins}, {Csubry}, {Evans}, {Jensen}, {Lovis}, {Marmier},
  {Nielsen}, {Osip}, {Pepe}, {Relles}, {S{\'e}gransan}, {Shporer}, {Stockdale},
  {Suc}, {Turner}, \& {Udry}}]{Kossakowski:2019}
{Kossakowski}, D., {Espinoza}, N., {Brahm}, R., {et~al.} 2019, \mnras, 490,
  1094

\bibitem[{{Laughlin} {et~al.}(2011){Laughlin}, {Crismani}, \&
  {Adams}}]{Laughlin:2011}
{Laughlin}, G., {Crismani}, M., \& {Adams}, F.~C. 2011, \apjl, 729, L7

\bibitem[{{Lazio}(2018)}]{Lazio:2018}
{Lazio}, T. J.~W. 2018, {Radio Observations as an Exoplanet Discovery Method},
  9

\bibitem[{{Li} {et~al.}(2012){Li}, {Baines}, {Smith}, {West},
  {P{\'e}rez-Hoyos}, {Trammell}, {Simon-Miller}, {Conrath}, {Gierasch},
  {Orton}, {Nixon}, {Filacchione}, {Fry}, \& {Momary}}]{Li:2012}
{Li}, L., {Baines}, K.~H., {Smith}, M.~A., {et~al.} 2012, Journal of
  Geophysical Research (Planets), 117, E11002

\bibitem[{{Linder} {et~al.}(2019){Linder}, {Mordasini}, {Molli{\`e}re},
  {Marleau}, {Malik}, {Quanz}, \& {Meyer}}]{Linder:2019}
{Linder}, E.~F., {Mordasini}, C., {Molli{\`e}re}, P., {et~al.} 2019, \aap, 623,
  A85

\bibitem[{{Lopez} \& {Fortney}(2016)}]{Lopez:2016}
{Lopez}, E.~D. \& {Fortney}, J.~J. 2016, \apj, 818, 4

\bibitem[{{Loredo} \& {Hendry}(2019)}]{Loredo:2019}
{Loredo}, T.~J. \& {Hendry}, M.~A. 2019, arXiv e-prints, arXiv:1911.12337

\bibitem[{{Marleau} {et~al.}(2019){Marleau}, {Coleman}, {Leleu}, \&
  {Mordasini}}]{Marleau:2019}
{Marleau}, G.-D., {Coleman}, G. A.~L., {Leleu}, A., \& {Mordasini}, C. 2019,
  \aap, 624, A20

\bibitem[{{Marleau} \& {Cumming}(2014)}]{Marleau:2014}
{Marleau}, G.~D. \& {Cumming}, A. 2014, \mnras, 437, 1378

\bibitem[{{Maxted} {et~al.}(2013){Maxted}, {Anderson}, {Doyle}, {Gillon},
  {Harrington}, {Iro}, {Jehin}, {Lafreni{\`e}re}, {Smalley}, \&
  {Southworth}}]{Maxted:2013}
{Maxted}, P.~F.~L., {Anderson}, D.~R., {Doyle}, A.~P., {et~al.} 2013, \mnras,
  428, 2645

\bibitem[{{Menou}(2012)}]{Menou:2012}
{Menou}, K. 2012, \apj, 745, 138

\bibitem[{{Miller} \& {Fortney}(2011)}]{Miller:2011}
{Miller}, N. \& {Fortney}, J.~J. 2011, \apj, 736, L29

\bibitem[{{Mol Lous} \& {Miguel}(2020)}]{MolLous:2020}
{Mol Lous}, M. \& {Miguel}, Y. 2020, \mnras, 495, 2994

\bibitem[{{Molli{\`e}re} {et~al.}(2017){Molli{\`e}re}, {van Boekel}, {Bouwman},
  {Henning}, {Lagage}, \& {Min}}]{Molliere:2017}
{Molli{\`e}re}, P., {van Boekel}, R., {Bouwman}, J., {et~al.} 2017, \aap, 600,
  A10

\bibitem[{{Molli{\`e}re} {et~al.}(2015){Molli{\`e}re}, {van Boekel},
  {Dullemond}, {Henning}, \& {Mordasini}}]{Molliere:2015}
{Molli{\`e}re}, P., {van Boekel}, R., {Dullemond}, C., {Henning}, T., \&
  {Mordasini}, C. 2015, \apj, 813, 47

\bibitem[{{Molli{\`e}re} {et~al.}(2019){Molli{\`e}re}, {Wardenier}, {van
  Boekel}, {Henning}, {Molaverdikhani}, \& {Snellen}}]{mollierewardenier2019}
{Molli{\`e}re}, P., {Wardenier}, J.~P., {van Boekel}, R., {et~al.} 2019, \aap,
  627, A67

\bibitem[{{Mordasini}(2018)}]{Mordasini:2018}
{Mordasini}, C. 2018, {Planetary Population Synthesis}, 143

\bibitem[{{Mordasini}(2020)}]{Mordasini:2020}
{Mordasini}, C. 2020, arXiv e-prints, arXiv:2002.02455

\bibitem[{{Mordasini} {et~al.}(2012){Mordasini}, {Alibert}, {Klahr}, \&
  {Henning}}]{Mordasini:2012}
{Mordasini}, C., {Alibert}, Y., {Klahr}, H., \& {Henning}, T. 2012, \aap, 547,
  A111

\bibitem[{{Mordasini} {et~al.}(2017){Mordasini}, {Marleau}, \&
  {Molli{\`e}re}}]{Mordasini:2017}
{Mordasini}, C., {Marleau}, G.~D., \& {Molli{\`e}re}, P. 2017, \aap, 608, A72

\bibitem[{{Muller} {et~al.}(2020){Muller}, {Helled}, \&
  {Cumming}}]{Muller:2020}
{Muller}, S., {Helled}, R., \& {Cumming}, A. 2020, arXiv e-prints,
  arXiv:2004.13534

\bibitem[{{Owen} \& {Jackson}(2012)}]{Owen:2012}
{Owen}, J.~E. \& {Jackson}, A.~P. 2012, \mnras, 425, 2931

\bibitem[{Pedregosa {et~al.}(2011)Pedregosa, Varoquaux, Gramfort, Michel,
  Thirion, Grisel, Blondel, Prettenhofer, Weiss, Dubourg, Vanderplas, Passos,
  Cournapeau, Brucher, Perrot, \& Duchesnay}]{scikit}
Pedregosa, F., Varoquaux, G., Gramfort, A., {et~al.} 2011, Journal of Machine
  Learning Research, 12, 2825

\bibitem[{{Perna} {et~al.}(2010{\natexlab{a}}){Perna}, {Menou}, \&
  {Rauscher}}]{Perna:2010a}
{Perna}, R., {Menou}, K., \& {Rauscher}, E. 2010{\natexlab{a}}, \apj, 719, 1421

\bibitem[{{Perna} {et~al.}(2010{\natexlab{b}}){Perna}, {Menou}, \&
  {Rauscher}}]{Perna:2010b}
{Perna}, R., {Menou}, K., \& {Rauscher}, E. 2010{\natexlab{b}}, \apj, 724, 313

\bibitem[{{Pollacco} {et~al.}(2006){Pollacco}, {Skillen}, {Collier Cameron},
  {Christian}, {Hellier}, {Irwin}, {Lister}, {Street}, {West}, {Anderson},
  {Clarkson}, {Deeg}, {Enoch}, {Evans}, {Fitzsimmons}, {Haswell}, {Hodgkin},
  {Horne}, {Kane}, {Keenan}, {Maxted}, {Norton}, {Osborne}, {Parley}, {Ryans},
  {Smalley}, {Wheatley}, \& {Wilson}}]{Pollacco:2006}
{Pollacco}, D.~L., {Skillen}, I., {Collier Cameron}, A., {et~al.} 2006, \pasp,
  118, 1407

\bibitem[{{Price-Whelan} {et~al.}(2018){Price-Whelan}, {Hogg}, {Rix}, {De Lee},
  {Majewski}, {Nidever}, {Troup}, {Fern{\'a}ndez-Trincado},
  {Garc{\'\i}a-Hern{\'a}ndez}, {Longa-Pe{\~n}a}, {Nitschelm}, {Sobeck}, \&
  {Zamora}}]{Price-Whelan:2018}
{Price-Whelan}, A.~M., {Hogg}, D.~W., {Rix}, H.-W., {et~al.} 2018, \aj, 156, 18

\bibitem[{{Rauscher} \& {Showman}(2014)}]{Rauscher:2014}
{Rauscher}, E. \& {Showman}, A.~P. 2014, \apj, 784, 160

\bibitem[{{Reiners} \& {Christensen}(2010)}]{Reiners:2010}
{Reiners}, A. \& {Christensen}, U.~R. 2010, \aap, 522, A13

\bibitem[{{Ricker} {et~al.}(2015){Ricker}, {Winn}, {Vanderspek}, {Latham},
  {Bakos}, {Bean}, {Berta-Thompson}, {Brown}, {Buchhave}, {Butler}, {Butler},
  {Chaplin}, {Charbonneau}, {Christensen-Dalsgaard}, {Clampin}, {Deming},
  {Doty}, {De Lee}, {Dressing}, {Dunham}, {Endl}, {Fressin}, {Ge}, {Henning},
  {Holman}, {Howard}, {Ida}, {Jenkins}, {Jernigan}, {Johnson}, {Kaltenegger},
  {Kawai}, {Kjeldsen}, {Laughlin}, {Levine}, {Lin}, {Lissauer}, {MacQueen},
  {Marcy}, {McCullough}, {Morton}, {Narita}, {Paegert}, {Palle}, {Pepe},
  {Pepper}, {Quirrenbach}, {Rinehart}, {Sasselov}, {Sato}, {Seager},
  {Sozzetti}, {Stassun}, {Sullivan}, {Szentgyorgyi}, {Torres}, {Udry}, \&
  {Villasenor}}]{Ricker:2015}
{Ricker}, G.~R., {Winn}, J.~N., {Vanderspek}, R., {et~al.} 2015, Journal of
  Astronomical Telescopes, Instruments, and Systems, 1, 014003

\bibitem[{{Rogers}(2015)}]{Rogers:2015}
{Rogers}, L.~A. 2015, \apj, 801, 41

\bibitem[{{Rogers} \& {Komacek}(2014)}]{Rogers:2014}
{Rogers}, T.~M. \& {Komacek}, T.~D. 2014, \apj, 794, 132

\bibitem[{{Sainsbury-Martinez} {et~al.}(2019){Sainsbury-Martinez}, {Wang},
  {Fromang}, {Tremblin}, {Dubos}, {Meurdesoif}, {Spiga}, {Leconte}, {Baraffe},
  {Chabrier}, {Mayne}, {Drummond}, \& {Debras}}]{Sainsbury-Martinez:2019}
{Sainsbury-Martinez}, F., {Wang}, P., {Fromang}, S., {et~al.} 2019, \aap, 632,
  A114

\bibitem[{{Saumon} {et~al.}(1995){Saumon}, {Chabrier}, \& {van
  Horn}}]{Saumon:1995}
{Saumon}, D., {Chabrier}, G., \& {van Horn}, H.~M. 1995, The Astrophysical
  Journal Supplement Series, 99, 713

\bibitem[{{Sestovic} {et~al.}(2018){Sestovic}, {Demory}, \&
  {Queloz}}]{Sestovic:2018}
{Sestovic}, M., {Demory}, B.-O., \& {Queloz}, D. 2018, Astronomy and
  Astrophysics, 616, A76

\bibitem[{{Showman} \& {Guillot}(2002)}]{Showman:2002}
{Showman}, A.~P. \& {Guillot}, T. 2002, \aap, 385, 166

\bibitem[{{Sing} {et~al.}(2019){Sing}, {Lavvas}, {Ballester}, {Lecavelier des
  Etangs}, {Marley}, {Nikolov}, {Ben-Jaffel}, {Bourrier}, {Buchhave}, {Deming},
  {Ehrenreich}, {Mikal-Evans}, {Kataria}, {Lewis}, {L{\'o}pez-Morales},
  {Garc{\'\i}a Mu{\~n}oz}, {Henry}, {Sanz-Forcada}, {Spake}, {Wakeford}, \&
  {The PanCET collaboration}}]{Sing:2019}
{Sing}, D.~K., {Lavvas}, P., {Ballester}, G.~E., {et~al.} 2019, \aj, 158, 91

\bibitem[{{Socrates}(2013)}]{Socrates:2013}
{Socrates}, A. 2013, arXiv e-prints, arXiv:1304.4121

\bibitem[{{Southworth}(2011)}]{Southworth:2011}
{Southworth}, J. 2011, \mnras, 417, 2166

\bibitem[{{Spiegel} \& {Burrows}(2013)}]{Spiegel:2013}
{Spiegel}, D.~S. \& {Burrows}, A. 2013, \apj, 772, 76

\bibitem[{{Thompson}(1990)}]{Thompson:1990}
{Thompson}, S.~L. 1990, Sandia Natl. Lab. Doc.

\bibitem[{{Thorngren} {et~al.}(2019){Thorngren}, {Gao}, \&
  {Fortney}}]{Thorngren:2019}
{Thorngren}, D., {Gao}, P., \& {Fortney}, J.~J. 2019, \apjl, 884, L6

\bibitem[{{Thorngren} \& {Fortney}(2018)}]{Thorngren:2018}
{Thorngren}, D.~P. \& {Fortney}, J.~J. 2018, The Astronomical Journal, 155, 214

\bibitem[{{Thorngren} {et~al.}(2016){Thorngren}, {Fortney}, {Murray-Clay}, \&
  {Lopez}}]{Thorngren:2016}
{Thorngren}, D.~P., {Fortney}, J.~J., {Murray-Clay}, R.~A., \& {Lopez}, E.~D.
  2016, The Astrophysical Journal, 831, 64

\bibitem[{{Tremblin} {et~al.}(2017){Tremblin}, {Chabrier}, {Mayne}, {Amundsen},
  {Baraffe}, {Debras}, {Drummond}, {Manners}, \& {Fromang}}]{Tremblin:2017}
{Tremblin}, P., {Chabrier}, G., {Mayne}, N.~J., {et~al.} 2017, \apj, 841, 30

\bibitem[{{Valencia} {et~al.}(2013){Valencia}, {Guillot}, {Parmentier}, \&
  {Freedman}}]{Valencia:2013}
{Valencia}, D., {Guillot}, T., {Parmentier}, V., \& {Freedman}, R.~S. 2013,
  \apj, 775, 10

\bibitem[{{van der Walt} {et~al.}(2011){van der Walt}, {Colbert}, \&
  {Varoquaux}}]{Walt:2011}
{van der Walt}, S., {Colbert}, S.~C., \& {Varoquaux}, G. 2011, Computing in
  Science and Engineering, 13, 22

\bibitem[{{Venturini} {et~al.}(2016){Venturini}, {Alibert}, \&
  {Benz}}]{Venturini:2016}
{Venturini}, J., {Alibert}, Y., \& {Benz}, W. 2016, \aap, 596, A90

\bibitem[{{Virtanen} {et~al.}(2020){Virtanen}, {Gommers}, {Oliphant},
  {Haberland}, {Reddy}, {Cournapeau}, {Burovski}, {Peterson}, {Weckesser},
  {Bright}, {van der Walt}, {Brett}, {Wilson}, {Jarrod Millman}, {Mayorov},
  {Nelson}, {Jones}, {Kern}, {Larson}, {Carey}, {Polat}, {Feng}, {Moore}, {Vand
  erPlas}, {Laxalde}, {Perktold}, {Cimrman}, {Henriksen}, {Quintero}, {Harris},
  {Archibald}, {Ribeiro}, {Pedregosa}, {van Mulbregt}, \&
  {Contributors}}]{scipy}
{Virtanen}, P., {Gommers}, R., {Oliphant}, T.~E., {et~al.} 2020, Nature
  Methods, 17, 261

\bibitem[{{Wahl} {et~al.}(2017){Wahl}, {Hubbard}, {Militzer}, {Guillot},
  {Miguel}, {Movshovitz}, {Kaspi}, {Helled}, {Reese}, {Galanti}, {Levin},
  {Connerney}, \& {Bolton}}]{Wahl:2017}
{Wahl}, S.~M., {Hubbard}, W.~B., {Militzer}, B., {et~al.} 2017, \grl, 44, 4649

\bibitem[{{Wang} {et~al.}(2019){Wang}, {Jones}, {Shporer}, {Fulton}, {Paredes},
  {Trifonov}, {Kossakowski}, {Eastman}, {Redfield}, {G{\"u}nther}, {Kreidberg},
  {Huang}, {Millholland}, {Seligman}, {Fischer}, {Brahm}, {Wang}, {Cruz},
  {Henry}, {James}, {Addison}, {Liang}, {Davis}, {Tronsgaard}, {Worku},
  {Brewer}, {K{\"u}rster}, {Zhang}, {Beichman}, {Bieryla}, {Brown},
  {Christiansen}, {Ciardi}, {Collins}, {Esquerdo}, {Howard}, {Isaacson},
  {Latham}, {Mazeh}, {Petigura}, {Quinn}, {Shahaf}, {Siverd}, {Rodler},
  {Reffert}, {Zakhozhay}, {Ricker}, {Vanderspek}, {Seager}, {Winn}, {Jenkins},
  {Boyd}, {F{\H{u}}r{\'e}sz}, {Henze}, {Levine}, {Morris}, {Paegert},
  {Stassun}, {Ting}, {Vezie}, \& {Laughlin}}]{Wang:2019}
{Wang}, S., {Jones}, M., {Shporer}, A., {et~al.} 2019, \aj, 157, 51

\bibitem[{{Weiss} {et~al.}(2013){Weiss}, {Marcy}, {Rowe}, {Howard}, {Isaacson},
  {Fortney}, {Miller}, {Demory}, {Fischer}, {Adams}, {Dupree}, {Howell},
  {Kolbl}, {Johnson}, {Horch}, {Everett}, {Fabrycky}, \& {Seager}}]{Weiss:2013}
{Weiss}, L.~M., {Marcy}, G.~W., {Rowe}, J.~F., {et~al.} 2013, \apj, 768, 14

\bibitem[{{W}es {M}c{K}inney(2010)}]{pandas:2010}
{W}es {M}c{K}inney. 2010, in {P}roceedings of the 9th {P}ython in {S}cience
  {C}onference, ed. {S}t\'efan van~der {W}alt \& {J}arrod {M}illman, 56 -- 61

\bibitem[{{Wolfgang} \& {Lopez}(2015)}]{Wolfgang:2015}
{Wolfgang}, A. \& {Lopez}, E. 2015, \apj, 806, 183

\bibitem[{{Wolfgang} {et~al.}(2016){Wolfgang}, {Rogers}, \&
  {Ford}}]{Wolfgang:2016}
{Wolfgang}, A., {Rogers}, L.~A., \& {Ford}, E.~B. 2016, \apj, 825, 19

\bibitem[{{Wu} \& {Lithwick}(2013)}]{Wu:2013}
{Wu}, Y. \& {Lithwick}, Y. 2013, \apj, 763, 13

\bibitem[{{Yadav} \& {Thorngren}(2017)}]{Yadav:2017}
{Yadav}, R.~K. \& {Thorngren}, D.~P. 2017, \apjl, 849, L12

\bibitem[{{Youdin} \& {Mitchell}(2010)}]{Youdin:2010}
{Youdin}, A.~N. \& {Mitchell}, J.~L. 2010, \apj, 721, 1113

\end{thebibliography}

% -----------------------------------------------------------
\begin{appendix}

\section{Supplemental Information}
\label{sec:appendix-info}

\begin{figure*}[ht!]
	\includegraphics[width=\textwidth]{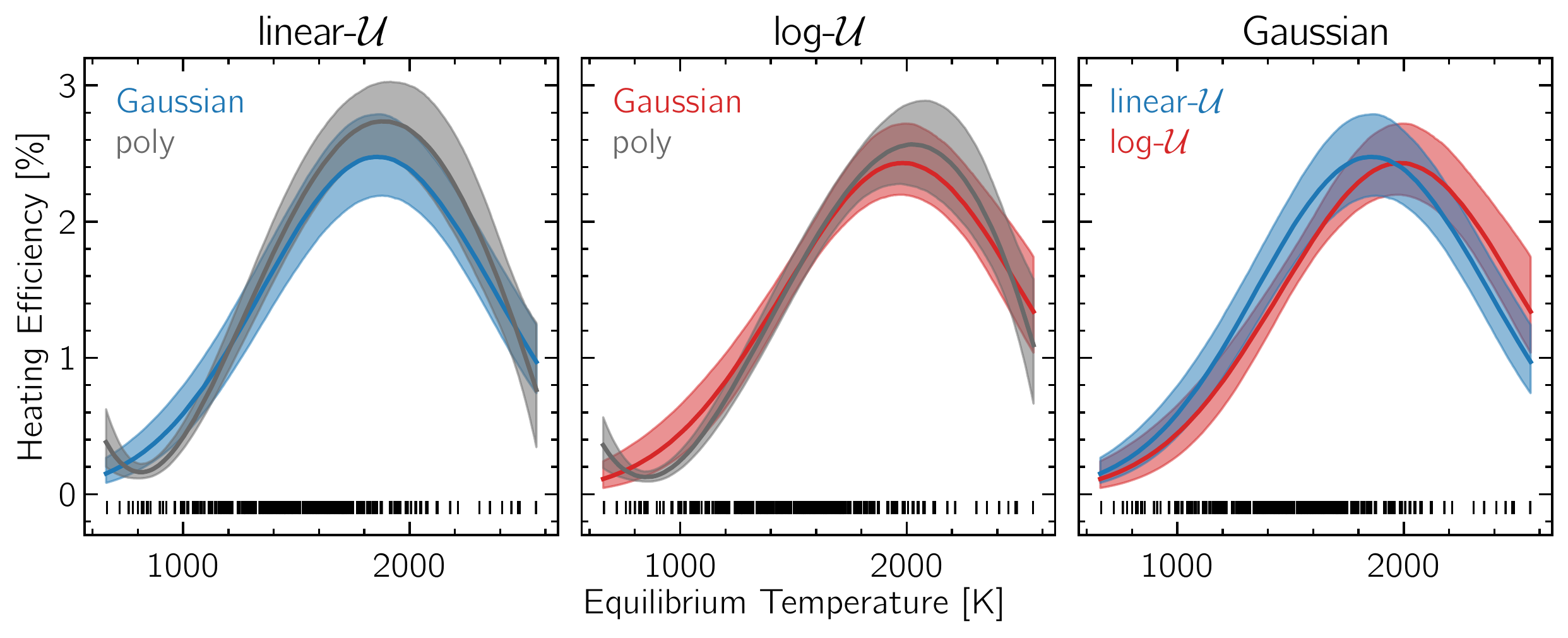}
	\caption{HEET posterior distribution under the 
		linear--uniform ({\it left}) and the log--uniform ({\it middle}) priors
		using a Gaussian and $4^{\rm th}$ degree polynomial.
		The shaded region shows the 68\% credible interval.
		There is a good agreement between both models using the same prior.
		To better compare the same model using different priors,
		the {\it right} panel shows the Gaussian models using 
		log (red) and linear (blue) uniform priors.
	\label{fig:HEET_appendix}} 
\end{figure*}

\begin{figure*}[t!]
	\includegraphics[width=\textwidth]{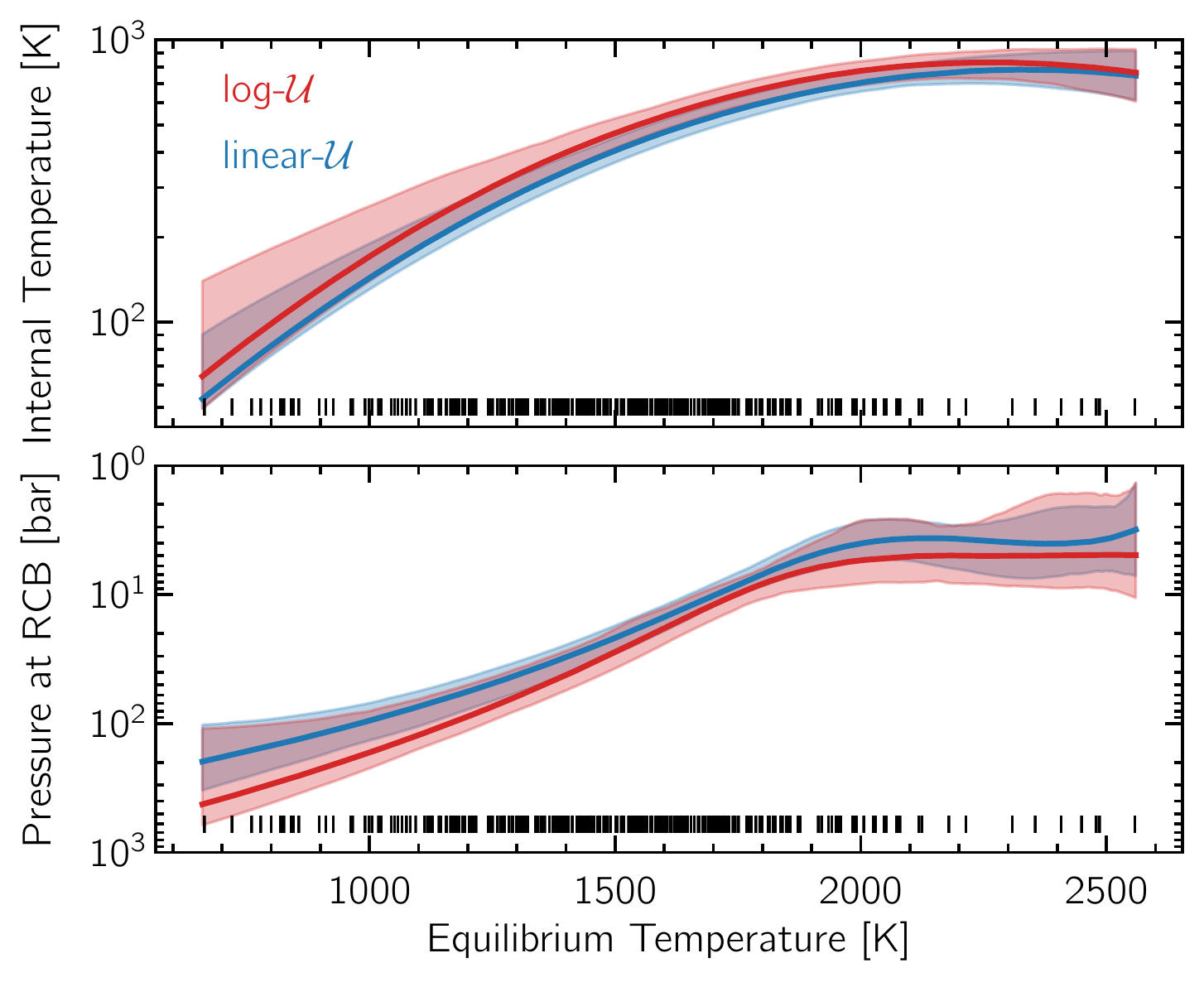}
	\caption{\tint--\teq\ and \prcb--\teq\ diagrams in the upper and lower panel, respectively.
				The shaded regions show the 95\% credible interval.
				Both distributions are similar at the 95\% level using the linear--uniform (red) and the log-uniform (blue) priors.
	\label{fig:TT_PT_results_compare_log_linear}}
\end{figure*}

In Section~\ref{sec:mlr}, we showed that the 
mass--luminosity--radius (MLR) posterior distribution 
is similar when assuming \lint\ follows either a linear-uniform or a log-uniform prior distribution.
In this Appendix we show that the 
heating-efficiency--equilibrium temperature (HEET), 
\tint\ -- \teq, and \prcb\ -- \teq\ 
distributions are also similar using both priors.
Figure~\ref{fig:HEET_appendix} and Figure~\ref{fig:TT_PT_results_compare_log_linear} 
show the HEET and both the \tint\ -- \teq\ and \prcb\ -- \teq\ distributions, respectively.
Tables~\ref{tab:results-mlr-linear} and \ref{tab:results-mlr-log}
present the 68\% credible interval values for the model parameters
for the MLR distribution assuming linear-uniform and log-uniform priors.
Similarly, Table~\ref{tab:results-heet-poly} for the HEET distribution
using a $4^{\rm th}$ degree polynomial, 
Table~\ref{tab:results-tt-gauss} for the \tint\ -- \teq\ distribution using a Gaussian function,
and finally Table~\ref{tab:results-pt-poly} for the \prcb\ -- \teq\ distribution
using a polynomial function.

\begin{table*}[ht!]
\caption{%
68\% credible interval values of the parameters for the mass--luminosity--radius (MLR) distribution 
for the {\bf linear--$\mathcal{U}$} case
modelled as a $4^{\rm th}$ degree polynomial
$ g_p \left(x \right) = a_0  + a_1 x + a_2 x^2 + a_3 x^3 + a_4 x^4 $
where $x=\rp$.
\label{tab:results-mlr-linear}}
\begin{center}
\begin{tabular}{cccccc}
\hline
\hline
\taw    &  $a_0$  &  $a_1$ & $a_2$ & $a_3$ & $a_4$ \\
\hline
 & & & & & \\

$0.37 - 0.7 \, \mjup$ & $28_{-10}^{+11}$ & $-85_{-31}^{+28}$ & $92_{-29}^{+29}$ & $-39_{-12}^{+12}$ & $6_{-2}^{+1}$ \\
 & & & & & \\
$0.7 - 0.98 \, \mjup$ & $27_{-13}^{+14}$ & $-91_{-44}^{+38}$ & $106_{-43}^{+46}$ & $-48_{-22}^{+20}$ & $8_{-4}^{+3}$ \\
 & & & & & \\
$0.98 - 2.5 \, \mjup$ & $48_{-13}^{+12}$ & $-160_{-38}^{+39}$ & $186_{-45}^{+40}$ & $-88_{-19}^{+21}$ & $15_{-4}^{+3}$ \\
 & & & & & \\
$> 2.5 \, \mjup$ & $72_{-54}^{+52}$ & $-224_{-198}^{+200}$ & $244_{-274}^{+276}$ & $-102_{-165}^{+161}$ & $14_{-36}^{+36}$ \\
 & & & & & \\

\hline
\end{tabular}
\end{center}
\end{table*}

\begin{table*}[ht!]
\caption{%
68\% credible interval values of the parameters for the mass--luminosity--radius (MLR) distribution  
for the {\bf log--$\mathcal{U}$} case
modelled as a $4^{\rm th}$ degree polynomial
$ g_p \left(x \right) = a_0  + a_1 x + a_2 x^2 + a_3 x^3 + a_4 x^4 $
where $x=\rp$.
\label{tab:results-mlr-log}}
\begin{center}
\begin{tabular}{cccccc}
\hline
\hline
\taw    &  $a_0$  &  $a_1$ & $a_2$ & $a_3$ & $a_4$ \\
\hline
 & & & & & \\
 
$0.37 - 0.7 \, \mjup$ & $20_{-7}^{+8}$ & $-66_{-22}^{+19}$ & $73_{-21}^{+22}$ & $-32_{-10}^{+9}$ & $5_{-1}^{+1}$ \\
 & & & & & \\
$0.7 - 0.98 \, \mjup$ & $23_{-10}^{+12}$ & $-79_{-39}^{+32}$ & $94_{-39}^{+43}$ & $-44_{-21}^{+19}$ & $7_{-4}^{+3}$ \\
 & & & & & \\
$0.98 - 2.5 \, \mjup$ & $50_{-14}^{+17}$ & $-166_{-51}^{+42}$ & $195_{-48}^{+54}$ & $-94_{-26}^{+22}$ & $16_{-4}^{+4}$ \\
 & & & & & \\
$> 2.5 \, \mjup$ & $86_{-70}^{+70}$ & $-272_{-269}^{+263}$ & $306_{-366}^{+370}$ & $-135_{-223}^{+219}$ & $19_{-48}^{+49}$ \\
 & & & & & \\
 
 \hline
\end{tabular}
\end{center}
\end{table*}

\begin{table*}[ht!]
\caption{%
68\% credible interval values of the parameters for the 
heating-efficiency--equilibrium temperature (HEET) distribution 
for the {\bf linear--$\mathcal{U}$} and  {\bf log--$\mathcal{U}$} cases
using the $4^{\rm th}$ degree polynomial model
$ g_p \left(x \right) = a_0  + a_1 x + a_2 x^2 + a_3 x^3 + a_4 x^4 $,
where $x= \teq/1000$.
\label{tab:results-heet-poly}}
\begin{center}
\begin{tabular}{cccccc}
\hline
\hline
\taw    &  $a_0$  &  $a_1$ & $a_2$ & $a_3$ & $a_4$ \\
\hline
 & & & & & \\
 
log-$\mathcal{U}$ & $11_{-5}^{+5}$ & $-33_{-17}^{+15}$ & $35_{-18}^{+19}$ & $-14_{-9}^{+8}$ & $2_{-1}^{+1}$ \\
 & & & & & \\
linear-$\mathcal{U}$ & $7_{-4}^{+4}$ & $-21_{-15}^{+13}$ & $19_{-16}^{+16}$ & $-6_{-8}^{+7}$ & $1_{-1}^{+1}$ \\
 & & & & & \\

\hline
\end{tabular}
\end{center}
\end{table*}

\begin{table}[ht!]
\caption{%
68\% credible interval values of the parameters for the \tint--\teq\ distribution 
for the {\bf linear--$\mathcal{U}$} and  {\bf log--$\mathcal{U}$} cases
using the Gaussian function Equation~(\ref{eq:gauss}),
where $x= \teq$ and \tint\ is in K.
\label{tab:results-tt-gauss}}
\begin{center}
\begin{tabular}{cccc}
\hline
\hline
\taw    &  \tint${_{\rm , max}}$  &  \tint$_{0}$ & $s$  \\
\hline
 & & & \\
  
log-$\mathcal{U}$ & $835_{-58}^{+69}$ & $2270_{-120}^{+202}$ & $709_{-70}^{+154}$ \\
 & & & \\
linear-$\mathcal{U}$ & $786_{-56}^{+85}$ & $2333_{-109}^{+149}$ & $723_{-46}^{+75}$ \\
 & & & \\
 
\hline
\end{tabular}
\end{center}
\end{table}

\begin{table}[ht!]
\caption{%
95\% credible interval values of the parameters for the \prcb--\teq\ distribution 
for the {\bf linear--$\mathcal{U}$} and  {\bf log--$\mathcal{U}$} cases
under the polynomial function Equation~(\ref{eq:poly}),
where $x= \teq/1000$ and $g_p / 100$ in bar.
\label{tab:results-pt-poly}}
\begin{center}
\begin{tabular}{cccccc}
\hline
\hline
\taw    &  $a_0$  &  $a_1$ & $a_2$ & $a_3$ & $a_4$ \\
\hline
 & & & \\
  
log-$\mathcal{U}$ & $16_{-13}^{+6}$ & $-29_{-15}^{+26}$ & $19_{-20}^{+12}$ & $-5_{-4}^{+6}$ & $0.59_{-0.85}^{+0.57}$ \\
 & & & & & \\
linear-$\mathcal{U}$ & $6_{-4}^{+6}$ & $-8_{-14}^{+9}$ & $4_{-8}^{+10}$ & $-1_{-4}^{+2}$ & $0.03_{-0.37}^{+0.48}$ \\
 & & & & & \\

\hline
\end{tabular}
\end{center}
\end{table}

\end{appendix}

\end{document}